\documentclass[floats,floatfix,
amssymb,prd,onecolumn,notitlepage,
superscriptaddress,nofootinbib]{revtex4-1}

\usepackage{amssymb,amsmath,verbatim,mathtools,
needspace,enumitem,etoolbox,graphicx,physics,microtype,afterpage,bm}

\usepackage[normalem]{ulem}

\usepackage[dvipsnames, usenames]{xcolor}

\definecolor{linkcolor}{rgb}{0.0,0.3,0.5}

\usepackage{hyperref}

\usepackage{graphics,appendix,afterpage,makecell}

\usepackage{bbold}
\usepackage{tikz}
\usepackage{adjustbox}
\usepackage{tikz-feynman}
\usepackage{tcolorbox}

\def\Im{\mbox{Im}\,}

\def\RE{{\rm Re}}
\def\IM{{\rm Im}}

\def\llp{\left [}

\def\rrp{\right ]}

\definecolor{oucrimsonred}{rgb}{0.6, 0.0, 0.0}
\definecolor{persianblue}{rgb}{0.11, 0.22, 0.73}
\definecolor{forestgreen}{rgb}{0.13,0.35,0.13}
\definecolor{lightgray}{rgb}{0.83, 0.83, 0.83}
 \hypersetup{colorlinks, citecolor=oucrimsonred, linkcolor=persianblue, urlcolor=oucrimsonred}
\definecolor{cornellred}{rgb}{0.7, 0.11, 0.11}
\definecolor{navyblue}{rgb}{0.0, 0.0, 0.5}
\definecolor{amethyst}{rgb}{0.6, 0.4, 0.8}
\definecolor{yellow}{rgb}{1.0, 1.0, 0.0}
\definecolor{firebrick}{rgb}{0.7, 0.13, 0.13}
\definecolor{tangerineyellow}{rgb}{1.0, 0.8, 0.0}
\definecolor{deepfuchsia}{rgb}{0.76, 0.33, 0.76}
\definecolor{amber}{rgb}{1.0, 0.75, 0.0}
\definecolor{VioletRed4}{rgb}{0.55, 0.13, .32}
\definecolor{indiagreen}{rgb}{0.07, 0.53, 0.03}
\definecolor{VioletRed4}{rgb}{0.55, 0.13, .32}
\newcommand{\be}{\begin{equation}}
\newcommand{\ee}{\end{equation}}
\newcommand{\bea}{\begin{eqnarray}}
\newcommand{\eea}{\end{eqnarray}}
\newcommand{\nn}{\nonumber}

\definecolor{oucrimsonred}{rgb}{0.6, 0.0, 0.0}
\newcommand\vertarrowbox[3][6ex]{%
  \begin{array}[t]{@{}c@{}} #2 \\
  \left\uparrow\vcenter{\hrule height #1}\right.\kern-\nulldelimiterspace\\
  \makebox[0pt]{\scriptsize#3}
  \end{array}%
}

\definecolor{mtcolor}{rgb}{.8,.3,.1}

\definecolor{violachiaro}{rgb}{1,0.6,1}

\definecolor{gbcolor}{rgb}{.43,.22,.12}
 
\definecolor{gbcolor2}{rgb}{.9,.2,.6}
\definecolor{gbcolor3}{rgb}{.3,.2,.6}

\definecolor{verdechiaro}{rgb}{0.6,1,0.6}
\definecolor{giallochiaro}{rgb}{1,1,0.6}
\definecolor{bluscuro}{rgb}{0.15, 0.2, 0.9}
\definecolor{verdes}{rgb}{0.1, 0.5, 0.1}%
\definecolor{tangerineyellow}{rgb}{1.0, 0.8, 0.0}
\definecolor{smokyblack}{rgb}{0.06, 0.05, 0.03}

\definecolor{americanrose}{rgb}{1.0, 0.01, 0.24}
\definecolor{cobalt}{rgb}{0.0, 0.28, 0.67}
\definecolor{brandeisblue}{rgb}{0.0, 0.44, 1.0}
\definecolor{mycolor}{rgb}{0.0, 0.0, 0.5}
\definecolor{oxfordblue}{rgb}{0.0, 0.13, 0.28}
\definecolor{azure}{rgb}{0.0, 0.5, 1.0}
\definecolor{turquoiseblue}{rgb}{0.0, 1.0, 0.94}
\newtcolorbox{mynewbox}[1]{colback=white!5!white,colframe=azure!75!black,fonttitle=\bfseries,title=#1}
\newtcolorbox{mybox}{colback=mycolor!5!white,colframe=azure!75!black}
\newtcolorbox{mynamedbox}[1]{colback=mycolor!5!white,colframe=azure!75!black,title=#1}
\definecolor{venetianred}{rgb}{0.78, 0.03, 0.08}
\newtcolorbox{mynamedbox1}[1]{colback=venetianred!5!white,colframe=venetianred!80!black,title=#1}
\newtcolorbox{mynamedbox2}[1]{colback=azure!5!white,colframe=azure!80!black,title=#1}

\definecolor{rossocorsa}{rgb}{0.83, 0.0, 0.0}

\tikzset{->-/.style={decoration={
  markings,
  mark=at position #1 with {\arrow{>}}},postaction={decorate}}}
\tikzset{-<-/.style={decoration={
  markings,
  mark=at position #1 with {\arrow{<}}},postaction={decorate}}} 

\newcommand{\uniroma}{Dipartimento di Fisica, Sapienza Università 
	di Roma, Piazzale Aldo Moro 5, 00185, Roma, Italy}
\newcommand{\infn}{INFN, Sezione di Roma, Piazzale Aldo Moro 2, 00185, Roma, Italy}
\newcommand{\infnTorino}{Istituto Nazionale di Fisica Nucleare, Sezione di Torino, via P. Giuria 1, I–10125 Torino,
Italy}

\begin{document}

\title[]{
One loop to rule them all:
\\
Perturbativity in the presence of ultra slow-roll dynamics
}

\date{\today}

\author{Gabriele Franciolini}
\email{gabriele.franciolini@uniroma1.it}
\affiliation{\uniroma}
\affiliation{\infn}
\author{Antonio Junior Iovino}
\email{antoniojunior.iovino@uniroma1.it}
\affiliation{\uniroma}
\affiliation{\infn}
\author{Marco Taoso}
\email{marco.taoso@to.infn.it}
\affiliation{\infnTorino}
\author{Alfredo Urbano}
\email{alfredo.urbano@uniroma1.it}
\affiliation{\uniroma}
\affiliation{\infn}

\begin{abstract}
\noindent 
We discuss the issue of perturbativity in single-field inflationary models with a phase of ultra slow-roll (USR) tailor suited to generate an order-one abundance of primordial black holes (PBHs). 
More in detail, we impose the condition 
that loop corrections 
made up of short-wavelength modes enhanced by the USR dynamics do not alter the tree-level power spectrum of curvature perturbations. 
In our analysis, 
the USR phase is preceded and followed by two stages of ordinary slow-roll (SR), and we model the resulting SR/USR/SR dynamics using both instantaneous and smooth transitions.  
Focusing on scales relevant for CMB observations, we find that it is not possible, with these arguments, to rule out the scenario of PBH formation via USR, not even in the limit of instantaneous transition. 
However, we also find that loop corrections of short modes on the power spectrum of long modes, even though not large enough to violate perturbativity requirements, 
remain appreciable and, most importantly, are not tamed in realistic realisations of smooth SR/USR/SR transitions. This makes perturbativity a powerful theoretical tool to constrain USR dynamics.
We extend the analysis at any scale beyond those relevant for CMB observations. We find that loop corrections of short modes remain within the few percent if compared to the tree-level power spectrum. However, we also find one notable exception of phenomenological relevance: we show that the so-called dip in the power spectrum of curvature perturbation is significantly reduced beyond the tree-level computation.
\end{abstract}
\maketitle

{

  \hypersetup{linkcolor=black}
  \tableofcontents
  
}

\section{Introduction}

In this work, we consider the so-called standard scenario of primordial black hole (PBH) formation \cite{Hawking:1971ei,Zeldovich:1967lct,Carr:1974nx}.
In this scenario, 
the formation of PBHs during the early Universe is an exceptional phenomenon in which extremely dense regions of radiation energy are tightly packed to the point of gravitational collapse \cite{Ivanov:1997ia,Garcia-Bellido:1996mdl,PhysRevD.50.7173}. 
General relativity and the inflationary stage preceding the radiation epoch offer a mechanism to generate such over-densities: small-scale curvature perturbations, stretched way beyond the horizon by the inflationary expansion,  are transferred to the radiation fluid after the end of inflation at around the time of their horizon re-entry. 
In order to truly trigger a gravitational collapse of the radiation fluid, the amplitude of such small-scale curvature perturbations needs to be greatly enhanced by some dynamics during the inflationary stage. 
This statement can be made more quantitative by introducing the dimensionless power spectrum  $\mathcal{P}(k)$ which gives the contribution to the variance of the curvature perturbation field 
per bin of $\log k$, with $k$ the comoving wavenumber in Fourier space. 
At scales relevant for Cosmic Microwave Background (CMB) observations (that is,  
$0.005 \lesssim k\,[\textrm{Mpc}^{-1}] \lesssim 0.2$)  we typically have $\mathcal{P}(k) = O(10^{-9})$; at smaller scales ($1.5\times 10^{13} \lesssim k\,[\textrm{Mpc}^{-1}] \lesssim 1.5\times 10^{14}$) asteroid-mass PBHs may comprise the totality of dark matter (DM) observed in the Universe but their formation requires  $\mathcal{P}(k) = O(10^{-2})$ (for recent reviews see \cite{Sasaki:2018dmp,Escriva:2022duf,Ozsoy:2023ryl,Carr:2020xqk,Green:2020jor}). 
Theory-side, therefore, what makes the formation of PBHs an exceptional phenomenon is the fact that it requires a seven-order-of-magnitude enhancement of the small-scale power spectrum with respect to the value observed at large scales.

In the context of single-field models of inflation, the above-mentioned enhancement can be dynamically realized by introducing a phase of ultra slow-roll (USR) \cite{Ballesteros:2020qam,Bhaumik:2020dor,Inomata:2016rbd,Iacconi:2021ltm,Kawai:2021edk,Bhaumik:2019tvl,Cheong:2019vzl,Inomata:2018cht,Dalianis:2018frf,Kannike:2017bxn,Hertzberg:2017dkh,Ballesteros:2017fsr,Garcia-Bellido:2017mdw,Karam:2022nym,Rasanen:2018fom,Balaji:2022rsy,Zhao:2023xnh,Frolovsky:2023hqd,Dimopoulos:2017ged,Germani:2017bcs,Choudhury:2013woa,Ragavendra:2023ret,Cheng:2021lif} during which the inflaton field, after the first conventional phase of slow-roll (SR) that is needed to fit large-scale cosmological observations, almost stops the descent along its potential (typically because of the presence of a quasi-stationary inflection point) before starting rolling down again in a final stage of SR dynamics that eventually ends inflation. In this work, we shall refer to this three-stage dynamics as SR/USR/SR.

A very legitimate question is whether the 
USR dynamics is consistent with perturbativity.  
Technically speaking, the dimensionless power spectrum of curvature perturbation $\mathcal{P}(k)$ is typically computed within the free theory. However, curvature perturbations, being gravitational in nature, feature an intricate architecture of non-linear interactions. The effect of non-linear interactions is twofold. 
On the one hand, they generate, in addition to the variance, non-zero higher-order cumulants that may leave a peculiar non-Gaussian pattern to the statistics of the curvature field. 
On the other hand, the variance itself gets corrected with respect to the value computed within the free theory. 
In this paper, we focus on the second effect and, on very general grounds, we define the perturbative criterion
\begin{align}
\mathcal{P}(k) \equiv \mathcal{P}_\text{tree}(k)
\left[1 + 
\Delta {\cal P}_\text{1-loop}(k) \right] ~~~\Longrightarrow~~~
\Delta {\cal P}_\text{1-loop}(k)\overset{!}{<} 1\,,\label{eq:Pertu}
\end{align}
meaning that the power spectrum computed within the free theory (the ``tree-level'' power spectrum in the above equation) must be larger than the corrections $\Delta\mathcal{P}$ introduced by the presence of interactions. Such corrections can be organized in a formal series expansion, and  
we will focus in particular on the first-order term, dubbed ``1-loop'' in the  above equation.

The gut feeling is that, unless one is led to consider $\mathcal{P}_{\rm tree}(k) = O(1)$, perturbativity should be under control.  
However, 
ref.\,\cite{Kristiano:2022maq} made the bold claim that, in the presence of USR, the perturbativity condition in eq.\,(\ref{eq:Pertu}) could be violated at scales $k$ relevant for 
CMB observations; even more strikingly, ref.\,\cite{Kristiano:2022maq}   
argues that USR dynamics  
tailored to generate a sizable abundance of asteroid- or solar-mass PBHs are ruled-out. 
What makes the claim of ref.\,\cite{Kristiano:2022maq} so hard to accept is that it basically says that 
loop of short modes alters the correlation of long CMB modes. 
This is counter-intuitive since it clashes with the intuition that physics at different scales should be decoupled. 

Given the above, it is not surprising that ref.\,\cite{Kristiano:2022maq} sparked an intense debate mostly exclusively polarized on defending or disproving the claim that PBH formation from single-field inflation is ruled out\,\cite{Riotto:2023hoz,Kristiano:2023scm,Riotto:2023gpm,Firouzjahi:2023aum,Firouzjahi:2023ahg,Choudhury:2023vuj,Choudhury:2023jlt,Choudhury:2023rks,Choudhury:2023hvf}. 
In this paper, we bury the hatchet and  critically examine the consequences of  
eq.\,(\ref{eq:Pertu}) in the presence of single-field inflation with USR dynamics. 

Our analysis is structured as follows.
In section\,\ref{sec:Compu}, we set the ground for the one-loop computation; 
in particular, we define all our conventions in section\,\ref{sec:Conve}, 
the SR/USR/SR background dynamics in section\,\ref{sec:MinDynUSR} and the interaction Hamiltonian in section\,\ref{sec:Cubo}.
In section\,\ref{sec:1LoopSec}, we compute the one-loop correction to the curvature power spectrum within the setup described in  section\,\ref{sec:Conve}; 
in particular,  in section\,\ref{sec:LoopHier},  we focus on the case in which there is a large hierarchy between the momenta running in the loop and the external ones while in section\,\ref{sec:any} we consider the case in which the external momenta are generic.   
In section\,\ref{sec:Pheno}, we discuss the implications of the perturbative bound in eq.\,(\ref{eq:Pertu}). In particular, in section\,\ref{sec:LoopCMB}, we consider the case in which the external momenta are long CMB modes. In this section, we critically compare our result with those of ref.\,\cite{Kristiano:2022maq}, and discuss a number of crucial generalization. In  section\,\ref{sec:LoopUSR}, we extend the computation to the case in which the external momenta are short modes.
Finally, we conclude in section\,\ref{sec:Discu}.

\section{Set-up of the computation using the ``in-in'' formalism}\label{sec:Compu}

\subsection{Conventions}\label{sec:Conve}

First, we set our conventions.
We set the reduced Planck mass to one; 
$t$ is the cosmic time (with $\dot{} \equiv d/dt$) and $\tau$ the conformal time (with ${}^\prime \equiv d/d\tau$) with $dt/d\tau = a$ being $a$ the scale factor of the flat Friedmann-Lema\^{\i}tre-Robertson-Walker (FLRW) metric $ds^2 = dt^2 - a^2(t)d\vec{x}^2$, with $\vec{x}$ comoving coordinates. The Hubble rate is $H\equiv \dot{a}/a$. The $e$-fold time $N$ is defined by $dN=Hdt$ from which we also have $dN/d\tau = aH$.  
The Hubble-flow parameters $\epsilon_{i}$ (for $i\geqslant 1$) are defined by the recursive relation
\begin{align}
\epsilon_{i} \equiv \frac{\dot{\epsilon}_{i-1}}{H\epsilon_{i-1}}\,,~~~~~\textrm{with:}~~~
\epsilon_0 \equiv \frac{1}{H}\,.\label{eq:HubblePar1}
\end{align}
As customary, we simply indicate as $\epsilon$ the first Hubble parameter, $\epsilon \equiv \epsilon_1 = -\dot{H}/H^2$. 
Instead of the second Hubble parameter $\epsilon_2$, sometimes it is useful to introduce the Hubble parameter $\eta$ 
defined by\footnote{We remark that in ref.\,\cite{Kristiano:2022maq} the symbol $\eta$ refers to the second Hubble parameter $\epsilon_2$.}
\begin{align}
\eta \equiv - \frac{\ddot{H}}{2H\dot{H}} 
= \epsilon - 
\frac{1}{2}\frac{d\log\epsilon}{dN}\,,~~~~~~
\textrm{with:}~~~~~\epsilon_2 = 2\epsilon - 2\eta\,.\label{eq:HubblePar2}
\end{align}

We consider the theory described by the action 
\begin{align}
\mathcal{S} =
\int d^4x \sqrt{-g}\bigg[
\frac{1}{2}R(g)
 + \frac{1}{2}g^{\mu\nu}
 (\partial_{\mu}\phi)
 (\partial_{\nu}\phi) - 
 V(\phi)
\bigg]
\,.\label{eq:BackAction}
\end{align}
$R(g)$ is the scalar curvature associated with the space-time whose geometry is described by the metric $g$ with line element 
$ds^2 = g_{\mu\nu}dx^{\mu}dx^{\nu}$.
The classical background evolves in the flat FLRW universe and the background value of the scalar field is a function of time, $\phi(t)$.  
We tacitly assume that the scalar potential features an approximate stationary inflection point so as to trigger the transition SR/USR/SR.

We focus on scalar perturbations. 
We consider the perturbed metric in the following generic form
\begin{align}
ds^2 = N^2dt^2 -
h_{ij}(N^i dt + dx^i)
(N^j dt + dx^j)\,,
\end{align}
and choose the gauge in which
\begin{align}
N = 1 + \delta N(\vec{x},t)\,,~~~~~~~~~
N^i = \delta^{ij}\partial_jB(\vec{x},t)\,,~~~~~~~~~
h_{ij} 
= a^2(t)e^{2\zeta(\vec{x},t)}\delta_{ij}\,,~~~~~~~~~
\delta\phi(\vec{x},t) = 0\,.\label{eq:Gauge}
\end{align}
The field $\zeta(\vec{x},t)$ is the only independent scalar degree of freedom since 
$N$ and $N^i$ are Lagrange multipliers subject to the momentum and Hamiltonian constraints. 
It is important to stress that the variable $\zeta$ as defined in eq.\,(\ref{eq:Gauge}) is constant outside the horizon (more in general, outside the horizon and after the end of possible non-adiabatic phases) and represents the correct non-linear generalization of the Bardeen variable\,\cite{Maldacena:2002vr}.

At the quadratic order in the fluctuations, the action is 
\begin{align}
\mathcal{S}_2 = 
\int d^4 x\,\epsilon\,a^3
\left[
\dot{\zeta}^2 - \frac{(\partial_k\zeta)(\partial^k\zeta)}{a^2}
\right]\,.\label{eq:MainQuad}
\end{align}
Comoving curvature perturbations are quantized by introducing the free operator 
\begin{align}
\hat{\zeta}(\vec{x},\tau) = 
\int\frac{d^3\vec{k}}{(2\pi)^3}
\hat{\zeta}(\vec{k},\tau)e^{i\vec{x}\cdot\vec{k}}\,,
~~~~~\textrm{with:}~~~
\hat{\zeta}(\vec{k},\tau)=
\zeta_{k}(\tau)a_{\vec{k}}  + 
\zeta_{k}^*(\tau)a^{\dag}_{-\vec{k}}\,,\label{eq:MainDef}
\end{align}
and 
\begin{equation}\label{eq:Anni}
[a_{\vec{k}},a_{\vec{k}^{\prime}}] =[a_{\vec{k}}^{\dag},a_{\vec{k}^{\prime}}^{\dag}] =0\,,~~~~~~~[a_{\vec{k}},a_{\vec{k}^{\prime}}^{\dag}] = (2\pi)^3\delta^{(3)}(\vec{k}-\vec{k}^{\prime})\,,~~~~~~~a_{\vec{k}}|0\rangle = 0\,,
\end{equation}
where the last condition defines the vacuum of the free theory $|0\rangle$. We define the comoving wavenumber $k\equiv |\vec{k}|$. 
The scale factor in the FLRW universe corresponds to a rescaling of the spatial coordinate; consequently, physically sensible results should be invariant under 
the rescaling\,\cite{Senatore:2009cf}
\begin{align}
a \to \lambda a\,,~~~~~~~
\vec{x} \to \vec{x}/\lambda\,,~~~~~~~
\vec{k} \to \lambda\vec{k}\,,~~~~~~
k \to |\lambda|k\,,~~~~~~
{\textrm{with}}~~\lambda \in \mathbb{R}\,.\label{eq:Rescaling}
\end{align}
Furthermore, if we consider the conformal time $\tau$ instead of the cosmic time $t$, we also have
\begin{align}
  \tau \to \tau/\lambda\,.  
\end{align}
Notice that, under the above rescaling, we have $a_{\vec{k}} \to a_{\vec{k}}/|\lambda|^{3/2}$ (from the scaling property of the three-dimensional $\delta$ function) and, consequently, $\zeta_k \to \zeta_k/|\lambda|^{3/2}$ so that $\hat{\zeta}(\vec{k},\tau) \to \hat{\zeta}(\vec{k},\tau)/|\lambda|^3$ and   $\hat{\zeta}(\vec{x},\tau)$ invariant.
In the case of free fields, we have 
\begin{align}
\langle 0|\hat{\zeta}(\vec{k}_1,\tau_1)
\hat{\zeta}(\vec{k}_2,\tau_2)|0\rangle =  
(2\pi)^3\delta(\vec{k}_1+\vec{k}_2)
\zeta_{k_1}(\tau_1)\zeta_{k_2}^*(\tau_2)\,.
\end{align}
In the presence of a time-derivative, we simply have 
\begin{align}
\langle 0|\hat{\zeta}^{\prime}(\vec{k}_1,\tau_1)
\hat{\zeta}(\vec{k}_2,\tau_2)|0\rangle =  
(2\pi)^3\delta(\vec{k}_1+\vec{k}_2)
\zeta^{\prime}_{k_1}(\tau_1)\zeta_{k_2}^*(\tau_2)\,.
\end{align}
Note that the time dependence occurs in the
mode function, not in the raising/lowering 
operator. The mode function $\zeta_{k}(\tau)$ is related to the linear-order Mukhanov-Sasaki (M-S) equation. More in detail, if we define $\zeta_k(\tau) = u_k(\tau)/z(\tau)$ with $z(\tau)\equiv a(\tau)\sqrt{2\epsilon(\tau)}$, the mode $u_k(\tau)$ verifies the equation 
\begin{align}
    \frac{d^2u_k}{d\tau^2} + \left(k^2 - \frac{1}{z}\frac{d^2z}{d\tau^2}\right)u_k = 0\,.\label{eq:EoM}
\end{align}
We are interested in the computation of the quantity 
\begin{align}
\lim_{\tau \to 0^-}\langle\hat{\zeta}(\vec{x}_1,\tau)
\hat{\zeta}(\vec{x}_2,\tau)\rangle = 
\int \frac{d^3\vec{k}}{(2\pi)^3}
P(k)e^{i\vec{k}\cdot (\vec{x}_1 - \vec{x}_2)}\,,\label{eq:TwoPointCorr}\end{align}
at some late time $\tau \to 0^-$ at which curvature perturbations become constant at super-horizon scales. Equivalently, we write
\begin{align}
\lim_{\tau \to 0^-}\langle\hat{\zeta}(\vec{x},\tau)
\hat{\zeta}(\vec{x},\tau)\rangle = 
\int \frac{dk}{k}\underbrace{\left[\frac{k^3}{2\pi^2}
P(k)\right]}_{\equiv \mathcal{P}(k)} 
= \int \frac{dk}{k}\mathcal{P}(k)\,,
\label{eq:CompaPS}
\end{align}
where $\mathcal{P}(k)$ is the a-dimensional power spectrum. 
At the level of the quadratic action, we find 
\begin{align}
\langle0|\hat{\zeta}(\vec{x}_1,\tau)
\hat{\zeta}(\vec{x}_2,\tau)|0\rangle & = 
\int\frac{d^3\vec{k}_1}{(2\pi)^3}\,
d^3\vec{k}_2 \delta(\vec{k}_1 + \vec{k}_2)\,
\zeta_{k_1}(\tau)\zeta_{k_2}^*(\tau)
e^{i(\vec{x}_1\cdot k_1 + \vec{x}_2\cdot k_2)} = \int\frac{d^3\vec{k}}{(2\pi)^3}
|\zeta_{k}(\tau)|^2 e^{i\vec{k}\cdot(\vec{x}_1 - \vec{x}_2)}\,,\label{eq:PSQuad}
\end{align}
which of course gives the familiar result
\begin{align}
\mathcal{P}(k) = 
\lim_{\tau \to 0^-}\frac{k^3}{2\pi^2}|\zeta_{k}(\tau)|^2\,.\label{eq:TreeLevel}
\end{align}
The goal is to compute corrections that arise from the presence of interactions. 
This means that, in 
eq.\,(\ref{eq:TwoPointCorr}), the vacuum expectation value should refer to the vacuum $|\Omega\rangle$ of the interacting theory and the dynamics of the operator $\hat{\zeta}(\vec{x},\tau)$ is described by the full action that also includes interactions.

We compute the left-hand side of eq.\,(\ref{eq:TwoPointCorr}) by means of the ``{\it in}-{\it in}'' formalism (see e.g.~\cite{Calzetta:1986ey,Jordan:1986ug,Adshead:2009cb}). Correlators are given by
\begin{align}
\langle\Omega|\hat{\mathcal{O}}(\tau)|\Omega\rangle
\equiv \langle\hat{\mathcal{O}}(\tau)\rangle = 
\langle 0|\left\{
\bar{T}\exp\left[{i\int_{-\infty(1+i\epsilon)}^{\tau}
d\tau^{\prime}\hat{H}_{\textrm{int}}(\tau^{\prime})}\right]
\right\}\hat{\mathcal{O}}_I(\tau)\left\{
T\exp\left[{-i\int_{-\infty(1-i\epsilon)}^{\tau}
d\tau^{\prime}\hat{H}_{\textrm{int}}(\tau^{\prime})}\right]
\right\}|0\rangle\,,\label{eq:MainInIn}
\end{align}
where on the right-hand side all fields appearing in $\hat{\mathcal{O}}_I(\tau)$ and $\hat{H}_{\textrm{int}}(\tau^{\prime})$ are free fields in the interaction picture. 
We shall indicate free fields in the interaction picture with the additional subscript $_{I}$. 
It should be noted that the latter are nothing but the operators of the free theory that we quantized in eq.\,(\ref{eq:MainDef}).
 
$T$ and $\bar{T}$ are the time and anti-time ordering operator, respectively.
As customary, the small imaginary deformation in the integration contour 
guarantees that $|\Omega\rangle\to |0\rangle$ as $\tau\to -\infty$ where $|\Omega\rangle$ is the vacuum of the interacting theory.
On the left-hand side, the operator $\hat{\mathcal{O}}(\tau)$ is the equal-time product of operators at different space
points, precisely like in eq.\,(\ref{eq:TwoPointCorr}). 
We expand in the interaction Hamiltonian, so we use the Dyson series
\begin{align}
T\exp\left[-i\int_{-\infty_-}^{\tau}
d\tau^{\prime}\hat{H}_{\textrm{int}}(\tau^{\prime})\right] = 1 -i\int_{-\infty_-}^{\tau}
d\tau^{\prime}\hat{H}_{\textrm{int}}(\tau^{\prime}) +
i^2
\int_{-\infty_-}^{\tau}d\tau^{\prime}
\int_{-\infty_-}^{\tau^{\prime}}d\tau^{\prime\prime}\hat{H}_{\textrm{int}}(\tau^{\prime})\hat{H}_{\textrm{int}}(\tau^{\prime\prime}) + \dots\,,\label{eq:DysonSeries}
\end{align}
where, for simplicity, we introduce the short-hand notation $\infty_{\pm}\equiv \infty(1\pm i\epsilon)$.
Each order in $\hat{H}_{\textrm{int}}$ is an interaction vertex, and carries both a time integral and the space integral (enclosed in the definition of $\hat{H}_{\textrm{int}}$) which in Fourier space enforces momentum conservation.

It is crucial to correctly identify the interaction Hamiltonian. Before proceeding in this direction, let us clarify our notation. We expand the action in the form
\begin{align}
\mathcal{S} = 
\int d^3\vec{x}dt\,
\underbrace{\mathcal{L}[\zeta(\vec{x},t),\dot{\zeta}(\vec{x},t),
\partial_k\zeta(\vec{x},t)]}_{
\equiv\,
\mathcal{L}[\zeta(\vec{x},t)]
} = 
\underbrace{\int d^3\vec{x}dt\,
\mathcal{L}_2(\vec{x},t)}_{\equiv\,\mathcal{S}_2} + 
\underbrace{\int d^3\vec{x}dt\,
\mathcal{L}_3[\zeta(\vec{x},t)]}_{\equiv\,\mathcal{S}_3} + 
\underbrace{\int d^3\vec{x}dt\,
\mathcal{L}_4[\zeta(\vec{x},t)]}_{\equiv\,\mathcal{S}_4}+\dots\,,\label{eq:DDefL}
\end{align}
with $\mathcal{S}_2$ defined in eq.\,(\ref{eq:MainQuad}). 
We also define (as a function of conformal time) 
\begin{align}
H_{\textrm{int}}^{(k)}(\tau) 
\equiv 
 \int d^3\vec{x}
 \,\mathcal{H}_k[\zeta(\vec{x},\tau)]
 ~~~~~\Longrightarrow~~~~~
 \hat{H}_{\textrm{int}}^{(k)}(\tau) 
\equiv 
 \int d^3\vec{x}
 \,\mathcal{H}_k[\hat{\zeta}_I(\vec{x},\tau)]
\,.\label{eq:Hk}
\end{align}
At the cubic order, we simply have
\begin{align}
\mathcal{H}_3[\zeta(\vec{x},\tau)] 
= - \mathcal{L}_3[\zeta(\vec{x},\tau)]\,.
\end{align}
We shall construct the relevant cubic interaction Hamiltonian in section\,\ref{sec:Cubo}.
At the quartic order, simply writing $\mathcal{H}_4 = - \mathcal{L}_4$ does not capture the correct result if the cubic Lagrangian features interactions that depend on the time derivative of $\zeta$ since the latter modify the definition of the conjugate momentum. 

Using, at the operator level, the notation introduced in eq.\,(\ref{eq:Hk}), we schematically write at the first order in the Dyson series expansion
 \begin{align}
\langle\hat{\zeta}&(\vec{x}_1,\tau)
\hat{\zeta}(\vec{x}_2,\tau)\rangle_{1^{\textrm{st}}} = \nn\\
&
\langle 0|\hat{\zeta}_I(\vec{x}_1,\tau)
\hat{\zeta}_I(\vec{x}_2,\tau)\bigg[
-i\int_{-\infty_-}^{\tau}
d\tau^{\prime}\hat{H}^{(4)}_{\textrm{int}}(\tau^{\prime})
\bigg]|0\rangle + 
\langle 0|
\bigg[
i\int_{-\infty_+}^{\tau}
d\tau^{\prime}\hat{H}^{(4)}_{\textrm{int}}(\tau^{\prime})
\bigg]
\hat{\zeta}_I(\vec{x}_1,\tau)
\hat{\zeta}_I(\vec{x}_2,\tau)|0\rangle\,.\label{eq:Shem1}
\end{align}
At the first order, therefore, the first non-zero quantum  correction involves the quartic Hamiltonian.  
At the second order in the Dyson series expansion and considering again terms with up to eight fields in the vacuum expectation values, we write schematically
\begin{align}
\langle\hat{\zeta}(\vec{x}_1,\tau)
\hat{\zeta}(\vec{x}_2,\tau)\rangle_{2^{\textrm{nd}}} & =  \langle 0|\hat{\zeta}_I(\vec{x}_1,\tau)
\hat{\zeta}_I(\vec{x}_2,\tau)\bigg[-
\int_{-\infty_-}^{\tau}d\tau^{\prime}
\int_{-\infty_-}^{\tau^{\prime}}d\tau^{\prime\prime}\hat{H}^{(3)}_{\textrm{int}}(\tau^{\prime})
\hat{H}^{(3)}_{\textrm{int}}(\tau^{\prime\prime})\bigg]|0\rangle 
\nn\\
& +
\langle 0|
\bigg[-
\int_{-\infty_+}^{\tau}d\tau^{\prime}
\int_{-\infty_+}^{\tau^{\prime}}d\tau^{\prime\prime}\hat{H}^{(3)}_{\textrm{int}}(\tau^{\prime\prime})
\hat{H}^{(3)}_{\textrm{int}}(\tau^{\prime})\bigg]
\hat{\zeta}_I(\vec{x}_1,\tau)
\hat{\zeta}_I(\vec{x}_2,\tau)|0\rangle
\nn\\
& +
\langle 0|
\bigg[
i\int_{-\infty_+}^{\tau}
d\tau^{\prime}\hat{H}^{(3)}_{\textrm{int}}(\tau^{\prime})
\bigg]
\hat{\zeta}_I(\vec{x}_1,\tau)
\hat{\zeta}_I(\vec{x}_2,\tau)
\bigg[
-i\int_{-\infty_-}^{\tau}
d\tau^{\prime\prime}\hat{H}^{(3)}_{\textrm{int}}(\tau^{\prime\prime})
\bigg]
|0\rangle\,.\label{eq:Shem2}
\end{align}
The vacuum expectation values of interacting-picture fields can be computed using Wick's theorem.
Schematically, eqs.\,(\ref{eq:Shem1},\,\ref{eq:Shem2}) 
give rise to the following 
connected diagrams.
\begin{align}
\begin{adjustbox}{max width=1\textwidth}
\raisebox{-8mm}{
 \begin{tikzpicture}
	\draw[thick,dashed][thick] (-3.5,0)--(3.5,0);
    \node at (4,0.2) {\scalebox{1}{$\tau$}};
    \node at (-3,0.5) {\scalebox{1}{$(\vec{x}_1,\tau)$}};
    \node at (+3,0.5) {\scalebox{1}{$(\vec{x}_2,\tau)$}};
    \draw[thick,dashed][thick] (-3.5,-2.19)--(3.5,-2.19);
    \node at (4,-2) {\scalebox{1}{$\tau_1$}};
    \draw[thick][thick] (-3,0)--(0.5,-2.19);
    \draw[thick][thick] (3,0)--(0.5,-2.19);
    \draw[black,thick] (0.5,-3.74)circle(45pt); 
    \draw[black,fill=venetianred,thick] (0.5,-2.19)circle(3pt);
        \node at (0.4,-2.8) {\scalebox{1}{{\color{venetianred}{$\hat{H}^{(4)}_{\textrm{int}}(\tau_1)$}}}};    
    \draw[black,fill=verdechiaro,thick] (-3,0)circle(3pt);
    \draw[black,fill=verdechiaro,thick] (+3,0)circle(3pt); 
	\end{tikzpicture}
    \hspace{1.cm}
	\begin{tikzpicture}
	\draw[thick,dashed][thick] (-3.5,0)--(3.5,0);
    \node at (4,0.2) {\scalebox{1}{$\tau$}};
    \node at (-3,0.5) {\scalebox{1}{$(\vec{x}_1,\tau)$}};
    \node at (+3,0.5) {\scalebox{1}{$(\vec{x}_2,\tau)$}};
    \draw[thick,dashed][thick] (-3.5,-2)--(3.5,-2);
    \node at (4,-2+0.28) {\scalebox{1}{$\tau_1$}};
    \draw[thick,dashed][thick] (-3.5,-4.5)--(3.5,-4.5);
    \node at (4,-4.5+0.28) {\scalebox{1}{$\tau_2$}};
    \draw[thick][thick] (-3,0)--(-2.1,-2);
    \draw[thick][thick] (3,0)--(+2.3,-4.5);
   \draw[thick](-2.1,-2)..controls(1,-2.5)and(1,-2.5)..(+2.3,-4.5);
   \draw[thick](-2.1,-2)..controls(-1,-4)and(-1,-4)..(+2.3,-4.5);
    \draw[black,fill=verdechiaro,thick] (-3,0)circle(3pt);
    \draw[black,fill=verdechiaro,thick] (+3,0)circle(3pt);
    \draw[black,fill=azure,thick] (-2.1,-2)circle(3pt);
    \draw[black,fill=azure,thick] (+2.3,-4.5)circle(3pt);   
    \node at (-2.65,-2.5) {\scalebox{1}{{\color{azure}{$\hat{H}^{(3)}_{\textrm{int}}(\tau_1)$}}}};
        \node at (2.5,-5) {\scalebox{1}{{\color{azure}{$\hat{H}^{(3)}_{\textrm{int}}(\tau_2)$}}}};
	\end{tikzpicture}
    \hspace{1.cm}
 	\begin{tikzpicture}
	\draw[thick,dashed][thick] (-3.5,0)--(3.5,0);
    \node at (4,0.2) {\scalebox{1}{$\tau$}};
    \node at (-3,0.5) {\scalebox{1}{$(\vec{x}_1,\tau)$}};
    \node at (+3,0.5) {\scalebox{1}{$(\vec{x}_2,\tau)$}};
    \draw[thick,dashed][thick] (-3.5,-1.75)--(3.5,-1.75);
    \node at (4,-2+0.28) {\scalebox{1}{$\tau_1$}};
    \draw[thick,dashed][thick] (-3.5,-3)--(3.5,-3);
    \node at (4,-3) {\scalebox{1}{$\tau_2$}}; 
    \draw[thick][thick] (-3,0)--(0,-1.75);
    \draw[thick][thick] (+3,0)--(0,-1.75);
    \draw[thick][thick] (0,-1.75)--(0,-3); 
    \draw[black,thick] (0,-4.2)circle(35pt);    
    \draw[black,fill=azure,thick] 
    (0,-1.75)circle(3pt);     \draw[black,fill=azure,thick] 
    (0,-3)circle(3pt); 
    \node at (-0.85,-2.1) {\scalebox{1}{{\color{azure}{$\hat{H}^{(3)}_{\textrm{int}}(\tau_1)$}}}};
    \node at (0.,-3.5) {\scalebox{1}{{\color{azure}{$\hat{H}^{(3)}_{\textrm{int}}(\tau_2)$}}}};
    \draw[black,fill=verdechiaro,thick] (-3,0)circle(3pt);
    \draw[black,fill=verdechiaro,thick] (+3,0)circle(3pt);
	\end{tikzpicture}
 }
\end{adjustbox} \label{eq:LoopsSchematic}
\end{align}
From the above classification, we see that, at the same loop order, we have three classes of connected diagrams that, in principle, should be discussed together. Notice that, contrary to the first two, the last diagram is not of 1-Particle-Irreducible (1PI) type since it consists of a tadpole attached to a two-point propagator. 

To proceed further, we need to specify the background dynamics, that shape the time evolution of the Hubble parameters, and the interaction Hamiltonian, that specifies which terms in the Dyson expansion contribute at a given perturbative order.   

\subsection{The (minimal) dynamics of ultra slow-roll}\label{sec:MinDynUSR}

We start with a discussion of the USR dynamics.
\begin{figure}[h]
\begin{center}
\includegraphics[width=.8\textwidth]{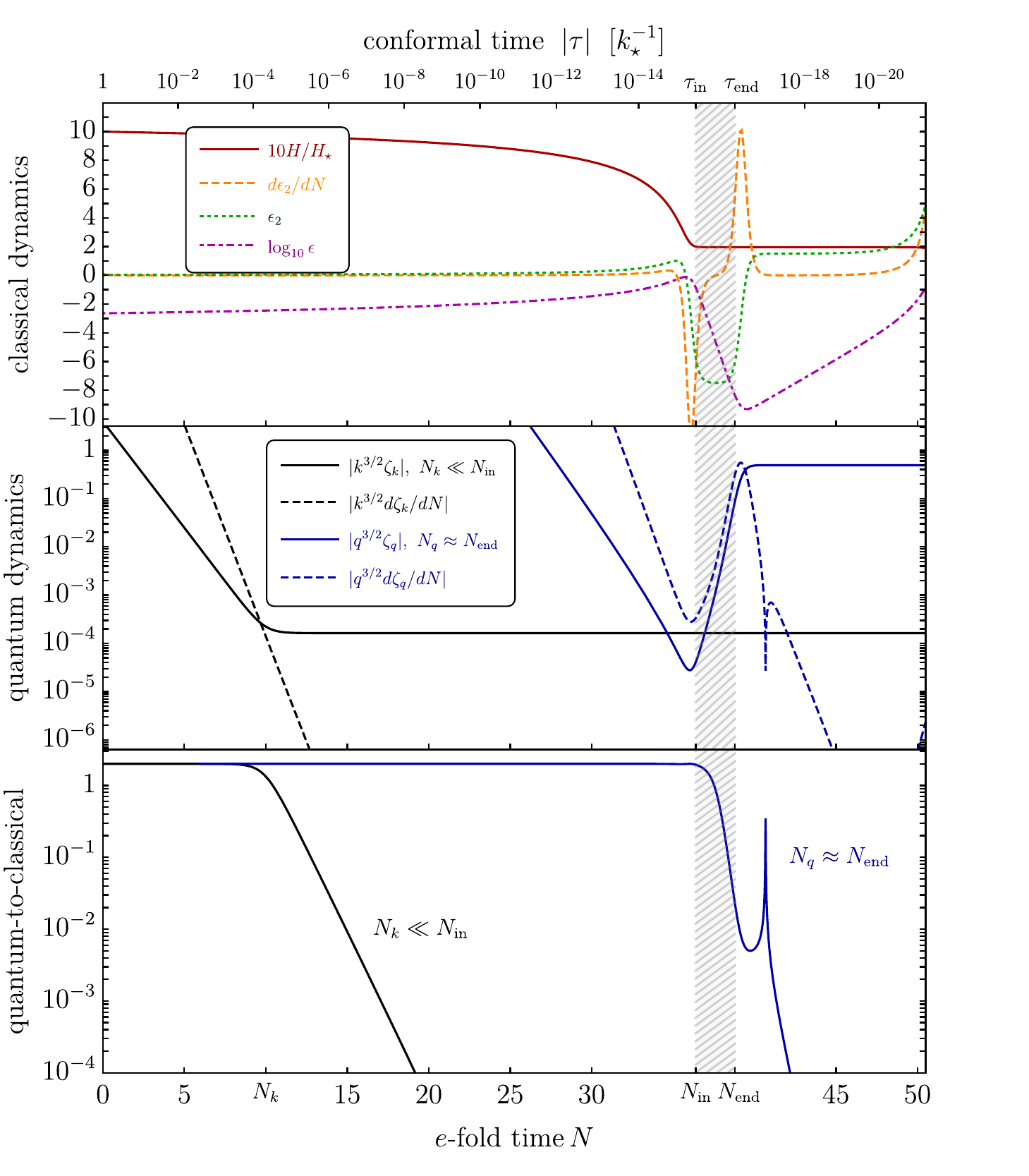}
\caption{\em 
Classical (top panel) and quantum (central panel) dynamics in the context of an explicit single-field model of inflation  that exhibits the presence of a phase of USR in between the time interval $N_{\textrm{\rm in}} < N < N_{\textrm{\rm end}}$ (cf. ref.\,\cite{Ballesteros:2020qam}).  In this specific realization, 
we have $N_{\textrm{\rm in}} = 36.3$ and
$N_{\textrm{\rm end}} = 38.8$. 
\textbf{\textit{Top panel:}} we plot the evolution of the background quantities $\epsilon$, $\epsilon_2$ and $\epsilon_2^{\prime}$ (cf. eqs.\,(\ref{eq:HubblePar1},\,\ref{eq:HubblePar2})) together with the evolution of the Hubble rate (normalized with respect to the value $H_*\equiv H(N_*)$ and scaled by a factor 10 for ease of comparison).
\textbf{\textit{Central panel:}} we plot the solutions of the M-S equation in eq.\,(\ref{eq:M-S}) for two different curvature modes. 
The mode in black (blue) exits the Hubble horizon well before (during) the USR phase. 
\textbf{\textit{Bottom panel:}} we plot the  classicality parameter $C_k$ for the same two modes (cf. the main text for details). 
In the case of the black mode ($N_k\ll N_{\rm in}$) the classicality parameter quickly vanishes after horizon crossing, and remains negligible also during the USR phase. In the case of the blue mode ($N_q\approx N_{\rm end}$) the classicality parameter remains sizable during the USR phase, signalling that this mode retains its quantum nature during USR.   
 }\label{fig:Schematic}  
\end{center}
\end{figure}
In order to make our discussion more concrete, in the following we shall refer to fig.\,\ref{fig:Schematic}, see the caption for details. In this figure, we plot both the classical (top panel) and quantum (bottom panel) dynamics that characterize a realistic model of single-field inflation that features a phase of USR because of the presence of an approximate stationary inflection point in the inflationary potential. We refer to ref.\,\cite{Ballesteros:2020qam} for more details about the model; we remark that this model, without including loop corrections to the computation of the curvature power spectrum, is compatible with CMB constraints and gives $\approx 100\%$ of dark matter in the form of asteroid-mass PBHs.\footnote{The model predicts the  tensor-to-scalar ratio $r\simeq 0.037$ which is still (barely) compatible, at  95\% confidence level, with the latest results released by the BICEP and Keck collaboration\,\cite{BICEP:2021xfz}.} 
In fig.\,\ref{fig:Schematic}, the relation between $e$-fold time $N$ (bottom $x$-axis) and conformal time $\tau$ (top $x$-axis) is given by the integral
\begin{align}
\tau = - \frac{1}{k_{\star}}
\int_{N}^{N_{0}}
\frac{H(N_{\star})}{H(N^{\prime})}e^{N_{\star}-N^{\prime}}dN^{\prime}\,,
\end{align}
where $N_0$ indicates the end of inflation, with $\tau$ conventionally set to $0$ at this time, and $N_{\star}$ the instant of time at which the comoving scale $k_{\star} = 0.05$ Mpc$^{-1}$ crosses the Hubble horizon. In fig.\,\ref{fig:Schematic}, we set $N_{\star} = 0$, and the model gives $N_0 - N_{\star} \simeq 52$.
We can highlight few crucial properties of the dynamics presented above:
\begin{itemize}
    \item[{\it a)}] We start from the classical analysis. During USR, 
$\epsilon_2(\tau)$ changes according to the schematic 
\begin{align}
\epsilon_2\approx 0 
\ \ 
\overset{\textrm{SR/USR at time }\tau_{\textrm{in}}}{\Longrightarrow}
\ \ 
|\epsilon_2| > 3 
\ \ 
\overset{\textrm{USR/SR at time }\tau_{\textrm{end}}}
{\Longrightarrow} 
\ \ 
\epsilon_2\approx O(1)\,,
\end{align}
thus making $\epsilon_2^{\prime}(\tau)$ non-zero at around the two transitions at conformal times $\tau_{\textrm{in}}$ and $\tau_{\textrm{end}}$ (equivalently, at $e$-fold times $N_{\textrm{in}}$ and $N_{\textrm{end}}$). 
The evolution of $\epsilon_2$ and $\epsilon^{\prime}_2$ are shown in the top panel of fig.\,\ref{fig:Schematic}, with dotted green and dashed orange lines, respectively. 

\item[{\it b)}] During USR, 
the Hubble parameter $\epsilon$  decreases exponentially fast (the inflaton almost stops its classical motion). 
The evolution of $\epsilon$ is shown in the top panel of fig.\,\ref{fig:Schematic} (dot-dashed magenta line); in addition, we also plot the time evolution of the Hubble rate $H$. 

 \item[{\it c)}] We now consider the USR dynamics at the quantum level. 
 It is crucial to understand the typical behaviour of curvature modes 
 (solid lines in the central panel of fig.\,\ref{fig:Schematic}) and their time derivatives (dashed lines). 
 In the central panel of fig.\,\ref{fig:Schematic}, we plot two representative cases: the black lines correspond to the case of a mode $\zeta_k$ that exits the Hubble horizon at some time $N_k$ well before the USR phase (like a CMB mode) while the blue lines correspond to a curvature mode $\zeta_q$ that exits the Hubble horizon at some time $N_q$ during the USR phase.
We notice that the derivative $|d\zeta_k/dN|$ decays exponentially fast, 
 and, soon after horizon crossing, becomes negligible, while $|\zeta_k|$ settles to a constant value. 
 Consequently, we expect that interaction terms that involve the time derivative of CMB modes will be strongly  suppressed. 

 \item[{\it d)}] Finally, we consider the issue of the quantum-to-classical transition.  We define the so-called classicality parameter\,\cite{Assassi:2012et} 
$C_k = |\zeta_k\dot{\zeta}_k^* -  
\zeta_k^*\dot{\zeta}_k|/
|\zeta_k\dot{\zeta}_k|$ which goes to zero in the classical limit. 
In the case of conventional SR dynamics, 
the  classicality parameter scales according to $C_k \sim 2k\tau$, and it vanishes exponentially fast right after the horizon crossing time. 
In the bottom panel of fig.\,\ref{fig:Schematic}, we plot the classicality parameter for two representative modes of our dynamics. 
The black modes experiences its horizon crossing well before the USR phase ($N_k\ll N_{\rm in}$). Its classicality parameter quickly vanishes and remains $\ll 1$ during the subsequent USR phase. The blue line, on the contrary, represents the classicality parameter for a mode that experiences its horizon crossing during the USR phase. Its classicality parameter remains sizable during the USR phase, signalling that this short mode can not be treated classically during USR.

\end{itemize}

With the aim of facilitating the numerical computations of the following sections, instead of working with a numerical description of USR, we now introduce a simple semi-analytical model\,\cite{Byrnes:2018txb,Taoso:2021uvl,Franciolini:2022pav}. 
We define the hyperbolic tangent parametrization
\begin{align}
\eta(N)  & = \frac{1}{2}\left[
-\eta_{\rm II}+ \eta_{\rm II}\tanh\left(\frac{N-N_{\rm in}}{\delta N}\right)
\right] + \frac{1}{2}\left[
\eta_{\rm II} + \eta_{\rm III} + (\eta_{\rm III}-\eta_{\rm II})\tanh\left(\frac{N-N_{\rm end}}{\delta N}\right)
\right]\,,\label{eq:DynEta}
\end{align}
where the parameter $\delta N$ controls the width of the two transitions at $N_{\rm in}$ and $N_{\rm end}$. 
The limit $\delta N\to 0$ reproduces the step-function approximation. 
Using the definition
\begin{align}
\delta(x) = \lim_{\epsilon\to 0} \frac{1}{2\epsilon\cosh^2(x/\epsilon)}\,,
\end{align}
we find
\begin{align}
\lim_{\delta N \to 0}
\frac{d\eta}{dN} = 
(-\eta_{\rm II} 
+  \eta_{\rm III})\delta(N-N_{\rm end}) 
 +\eta_{\rm II}\delta(N-N_{\rm in})\,.\label{eq:DeltaDer}
\end{align}
Using $\eta \simeq  - (1/2)d\log\epsilon/dN$, we find the following expression
\begin{align}
&
\frac{\epsilon(N)}{\epsilon_{\textrm{ref}}} = 
e^{-\eta_{\rm III}(N-N_{\rm ref})}\left[
\cosh\left(\frac{N-N_{\rm end}}{\delta N}\right)\cosh\left(\frac{N-N_{\rm in}}{\delta N}\right)
\right]^{-\frac{\delta N\eta_{\rm III}}{2}}
\left[
\cosh\left(\frac{N_{\rm ref}-N_{\rm end}}{\delta N}\right)\cosh\left(\frac{N_{\rm ref}-N_{\rm in}}{\delta N}\right)
\right]^{\frac{\delta N\eta_{\rm III}}{2}} \nn\\
&
\left[
\cosh\left(\frac{N-N_{\rm end}}{\delta N}\right){\rm sech}\left(\frac{N-N_{\rm in}}{\delta N}\right)
\right]^{\delta N\left(\eta_{\rm II} - \frac{\eta_{\rm III}}{2}\right)}
\left[
\cosh\left(\frac{N_{\rm ref}-N_{\rm end}}{\delta N}\right){\rm sech}\left(\frac{N_{\rm ref}-N_{\rm in}}{\delta N}\right)
\right]^{\frac{\delta N}{2}(-2\eta_{\rm II} + \eta_{\rm III})},\,\label{eq:DynEps}
\end{align}
where $\epsilon_{\textrm{ref}}\ll 1$ is the value of $\epsilon$ at some initial reference time $N_{\textrm{ref}}$. 
For future reference, we define $\bar{\epsilon}(N)\equiv \epsilon(N)/\epsilon_{\textrm{ref}}$.
\begin{figure}[h]
\begin{center}
$$\includegraphics[width=.33\textwidth]{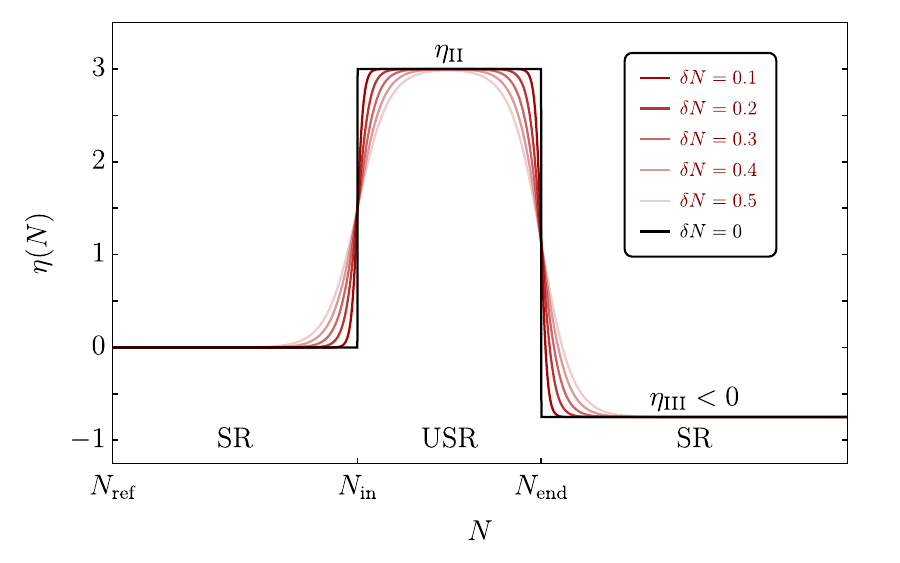}~
\includegraphics[width=.33\textwidth]{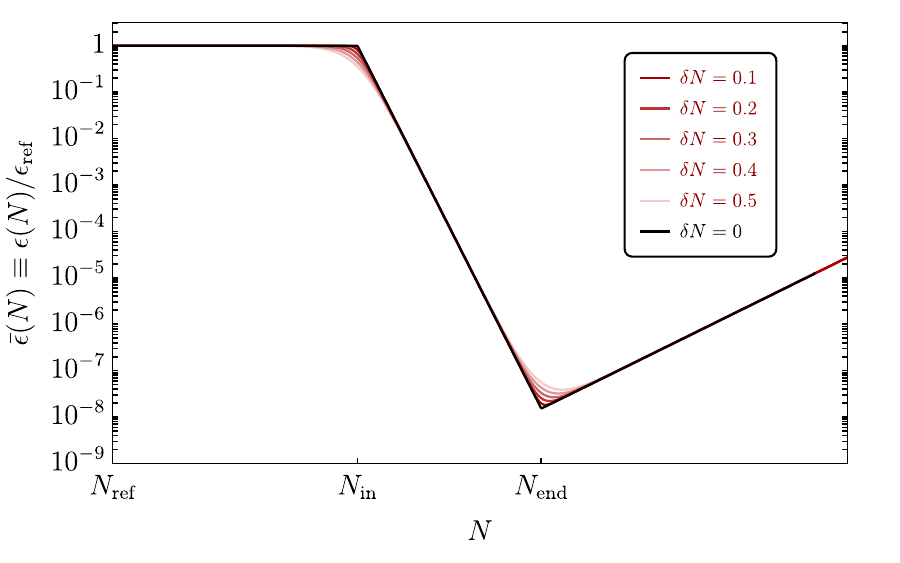}~
\includegraphics[width=.33\textwidth]{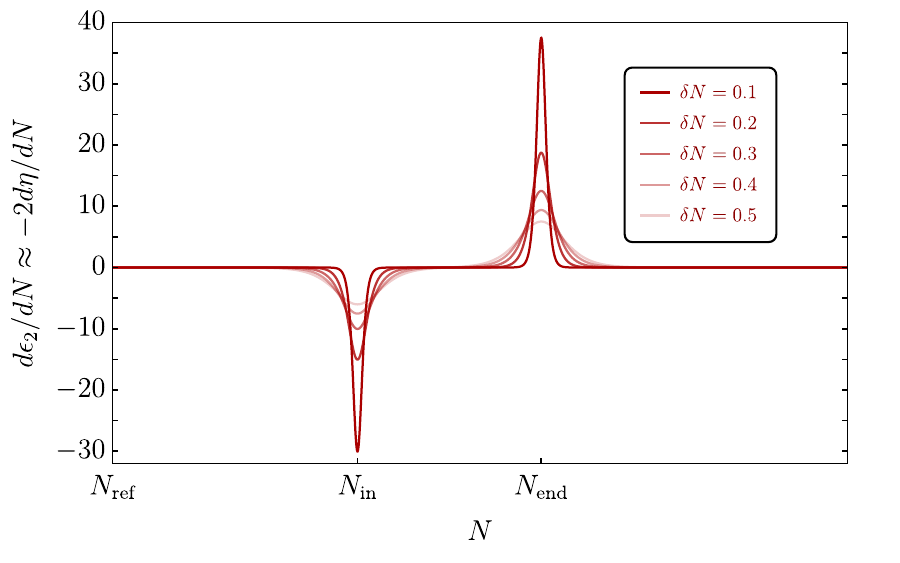}$$\vspace{-0.5cm}
\caption{\em 
Schematic evolution of $\eta(N)$ in eq.\,(\ref{eq:DynEta}) (\textbf{\textit{left panel}}), $\epsilon(N)$ in eq.\,(\ref{eq:DynEps}) (\textbf{\textit{central panel}}) 
and $d\epsilon_2/dN$ (\textbf{\textit{right panel}})
as function of the number 
of $e$-folds $N$. We explore different values of $\delta N$ with the limit $\delta N\to 0$ that corresponds to instantaneous transitions SR/USR at $N=N_{\textrm{\rm in}}$ and USR/SR at $N=N_{\textrm{\rm end}}$. 
In the right panel, the limit $\delta N \to 0$ corresponds to  $\delta$ function transitions at $N_{\textrm{\rm in}}$ and $N_{\textrm{\rm end}}$. Furthermore, notice that the lines corresponding to $d\epsilon_2/dN$ and $-2d\eta/dN$ superimpose, showing that $-2d\eta/dN$ is a perfect approximation of $d\epsilon_2/dN$. 
 }\label{fig:Dyn}  
\end{center}
\end{figure}
In this way we have an analytical description of the background dynamics; most importantly, eqs.\,(\ref{eq:DynEta},\,\ref{eq:DynEps}) are almost all that we need to know to solve
the M-S equation \cite{Sasaki:1986hm,Mukhanov:1988jd}
\begin{align}\label{eq:M-S}
\frac{d^2 u_k}{dN^2} &+ (1-\epsilon)\frac{du_k}{dN} + 
\left[
\frac{k^2}{(aH)^2} + (1+\epsilon-\eta)(\eta - 2) - \frac{d}{dN}(\epsilon - \eta)
\right]u_k = 0\,.
\end{align}
For consistency with eq.\,(\ref{eq:DynEps}), we consider the Hubble rate as a function of time according to 
\begin{align}
H(N) = H(N_{\textrm{ref}})\exp\left[
-\int_{N_{\textrm{ref}}}^{N}
\epsilon(N^{\prime})dN^{\prime}
\right]\,.\label{eq:HubbleRatewithTime}
\end{align}
We shall use the short-hand notation 
$a(N_{\textrm{i}}) \equiv a_{\textrm{i}}$ and 
$H(N_{\textrm{i}}) \equiv H_{\textrm{i}}$. 
Consequently, 
we rewrite eq.\,(\ref{eq:M-S}) in the form 
\begin{align}\label{eq:MSimpl}
\frac{d^2 u_k}{dN^2} &+ (1-\epsilon)\frac{du_k}{dN} + 
\left[
\bar{k}^2 \bigg(\frac{H_{\textrm{in}}}{H}\bigg)^2 e^{2(N_{\textrm{in}} - N)} + (1+\epsilon-\eta)(\eta - 2) - \frac{d}{dN}(\epsilon - \eta)
\right]u_k = 0\,,
\end{align}
with $\bar{k} \equiv 
{k}/({a_{\textrm{in}}H_{\textrm{in}}})$.
We solve the above equation for different $\bar{k}$ with Bunch-Davies initial conditions
\begin{align}
 \sqrt{k}\,u_k(N) = \frac{1}{\sqrt{2}}\,,~~~~~~~~~~~~  
 \sqrt{k}\,\frac{d u_k}{dN}(N) = 
 -\frac{i}{\sqrt{2}}\frac{k}{a(N)H(N)}
 \,,
\end{align}
at some arbitrary time $N \ll N_k$ with $k = a(N_k)H(N_k)$. 
Modes with $\bar{k} \approx O(1)$ are modes that exit the Hubble horizon at about the beginning of the USR phase,
modes with $\bar{k} \ll 1$ are modes that exit the Hubble horizon well before the beginning of the USR phase, 
modes with $\bar{k} \gg 1$ are modes that exit the Hubble horizon well after the beginning of the USR phase.
\begin{figure}[h]
\begin{center}
$$\includegraphics[width=.495\textwidth]{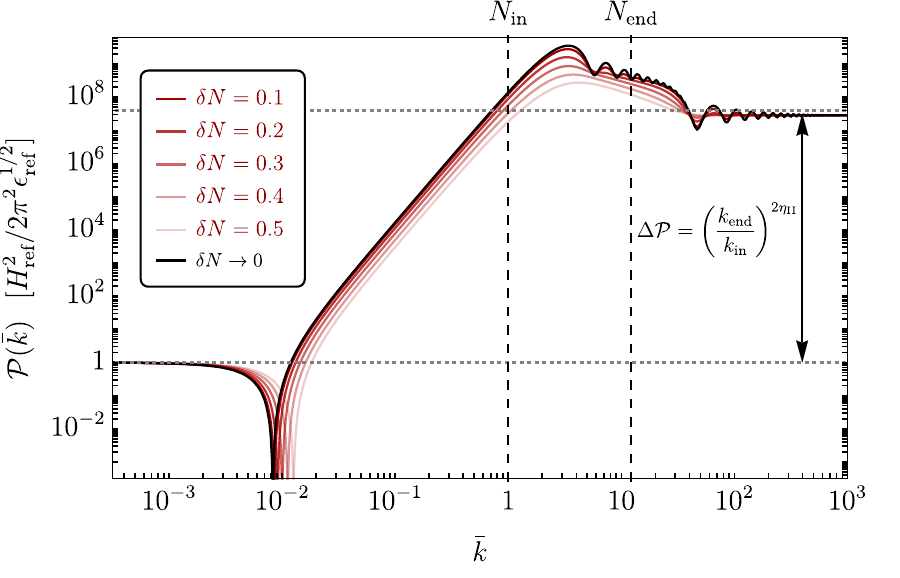}~~
\includegraphics[width=.495\textwidth]{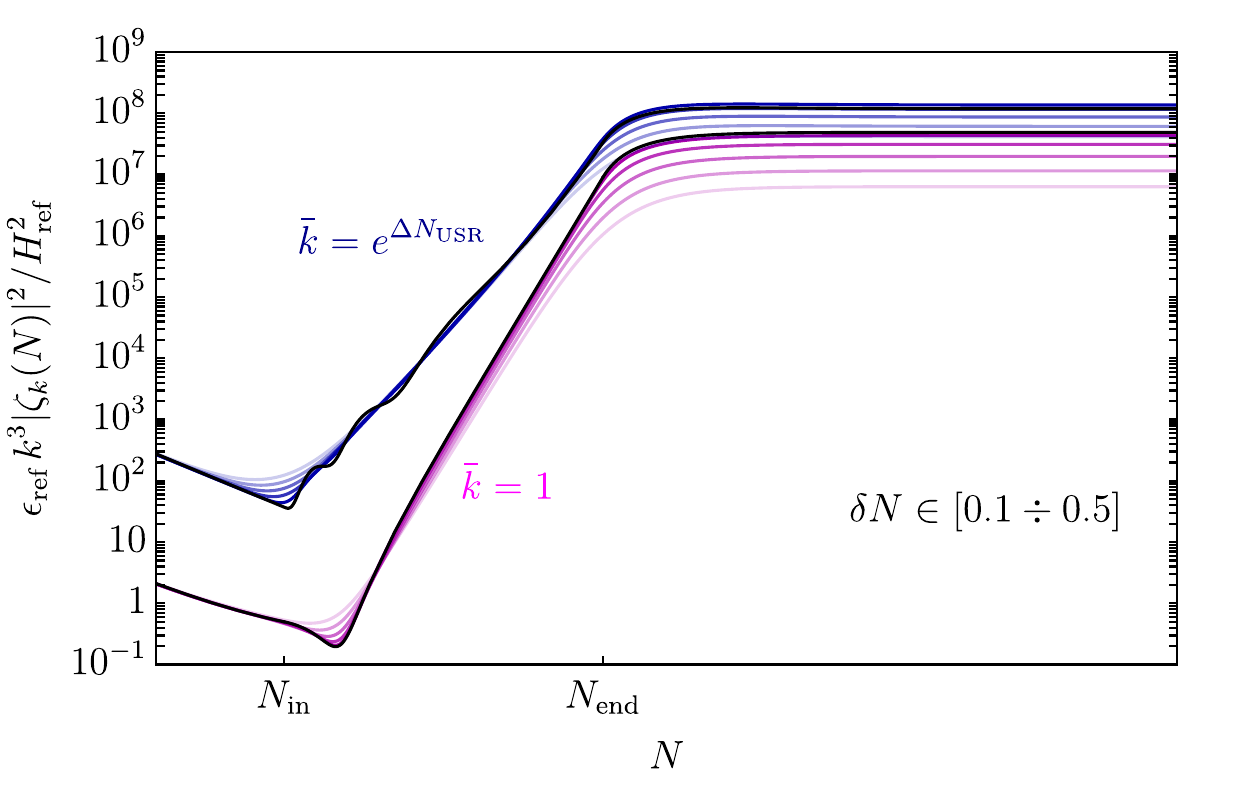}$$\vspace{-0.5cm}
\caption{\em 
\textit{\textbf{Left panel:}}
Tree-level power spectrum in the minimal dynamics of section\,\ref{sec:MinDynUSR}.  
The numerical values of the other parameters are 
$\eta_{\textrm{\rm II}} = 3.5$, 
$\eta_{\textrm{\rm III}} = 0$ and 
$N_{\textrm{\rm end}} - N_{\textrm{\rm in}} = 2.5$. 
In our parametrization, we go beyond the instantaneous transition approximation and 
we explore different values of $\delta N$.
The vertical double-arrow indicates the growth of the power spectrum 
given by the na\"{\i}ve scaling 
$\Delta\mathcal{P} = (k_{\rm end}/k_{\rm in})^{2\eta_{\rm II}} =  e^{2\eta_{\rm II}\Delta N_{\rm{USR}}}$.  
This scaling captures well the amplitude of the transition from the initial to the final SR phase but it does not give a reliable estimate of the peak amplitude of the power spectrum, which can easily be one order of magnitude larger.
\textit{\textbf{Right panel:}}
Time evolution of two representative modes with $\bar{k} = 1$ and $\bar{k} = e^{\Delta N_{\rm{USR}}}$ for $\delta N \in [0.1\divisionsymbol 0.5]$ (from darker to lighter colors, respectively). 
The black lines represent the limit $\delta N\to 0$.   
 }\label{fig:TestPS}  
\end{center}
\end{figure}
In the left panel of fig.\,\ref{fig:TestPS}, 
we show the tree-level power spectrum that we obtain by numerically solving eq.\,(\ref{eq:MSimpl}) and using eq.\,(\ref{eq:TreeLevel}). 
Thanks to our parametrization in eq.\,(\ref{eq:DynEta}) we control the sharpness of the transition varying $\delta N$. 

In order to make contact with the analysis of ref.\,\cite{Kristiano:2022maq}, we set $\eta_{\rm III} = 0$.
However,  it should be noted that 
in more realistic models we need $\eta_{\textrm{III}}\neq 0$ and negative so that the power spectrum decreases for modes with $\bar{k} \gg 1$. This feature is necessary if we want to  connect the USR phase to a subsequent SR phase that ends inflation.  
Since we are considering single-field models of inflation, in our analysis this is a necessary requirement.
Consequently, the power spectrum at small scales -- both before and after the  peak --  does not respect the property of scale invariance. 
Before the peak, the power spectrum 
of the short modes grows with a maximum slope given by $\mathcal{P}(\bar{k}) \sim \bar{k}^4$; after the peak, the power spectrum 
of the short modes decays approximately as $\mathcal{P}(\bar{k}) \sim \bar{k}^{2\eta_{\textrm{III}}}$. 
After the peak, therefore, the power spectrum becomes approximately scale invariant only 
if we take $\eta_{\textrm{III}} \approx 0$; 
however, in such case $\epsilon$ remains anchored to the tiny value reached during the USR phase and inflation never ends.

In ref.\,\cite{Kristiano:2022maq}, 
the loop integration is restricted to the interval of modes $\bar{k} \in [\bar{k}_{\textrm{in}},
\bar{k}_{\textrm{end}}]$  where 
$\bar{k}_{\textrm{in}} = 1$  
and 
$\bar{k}_{\textrm{end}} = e^{\Delta N_{\textrm{USR}}}(H_{\textrm{end}}/H_{\textrm{in}}) 
\simeq e^{\Delta N_{\textrm{USR}}}$ 
with $\Delta N_{\textrm{USR}} \equiv N_{\textrm{end}} - 
N_{\textrm{in}}$.  
This interval of modes is limited by the two vertical dashed lines in the left panel of fig.\,\ref{fig:TestPS}. 
In ref.\,\cite{Kristiano:2022maq}, limiting the integration to the range $\bar{k} \in [\bar{k}_{\textrm{in}},
\bar{k}_{\textrm{end}}]$ is justified by the fact that the power spectrum of short modes peaks in this window of modes.

For future reference, let us stress one more important point. In the left panel of fig.\,\ref{fig:TestPS} we indicate the growth of the power spectrum given by the scaling $\Delta\mathcal{P} = (k_{\rm end}/k_{\rm in})^{2\eta_{\rm II}}$.  
This result immediately follows from the application of the SR formula 
$\mathcal{P}(k) = H^2/8\pi^2\epsilon$ if one accounts for the exponential decay  $\epsilon \sim e^{-2\eta_{\rm II}N}$ during USR and converts $N$ into $k$ by means of  the horizon-crossing condition $k=aH$. 
Therefore, not surprisingly, the scaling 
$\Delta\mathcal{P} = (k_{\rm end}/k_{\rm in})^{2\eta_{\rm II}}$ captures well the growth of the power spectrum if one directly jumps from the initial to the final SR phase.  
However, as shown in the left panel of fig.\,\ref{fig:TestPS},  the above estimate does not accurately describe the amplitude of the power spectrum at the  position of its peak; the latter can easily be one order of magnitude larger than what suggested by $\Delta\mathcal{P} = (k_{\rm end}/k_{\rm in})^{2\eta_{\rm II}}$.
This features has important consequences when estimating the PBH abundance, which rather sensitive to the spectral amplitude. We will come back to this point in the next section. 

Finally, it is possible to check numerically that neglecting the time dependence of the Hubble rate as in eq.\,(\ref{eq:HubbleRatewithTime}) has a negligible impact. 
In the following, therefore, we shall keep $H$ constant (that is, $H=H_{\textrm{ref}}$ does not evolve in time). 
Furthermore, if we take $H$ constant and in the limit $\delta N = 0$, 
it is possible to get, for some special values of $\eta_{\textrm{II}}$ and $\eta_{\textrm{III}}$, a complete analytical description of the SR/USR/SR dynamics\,\cite{Byrnes:2018txb,Ballesteros:2020qam}.

\subsection{The cubic action}\label{sec:Cubo}

At the cubic order in the fluctuations, 
the action is
\begin{align}
\mathcal{S}_3 = 
\int d^4 x\bigg\{&
\epsilon^2a^3\dot{\zeta}^2\zeta 
+ \epsilon^2a \zeta(\partial_k\zeta)(\partial^k\zeta) 
-2\epsilon^2a^3\dot{\zeta}
(\partial_k\zeta)\partial^k(\partial^{-2}\dot{\zeta}) 
+ \frac{\epsilon\dot{\epsilon_2}}{2}a^3\dot{\zeta}\zeta^2
-\frac{a^3\epsilon^3}{2}
\big[
\dot{\zeta}^2\zeta 
- \zeta\partial_{k}\partial_l(\partial^{-2}\dot{\zeta})
\partial^{k}\partial^l
(\partial^{-2}\dot{\zeta})
\big] \nn\\
& +\bigg[
\frac{d}{dt}\left(\epsilon a^3 \dot{\zeta}\right)  - 
\epsilon a \partial_k\partial^k\zeta
\bigg]
\bigg[
\frac{\epsilon_2}{2}\zeta^2
+ \frac{2}{H}\dot{\zeta}\zeta 
 -\frac{1}{2a^2 H^2}(\partial_k\zeta)(\partial^k\zeta) 
+\frac{1}{2a^2 H^2}\partial^{-2}
\partial_k\partial_l(\partial^k\zeta
\partial^l\zeta) \nn\\
&\hspace{4.1cm}+
\frac{\epsilon}{H}
(\partial_k\zeta)\partial^k(\partial^{-2}\dot{\zeta}) 
-
\frac{\epsilon}{H}\partial^{-2}
\partial_{k}\partial_{l}
\partial^{k}\zeta 
\partial^{l}(\partial^{-2}\dot{\zeta})
\bigg]\bigg\}\,.\label{eq:Cubic}
\end{align}
As shown in ref.\,\cite{Maldacena:2002vr}, it is possible to simplify the cubic action by means of a field redefinition that introduces a non-linear shift in the original field. 
Concretely, if we define
\begin{align}
\zeta \equiv \zeta_n + f(\zeta_n)\,,\label{eq:FieldRed}  
\end{align}
with
\begin{align}
f(&\zeta)
\equiv
\frac{1}{2}\bigg[
\frac{\epsilon_2}{2}\zeta^2
+ \frac{2}{H}\dot{\zeta}\zeta 
 -\frac{(\partial_k\zeta)(\partial^k\zeta) }{2a^2 H^2}
+\frac{1}{2a^2 H^2}\partial^{-2}
\partial_k\partial_l(\partial^k\zeta
\partial^l\zeta)+
\frac{\epsilon}{H}
(\partial_k\zeta)\partial^k(\partial^{-2}\dot{\zeta}) 
-
\frac{\epsilon}{H}\partial^{-2}
\partial_{k}\partial_{l}
\partial^{k}\zeta 
\partial^{l}(\partial^{-2}\dot{\zeta})
\bigg]\,,\label{eq:FR1}
\end{align}
we find, by direct computation, that at the quadratic order the field 
$\zeta_n$ is described by the action 
\begin{align}
\mathcal{S}_2(\zeta_n) = 
\int d^4 x\,\epsilon\,a^3
\left[
\dot{\zeta}_n^2 - \frac{(\partial_k\zeta_n)(\partial^k\zeta_n)}{a^2}
\right]\,,\label{eq:QuadraticAction}
\end{align}
which has the same structure as the quadratic action for the original variable $\zeta$. 
However, at the cubic order, we find 
\begin{align}
\mathcal{S}_3(\zeta_n) = 
\int d^4 x\bigg\{&
\epsilon^2a^3\dot{\zeta}_n^2\zeta_n
+ \epsilon^2a \zeta_n(\partial_k\zeta_n)(\partial^k\zeta_n) 
-2\epsilon^2a^3\dot{\zeta}_n
(\partial_k\zeta_n)\partial^k(\partial^{-2}\dot{\zeta}_n) 
+ \frac{\epsilon\dot{\epsilon_2}}{2}a^3\dot{\zeta}_n\zeta_n^2\nn\\
&
-\frac{a^3\epsilon^3}{2}
\big[
\dot{\zeta}_n^2\zeta_n 
- \zeta_n\partial_{k}\partial_l(\partial^{-2}\dot{\zeta}_n)
\partial^{k}\partial^l
(\partial^{-2}\dot{\zeta}_n)
\big]\bigg\}\,,  \label{eq:Cubic2}
\end{align}
in which, thanks to the above field redefinition, the second and third lines in eq.\,(\ref{eq:Cubic}) cancel out.

If we neglect terms with spatial derivatives and interactions suppressed by two or more powers of the Hubble parameter $\epsilon$, we find 
\begin{align}
\mathcal{S}_3(\zeta_n)  \ni 
\int d^4x\,\frac{\epsilon\dot{\epsilon_2}}{2}a^3\dot{\zeta}_n\zeta_n^2\,.\label{eq:Yoko}
\end{align}
Notice that we do not count the coupling  $\epsilon_2$ as a slow-roll suppression since we are interested in the USR phase during which $|\epsilon_2| > 3$ and $\dot{\epsilon}_2 \neq 0$.
Eq.\,(\ref{eq:Yoko}) is the only interaction included in 
ref.\,\cite{Kristiano:2022maq}. 
This means that, implicitly, ref.\,\cite{Kristiano:2022maq} computes the two-point function for the field $\zeta_n$.  
This is because, in terms of the dynamical variable $\zeta$, there is another interaction of order $\epsilon\epsilon_2$ that should be included, that is the one in the second line of 
eq.\,(\ref{eq:Cubic}).

However, as stressed in ref.\cite{Maldacena:2002vr}, $\zeta_n$ is not the right dynamical variable to consider since it is not conserved outside the horizon. 
This is a trivial consequence of eq.\,(\ref{eq:FieldRed}). 
Since $\zeta$ is conserved outside the horizon, $\zeta_n$ can not be conserved simply because various coefficients in the non-linear relation in eq.\,(\ref{eq:FieldRed}) are time-dependent. 
Alternatively, as discussed in ref.\cite{Maldacena:2002vr}, 
the above fact is also evident from the very same structure of the interactions that appear in eq.\,(\ref{eq:Cubic2}). 
The interaction $\epsilon\dot{\epsilon}_2\dot{\zeta}_n\zeta_n^2$ only has one time-derivative acting on the field $\zeta_n$; consequently, 
it alters 
the value of $\zeta_n$ on super-horizon scales (if one computes the equation of motion for $\zeta_n$, it is easy to see that the constant solution is not stable).
Let us make the above considerations more concrete.
Eventually, we are interested in the computation of the two-point function for the original curvature field. 
Given the field redefinition in eq.\,(\ref{eq:FieldRed}), we write
\begin{align}
\langle\hat{\zeta}(\vec{x}_1,\tau)
\hat{\zeta}(\vec{x}_2,\tau)\rangle = &  
\langle
\big\{
\hat{\zeta}_n(\vec{x}_1,\tau) 
+ f[\hat{\zeta}_n(\vec{x}_1,\tau)]
\big\}
\big\{
\hat{\zeta}_n(\vec{x}_2,\tau) 
+ f[\hat{\zeta}_n(\vec{x}_2,\tau)]
\big\}
\rangle \nn\\
= & 
\langle\hat{\zeta}_n(\vec{x}_1,\tau)
\hat{\zeta}_n(\vec{x}_2,\tau)\rangle + \label{eq:Expa1}\\
& \langle\hat{\zeta}_n(\vec{x}_1,\tau)
f[\hat{\zeta}_n(\vec{x}_2,\tau)]\rangle +
\langle
f[\hat{\zeta}_n(\vec{x}_1,\tau)]
\hat{\zeta}_n(\vec{x}_1,\tau)
\rangle + \label{eq:Expa2}\\
& \langle
f[\hat{\zeta}_n(\vec{x}_1,\tau)]
f[\hat{\zeta}_n(\vec{x}_2,\tau)]
\rangle\,,\label{eq:Expa3}
\end{align}
The first term, eq.\,(\ref{eq:Expa1}), corresponds to the two-point function for the shifted curvature field whose  cubic action is given by eq.\,(\ref{eq:Cubic2}); 
$\langle\hat{\zeta}_n(\vec{x}_1,\tau)
\hat{\zeta}_n(\vec{x}_2,\tau)\rangle$ can be computed perturbatively by means of 
the ``{\it in}-{\it in}'' formalism sketched at the  end of section\,\ref{sec:Conve}.
Eqs.\,(\ref{eq:Expa2},\,\ref{eq:Expa3})  account for the difference between $\zeta$ and $\zeta_n$ at the non-linear level.
Notice that the first term in the 
functional form in eq.\,(\ref{eq:FR1}) does not die off in the late-time limit $\tau\to 0^-$ (in which the power spectrum must be eventually evaluated) 
if we consider the case in which 
$\epsilon_2 \neq 0$ after the USR phase (as expected in realistic single-field models, cf. section\,\ref{sec:MinDynUSR}). 
However, if we limit to the case in which 
$\eta_{\rm III} = 0$ the contribution from the field redefinition vanishes. 
This limit was considered in 
ref.\,\cite{Kristiano:2022maq}. 
In order to make contact with the analysis presented in ref.\,\cite{Kristiano:2022maq}, we shall also adopt in the bulk of this work the assumption $\eta_{\rm III}  = 0$.

Let us  now come back to the schematic in eq.\,(\ref{eq:LoopsSchematic}). 
The cubic Hamiltonian interaction that follows from eq.\,(\ref{eq:Yoko}) gives rise to the last two topologies of connected diagrams illustrated in eq.\,(\ref{eq:LoopsSchematic}). 
As in ref.\,\cite{Kristiano:2022maq},  we will only focus on the 1PI diagram, that is, the central diagram in eq.\,(\ref{eq:LoopsSchematic}).
The last diagram in eq.\,(\ref{eq:LoopsSchematic}) consists of a tadpole 
that is attached to a $\zeta$-propagator and affects at one-loop its two-point correlation function. 
The correct way to deal with tadpoles is by changing the background solution, cf. ref.\,\cite{Sloth:2006nu} for a discussion in the case of ordinary SR inflation and ref.\,\cite{Senatore:2009cf} for the case in which there are additional spectator fields.  Recently, ref.\,\cite{Inomata:2022yte} estimated the tadpole correction to the background evolution in the context of a model in which there is a resonant amplification of field fluctuations.  
Imposing the condition that such modification is negligible could give rise to an additional perturbativity bound. We postpone a comprehensive exploration of this issue in the context of realistic USR dynamics to future work, cf. section\,\ref{sec:Discu}.

\subsection{Beyond the cubic action}\label{sec:Qua}
Before proceeding, we comment about quartic interactions since, as qualitatively discussed in eq.\,(\ref{eq:LoopsSchematic}), they give rise to one-loop corrections which are of the same order if compared to those generated by cubic interaction terms.  
The derivation of the  fourth-order action has been discussed in ref.\,\cite{Jarnhus:2007ia}. 
Based on this result, ref.\,\cite{Kristiano:2022maq,Kristiano:2023scm} claims that the relevant quartic interaction in the case of USR dynamics (that is, the quartic interaction proportional to $\dot{\epsilon_2}$) gives a vanishing contribution when inserted in eq.\,(\ref{eq:Shem1}). 
Ref.\,\cite{Firouzjahi:2023aum} adopts an approach based on the effective field theory of inflation and includes cubic and quartic interactions. It finds that the latter gives a non-trivial contribution, and finds a loop-corrected power spectrum different from the one in ref.\,\cite{Kristiano:2022maq,Kristiano:2023scm}. It would be important to perform a consistent comparison between these two approaches, including the full cubic and quartic interactions in both cases.
Generally speaking, we expect cubic and quartic interactions to be inextricably linked. 
For instance, the quartic Hamiltonian receives a
contribution that arises from the modification of the conjugate momentum if there are cubic interactions which depend on $\dot{\zeta}$. 
Similarly, cubic interactions with spatial derivatives are paired with quartic interactions induced by a residual spatial conformal symmetry of the perturbed metric\,\cite{Senatore:2009cf}. 
En route, we notice  that  interactions with spatial derivatives are usually neglected for modes that are  super-horizon. However, in the spirit of the loop computation in ref.\,\cite{Kristiano:2022maq}, the momenta over which the loop is integrated cross the horizon during the USR phase, and, na\"{\i}vely, their spatial derivatives do not pay any  super-horizon suppression.  

In this work, as a preliminary step towards a more complete analysis and in order to compare our results with the claim made in refs.\,\cite{Riotto:2023hoz,Kristiano:2023scm,Riotto:2023gpm,Firouzjahi:2023aum,Firouzjahi:2023ahg,Choudhury:2023vuj,Choudhury:2023jlt,Choudhury:2023rks,Choudhury:2023hvf},  
we only  focus on the cubic interaction in eq.\,(\ref{eq:Yoko}).
However, we stress  that  
all the arguments listed above motivate the need for a more comprehensive analysis. We postpone this task to future work, cf. section\,\ref{sec:Discu}.

\section{One-loop computation}\label{sec:1LoopSec}

We consider in this section the cubic interaction Hamiltonian given by (we omit the subscript $_{I}$ in the interaction-picture fields)
\begin{align}
\hat{H}_{\rm int}^{(3)}(\tau) = 
\frac{1}{2}\int d^3\vec{x}\,
\epsilon(\tau)\epsilon_2^{\prime}(\tau)a^2(\tau)
\zeta^{\prime}(\vec{x},\tau)\zeta(\vec{x},\tau)^2\,.\label{eq:MainHami}
\end{align}
We consider 
eq.\,(\ref{eq:Shem2});
this can be written in the compact form 
\begin{align}
\langle\hat{\zeta}(\vec{x}_1,\tau)&
\hat{\zeta}(\vec{x}_2,\tau)\rangle_{2^{\textrm{nd}}} = 
\langle\hat{\zeta}(\vec{x}_1,\tau)
\hat{\zeta}(\vec{x}_2,\tau)\rangle_{2^{\textrm{nd}}}^{(1,1)} 
-2 \RE \llp \langle\hat{\zeta}(\vec{x}_1,\tau)
\hat{\zeta}(\vec{x}_2,\tau)\rangle_{2^{\textrm{nd}}}^{(0,2)} \rrp
\end{align}
where
\begin{align}
\langle\hat{\zeta}(\vec{x}_1,\tau)
\hat{\zeta}(\vec{x}_2,\tau)\rangle_{2^{\textrm{nd}}}^{(1,1)}  \equiv 
& \int_{-\infty(1+i\epsilon)}^{\tau}
d\tau_{1}
\int_{-\infty(1-i\epsilon)}^{\tau}
d\tau_2
\langle 0|
\hat{H}_{\textrm{int}}(\tau_1)
\hat{\zeta}_I(\vec{x}_1,\tau)
\hat{\zeta}_I(\vec{x}_2,\tau)
\hat{H}_{\textrm{int}}(\tau_2)|0\rangle\,,\label{eq:OneLoop11}
\\
\langle\hat{\zeta}(\vec{x}_1,\tau)
\hat{\zeta}(\vec{x}_2,\tau)\rangle_{2^{\textrm{nd}}}^{(0,2)}  \equiv &   
\int_{-\infty(1-i\epsilon)}^{\tau}d\tau_1
\int_{-\infty(1-i\epsilon)}^{\tau_1}d\tau_2\langle 0|
\hat{\zeta}_I(\vec{x}_1,\tau)
\hat{\zeta}_I(\vec{x}_2,\tau)\hat{H}_{\textrm{int}}(\tau_1)\hat{H}_{\textrm{int}}(\tau_2)|0\rangle\,.\label{eq:OneLoop02} 
\end{align}
This expansion is consistent with Eq.~(16) of Ref.~\cite{Senatore:2009cf}. 
Consider the first contribution in eq.\,(\ref{eq:OneLoop11}), one finds
\begin{align}
\langle\hat{\zeta}(\vec{x}_1,\tau)
\hat{\zeta}(\vec{x}_2,\tau)\rangle_{2^{\textrm{nd}}}^{(1,1)} =  \frac{1}{4}&\int_{-\infty_+}^{\tau}
d\tau_{1}\epsilon(\tau_1)\epsilon_2^{\prime}(\tau_1)
  a^2(\tau_1)
\int_{-\infty_-}^{\tau}
d\tau_2\epsilon(\tau_2)\epsilon_2^{\prime}(\tau_2)
  a^2(\tau_2)\int d^3\vec{y}d^3\vec{z}\nn\\
  &
  \int\left[\prod_{i=1}^8\frac{d^3\vec{k}_i}{(2\pi)^3}\right]
  e^{i\vec{y}\cdot (\vec{k}_1 + \vec{k}_2 + \vec{k}_3)}
  e^{i(\vec{x}_1\cdot \vec{k}_4+\vec{x}_2\cdot \vec{k}_5)}
  e^{i\vec{z}\cdot (\vec{k}_6 + \vec{k}_7 + \vec{k}_8)}\nn\\
  &
  \langle 0|
\hat{\zeta}^{\prime}_I(\vec{k}_1,\tau_1)
\hat{\zeta}_I(\vec{k}_2,\tau_1)
\hat{\zeta}_I(\vec{k}_3,\tau_1)
\hat{\zeta}_I(\vec{k}_4,\tau)
\hat{\zeta}_I(\vec{k}_5,\tau)
\hat{\zeta}^{\prime}_I(\vec{k}_6,\tau_2)
\hat{\zeta}_I(\vec{k}_7,\tau_2)
\hat{\zeta}_I(\vec{k}_8,\tau_2)
  |0\rangle\,.\label{eq:Main11}
\end{align}
The $36$ connected Wick contractions can be expressed as 
\begin{align}
\langle\hat{\zeta}(\vec{x}_1,\tau)
\hat{\zeta}(\vec{x}_2,\tau)\rangle_{2^{\textrm{nd}}}^{(1,1)}   = 
&
\int_{-\infty_+}^{\tau}
d\tau_{1}\epsilon(\tau_1)\epsilon_2^{\prime}(\tau_1)
  a^2(\tau_1)
\int_{-\infty_-}^{\tau}
d\tau_2\epsilon(\tau_2)\epsilon_2^{\prime}(\tau_2)
  a^2(\tau_2) 
\int\frac{d^3\vec{k}}{(2\pi)^3} 
\frac{d^3\vec{q}}{(2\pi)^3}\,e^{i(\vec{x}_1 - \vec{x}_2)\cdot(\vec{k}+\vec{q})}
\nn\\
&\big [ \left|\zeta_{k+q}(\tau)\right|^2
\big\{
\zeta_{k}(\tau_1)
\zeta^{\prime}_{k+q}(\tau_1)
\zeta_{q}(\tau_1)
\zeta_{k}^*(\tau_2)
\zeta_{k+q}^{\prime\,*}(\tau_2)
\zeta_q^*(\tau_2) + \nn\\
&\hspace{2cm}
\zeta_k(\tau_1)\zeta^{\prime}_{k+q}(\tau_1)\zeta_q(\tau_1)\zeta_{k+q}^*(\tau_2)\big[
\zeta_k^{\prime\,*}(\tau_2)\zeta_q^*(\tau_2) +
\zeta_k^*(\tau_2)\zeta_q^{\prime\,*}(\tau_2)
\big]+ \nn\\
&\hspace{2cm}
\zeta_k^*(\tau_2)\zeta^{\prime\,*}_{k+q}(\tau_2)\zeta_q^*(\tau_2)\zeta_{k+q}(\tau_1)\big[
\zeta_k^{\prime}(\tau_1)\zeta_q(\tau_1) +
\zeta_k(\tau_1)\zeta_q^{\prime}(\tau_1)
\big]+ \nn\\
&\hspace{2cm}
\zeta_k^{\prime}(\tau_1)\zeta_{k+q}(\tau_1)\zeta_q(\tau_1)\zeta_{k+q}^*(\tau_2)
\big[
\zeta_q^*(\tau_2)\zeta_k^{\prime\,*}(\tau_2) +
\zeta_k^*(\tau_2)\zeta_q^{\prime\,*}(\tau_2)
\big]+ \nn\\
&\hspace{2cm}
\zeta_k^{\prime\,*}(\tau_2)\zeta_{k+q}^*(\tau_2)\zeta_q^*(\tau_2)\zeta_{k+q}(\tau_1)
\big[
\zeta_q(\tau_1)\zeta_k^{\prime}(\tau_1) +
\zeta_k(\tau_1)\zeta_q^{\prime}(\tau_1)
\big]\big\}
\big ]\,.\label{eq:1stContra}
\end{align}

Consider now eq.\,(\ref{eq:OneLoop02}). 
One can write it in the form
\begin{align}
\langle\hat{\zeta}(\vec{x}_1,\tau)
\hat{\zeta}(\vec{x}_2,\tau)\rangle_{2^{\textrm{nd}}}^{(0,2)} =  
\frac{1}{4}
&\int_{-\infty_-}^{\tau}
d\tau_{1}\epsilon(\tau_1)\epsilon_2^{\prime}(\tau_1)
  a^2(\tau_1)
\int_{-\infty_-}^{\tau_1}
d\tau_2\epsilon(\tau_2)\epsilon_2^{\prime}(\tau_2)
  a^2(\tau_2)\int d^3\vec{y}d^3\vec{z}\nn\\
  &
  \int\left[\prod_{i=1}^8\frac{d^3\vec{k}_i}{(2\pi)^3}\right]
  e^{i\vec{y}\cdot (\vec{k}_1 + \vec{k}_2 + \vec{k}_3)}
  e^{i(\vec{x}_1\cdot \vec{k}_4+\vec{x}_2\cdot \vec{k}_5)}
  e^{i\vec{z}\cdot (\vec{k}_6 + \vec{k}_7 + \vec{k}_8)}\nn\\
  &
  \langle 0|
  \hat{\zeta}_I(\vec{k}_4,\tau)
\hat{\zeta}_I(\vec{k}_5,\tau)
\hat{\zeta}^{\prime}_I(\vec{k}_1,\tau_1)
\hat{\zeta}_I(\vec{k}_2,\tau_1)
\hat{\zeta}_I(\vec{k}_3,\tau_1)
\hat{\zeta}^{\prime}_I(\vec{k}_6,\tau_2)
\hat{\zeta}_I(\vec{k}_7,\tau_2)
\hat{\zeta}_I(\vec{k}_8,\tau_2)
  |0\rangle\,.\label{eq:Main02}
\end{align}
After Wick contractions, we find
\begin{align}
 \langle\hat{\zeta}(\vec{x}_1,\tau)
\hat{\zeta}(\vec{x}_2,\tau)\rangle_{2^{\textrm{nd}}}^{(0,2)} 
=
&\int_{-\infty_-}^{\tau}
d\tau_{1}\epsilon(\tau_1)\epsilon_2^{\prime}(\tau_1)
  a^2(\tau_1)
\int_{-\infty_-}^{\tau_1}
d\tau_2\epsilon(\tau_2)\epsilon_2^{\prime}(\tau_2)
  a^2(\tau_2) 
\int\frac{d^3\vec{k}}{(2\pi)^3} 
\frac{d^3\vec{q}}{(2\pi)^3}
e^{i(\vec{x}_1 - \vec{x}_2)\cdot(\vec{k}+\vec{q})}
\nn\\
&\big[
\zeta_{k+q}^2(\tau)
\big\{
\zeta_{k}(\tau_1)
\zeta^{\prime\,*}_{k+q}(\tau_1)
\zeta_{q}(\tau_1)
\zeta_{k}^*(\tau_2)
\zeta_{k+q}^{\prime\,*}(\tau_2)
\zeta_q^*(\tau_2) + \nn\\
& \hspace{1.5cm}
\zeta_k(\tau_1)\zeta^{\prime\,*}_{k+q}(\tau_1)\zeta_q(\tau_1)\zeta_{k+q}^*(\tau_2)\big[
\zeta_k^{\prime\,*}(\tau_2)\zeta_q^*(\tau_2) +
\zeta_k^*(\tau_2)\zeta_q^{\prime\,*}(\tau_2)
\big] + \nn\\
&\hspace{1.5cm}
\zeta_k^*(\tau_2)\zeta^{\prime\,*}_{k+q}(\tau_2)\zeta_q^*(\tau_2)\zeta_{k+q}^*(\tau_1)\big[
\zeta_k^{\prime}(\tau_1)\zeta_q(\tau_1) +
\zeta_k(\tau_1)\zeta_q^{\prime}(\tau_1)
\big]+ \nn\\
&\hspace{1.5cm}
\zeta_k^{\prime}(\tau_1)\zeta_{k+q}^*(\tau_1)\zeta_q(\tau_1)\zeta_{k+q}^*(\tau_2)
\big[
\zeta_q^*(\tau_2)\zeta_k^{\prime\,*}(\tau_2) +
\zeta_k^*(\tau_2)\zeta_q^{\prime\,*}(\tau_2)
\big]+ \nn\\
&\hspace{1.5cm}
\zeta_k^{\prime\,*}(\tau_2)\zeta_{k+q}^*(\tau_2)\zeta_q^*(\tau_2)\zeta_{k+q}^*(\tau_1)
\big[
\zeta_q(\tau_1)\zeta_k^{\prime}(\tau_1) +
\zeta_k(\tau_1)\zeta_q^{\prime}(\tau_1)
\big]\big\}
\big]\,.\label{eq:2nd3rdContra}
\end{align}
At this point we shift the momentum following the prescription $	k \rightarrow k-q$ in such a way that $k$ is identified with the external momentum.
The power spectrum at one loop can be therefore written as 
\begin{equation}
\mathcal{P}(k) = \lim_{\tau \to 0^-}\left(\frac{k^3}{2\pi^2}\right)
\left\{
\left|\zeta_k(\tau)\right|^2 + 
\frac{1}{(4\pi)^2}\left[\Delta P_1(k,\tau) + \Delta P_2(k,\tau)\right]
\right\}\,,\label{eq:MasterOne}
\end{equation}
with
\begingroup
\allowdisplaybreaks
\begin{align}
\Delta P_1(k,\tau) & \equiv 
4\int_{-\infty_+}^{\tau}
d\tau_{1}\epsilon(\tau_1)\epsilon_2^{\prime}(\tau_1)
  a^2(\tau_1)
\int_{-\infty_-}^{\tau}
d\tau_2\epsilon(\tau_2)\epsilon_2^{\prime}(\tau_2)
  a^2(\tau_2) 
\int_0^{\infty}dq\,q^2\,d(\cos\theta)
\left|\zeta_{k}(\tau)\right|^2 
\nn
\\
&\times \big\{
\zeta_{k-q}(\tau_1)
\zeta^{\prime}_{k}(\tau_1)
\zeta_{q}(\tau_1)
\zeta_{k-q}^*(\tau_2)
\zeta_{k}^{\prime\,*}(\tau_2)
\zeta_q^*(\tau_2) + 
\nn\\ 
& \hspace{.62cm}
\zeta_{k-q}(\tau_1)\zeta^{\prime}_{k}(\tau_1)\zeta_q(\tau_1)\zeta_{k}^*(\tau_2)\big[
\zeta_{k-q}^{\prime\,*}(\tau_2)\zeta_q^*(\tau_2) +
\zeta_{k-q}^*(\tau_2)\zeta_q^{\prime\,*}(\tau_2)
\big]+ \nn\\
&\hspace{.62cm}
\zeta_{k-q}^*(\tau_2)\zeta^{\prime\,*}_{k}(\tau_2)\zeta_q^*(\tau_2)\zeta_{k}(\tau_1)\big[
\zeta_{k-q}^{\prime}(\tau_1)\zeta_q(\tau_1) +
\zeta_{k-q}(\tau_1)\zeta_q^{\prime}(\tau_1)
\big]+ \nn\\
&\hspace{.62cm}
\zeta_{k-q}^{\prime}(\tau_1)\zeta_{k}(\tau_1)\zeta_q(\tau_1)\zeta_{k}^*(\tau_2)
\big[
\zeta_q^*(\tau_2)\zeta_{k-q}^{\prime\,*}(\tau_2) +
\zeta_{k-q}^*(\tau_2)\zeta_q^{\prime\,*}(\tau_2)
\big]+ \nn\\
&\hspace{.62cm}
\zeta_{k-q}^{\prime\,*}(\tau_2)\zeta_{k}^*(\tau_2)\zeta_q^*(\tau_2)\zeta_{k}(\tau_1)
\big[
\zeta_q(\tau_1)\zeta_{k-q}^{\prime}(\tau_1) +
\zeta_{k-q}(\tau_1)\zeta_q^{\prime}(\tau_1)
\big]\big\}\,,\label{eq:DeltaP1Full}
\\
\Delta P_2(k,\tau) & \equiv 
-8 \RE \Big [ \int_{-\infty_-}^{\tau}
d\tau_{1}\epsilon(\tau_1)\epsilon_2^{\prime}(\tau_1)
  a^2(\tau_1)
\int_{-\infty_-}^{\tau_1}
d\tau_2\epsilon(\tau_2)\epsilon_2^{\prime}(\tau_2)
  a^2(\tau_2) 
\int_0^{\infty}dq\,q^2\,d(\cos\theta) \zeta_{k}^2(\tau)
\nn \\
& \times 
\big\{
\zeta_{k-q}(\tau_1)
\zeta^{\prime\,*}_{k}(\tau_1)
\zeta_{q}(\tau_1)
\zeta_{k-q}^*(\tau_2)
\zeta_{k}^{\prime\,*}(\tau_2)
\zeta_q^*(\tau_2) + \nn\\
& \hspace{0.62cm}
\zeta_{k-q}(\tau_1)\zeta^{\prime\,*}_{k}(\tau_1)\zeta_q(\tau_1)\zeta_{k}^*(\tau_2)\big[
\zeta_{k-q}^{\prime\,*}(\tau_2)\zeta_q^*(\tau_2) +
\zeta_{k-q}^*(\tau_2)\zeta_q^{\prime\,*}(\tau_2)
\big] + \nn\\
&\hspace{0.62cm}
\zeta_{k-q}^*(\tau_2)\zeta^{\prime\,*}_{k}(\tau_2)\zeta_q^*(\tau_2)\zeta_{k}^*(\tau_1)\big[
\zeta_{k-q}^{\prime}(\tau_1)\zeta_q(\tau_1) +
\zeta_{k-q}(\tau_1)\zeta_q^{\prime}(\tau_1)
\big]+ \nn\\
&\hspace{0.62cm}
\zeta_{k-q}^{\prime}(\tau_1)\zeta_{k}^*(\tau_1)\zeta_q(\tau_1)\zeta_{k}^*(\tau_2)
\big[
\zeta_q^*(\tau_2)\zeta_{k-q}^{\prime\,*}(\tau_2) +
\zeta_{k-q}^*(\tau_2)\zeta_q^{\prime\,*}(\tau_2)
\big]+ \nn\\
&\hspace{0.62cm}
\zeta_{k-q}^{\prime\,*}(\tau_2)\zeta_{k}^*(\tau_2)\zeta_q^*(\tau_2)\zeta_{k}^*(\tau_1)
\big[
\zeta_q(\tau_1)\zeta_{k-q}^{\prime}(\tau_1) +
\zeta_{k-q}(\tau_1)\zeta_q^{\prime}(\tau_1)
\big]\big\}
\Big ]
\,.\label{eq:DeltaP2Full}
\end{align}
\endgroup

\subsection{Loop correction with a large hierarchy of scales}\label{sec:LoopHier}

First, we will be concerned with external momenta that describe the large CMB scales, while the USR takes place when modes $k_\text{\tiny USR} \gg k$ cross the horizon. 
The situation is summarized in the following schematic
\begin{align}
\begin{adjustbox}{max width=0.9\textwidth}
\raisebox{-8mm}{
	\begin{tikzpicture}
      \draw[thick,dotted,color=magenta][thick] (-5.12,1)--(-5.12,-6.2);  
\draw[draw=white,fill=magenta!25] (1,1) rectangle ++(0.7,-7);
\draw[draw=white,fill=blue!25] (-6,-4) rectangle ++(+12,-0.4);
	\draw[->,>=Latex,thick][thick] (-6,-6)--(6,-6);
 \draw[->,>=Latex,thick][thick] (-6,-6)--(-6,1);
  \draw[red,thick][thick] (-6,0)--(4.5,-6);
  	\draw[thick,dashed][thick] (-6,-0.5)--(6,-0.5);
   \draw[thick,dotted,color=magenta][thick] (2.6,1)--(2.6,-6.2);  
  	\draw[thick,dotted,color=blue][thick] (-6,-4)--(6,-4);  
     	\draw[thick,dotted,color=blue][thick] (-6,-4.4)--(6,-4.4);
    \node at (-6.5,1.5) {\scalebox{1}{$\textrm{comoving length}\,\lambda$}}; 
    \node at (7.2,-6.5) {\scalebox{1}{$\textrm{comoving time}$}};
 \node at (5.4,-0.1) {\scalebox{1}{$\textrm{long mode}\,\lambda =1/k$}};
  \node at (4.5,-4.2) {\scalebox{1}{{\color{blue}{
  $\textrm{short modes}$}}}};
    \node at (-6.4,-3.9) {\scalebox{1}{{\color{blue}{$q_{\textrm{in}}$}}}};
    \node at (-6.4,-4.4) {\scalebox{1}{{\color{blue}{$q_{\textrm{end}}$}}}};
    \node at (2.6,-6.5) {\scalebox{1}{{\color{magenta}{$\tau$}}}}; 
    \node at (4.62,-5.5) {\scalebox{1}{{\color{red}{$1/aH$}}}};   
    \node at (-6.75,-6.5) {\scalebox{1}{{\color{magenta}{$\tau\to -\infty$}}}};  
     \node at (-5.12,-6.5) {\scalebox{1}{{\color{magenta}{$\tau_k$}}}};  
     \node at (5,-6.5) {\scalebox{1}{{\color{magenta}{$\tau\to 0^-$}}}};
      \node at (1.35,-2.3) {\scalebox{0.9}{{\color{magenta}{$\textrm{USR}$}}}};
      \node at (0.97,-6.25) {\scalebox{1}{{\color{magenta}{$\tau_{\textrm{in}}$}}}}; 
    \node at (1.75,-6.25) {\scalebox{1}{{\color{magenta}{$\tau_{\textrm{end}}$}}}};
	\end{tikzpicture}
 }
\end{adjustbox} 
\label{eq:TimeEvo}
\end{align}
in which the blue horizontal band represents the interval of modes that cross the horizon during the USR phase, the vertical band shaded in magenta. 
In other words, as we will restrict the integration over momenta $q  \in [q_{\rm in},q_{\rm end}]$ that are enhanced by the USR phase, we can assume $q\gg k$.
 Consequently, as in ref.\,\cite{Kristiano:2022maq}, 
we approximate
\begin{align}
 k - q = \sqrt{k^2 + q^2 - 2kq\cos(\theta)} \approx q\,,~~~~~~~
 \int_{-1}^{+1} d(\cos\theta) = 2\,.\label{eq:ApproxQ}
\end{align}
With these assumptions, we can further simplify the expressions. 
We collect each contribution depending on the number of time derivatives acting on the long mode $\zeta_k$.
In each expression, the first line indicates terms with no derivative on the long modes, the second one those with one derivative, while the last with two.
One finds 
\begin{align}
\Delta P_1(k,\tau) & \equiv 
8\int_{\tau_{\rm in}}^{\tau}
d\tau_{1}\epsilon(\tau_1)\epsilon_2^{\prime}(\tau_1)
  a^2(\tau_1)
\int_{\tau_{\rm in}}^{\tau}
d\tau_2\epsilon(\tau_2)\epsilon_2^{\prime}(\tau_2)
  a^2(\tau_2) 
\int_{q_{\rm in}}^{q_{\rm end}}dq\,q^2\,
\left|\zeta_{k}(\tau)\right|^2 
 \nn\\
&
\times \big\{
4 \zeta_{k}(\tau_1)
\zeta_{k}^*(\tau_2) 
\zeta_q(\tau_1)
\zeta_{q}^{\prime}(\tau_1)
\zeta_q^*(\tau_2)
\zeta_{q}^{\prime\,*}(\tau_2) +
\nn\\ 
& \hspace{.62cm}
2 
\zeta^{\prime}_{k}(\tau_1)
\zeta_{k}^*(\tau_2)
\zeta_{q}(\tau_1)
\zeta_q(\tau_1)
\zeta_{q}^{\prime\,*}(\tau_2)
\zeta_q^*(\tau_2)
+
2
\zeta_{k}(\tau_1)
\zeta^{\prime\,*}_{k}(\tau_2)
\zeta_q(\tau_1) 
\zeta_{q}^{\prime}(\tau_1)
\zeta_{q}^*(\tau_2)
\zeta_q^*(\tau_2)
+
\nn
\\
& \hspace{.62cm}
\zeta^{\prime}_{k}(\tau_1)
\zeta_{k}^{\prime\,*}(\tau_2)
\zeta_{q}(\tau_1)
\zeta_{q}(\tau_1)
\zeta_{q}^*(\tau_2)
\zeta_q^*(\tau_2) 
\big\}\,,
\label{eq:int_1}
\\
\Delta P_2(k,\tau) & \equiv 
-16 \RE \Big [ \int_{\tau_{\rm in}}^{\tau}
d\tau_{1}\epsilon(\tau_1)\epsilon_2^{\prime}(\tau_1)
  a^2(\tau_1)
\int_{\tau_{\rm in}}^{\tau_1}
d\tau_2\epsilon(\tau_2)\epsilon_2^{\prime}(\tau_2)
  a^2(\tau_2) 
\int_{q_{\rm in}}^{q_{\rm end}}dq\,q^2\, \zeta_{k}(\tau)^2
\nn \\
& \times 
\big\{
4 
\zeta_{k}^*(\tau_1)
\zeta_{k}^*(\tau_2)
\zeta_{q}^{\prime}(\tau_1)
\zeta_q(\tau_1)
\zeta_q^*(\tau_2)
\zeta_{q}^{\prime\,*}(\tau_2)
+
\nn\\
& \hspace{0.62cm}
2 
\zeta^{\prime\,*}_{k}(\tau_1)
\zeta^*_{k}(\tau_2)
\zeta_{q}(\tau_1)^2
\zeta_{q}^{\prime\,*}(\tau_2)
\zeta_q^*(\tau_2) 
+ 
2 
\zeta^*_{k}(\tau_1)
\zeta^{\prime\,*}_{k}(\tau_2)
\zeta_{q}^{\prime}(\tau_1)
\zeta_q(\tau_1)
\zeta_{q}^*(\tau_2)^2
+ \nn\\
&\hspace{0.62cm}
\zeta^{\prime\,*}_{k}(\tau_1)
\zeta_{k}^{\prime\,*}(\tau_2)
\zeta_{q}(\tau_1)^2
\zeta_{q}^*(\tau_2)^2
\big\}
\Big ]
\,.
\label{eq:int_2}
\end{align}
We can combine the two contributions using the properties of symmetric integrals for holomorphic symmetric functions
$f(\tau_1,\tau_2) = f(\tau_2,\tau_1)$ \cite{Adshead:2008gk}
\begin{equation}
\int_{\tau_{\rm in}}^{\tau} d \tau_1 \int_{\tau_{\rm in}}^{\tau_1} d \tau_2 f\left(\tau_1, \tau_2\right)=
\frac{1}{2} \int_{\tau_{\rm in}}^{\tau} d \tau_1 \int_{\tau_{\rm in}}^{\tau} d \tau_2 f\left(\tau_1, \tau_2\right).\label{eq:InteIde}
\end{equation}
To shorten the notation, we introduce $\Delta P(k,\tau) = \Delta P_1(k,\tau)+\Delta P_2(k,\tau)$ and collect the individual contribution order by order in derivatives:
\begin{itemize}
\item {\bf 0th order in time derivatives of the long mode.}
For ease of reading, we introduce 
the short-hand notation 
$\epsilon(\tau)\epsilon_2^{\prime}(\tau)
  a^2(\tau) \equiv g(\tau)$.
Consider the sum of the two integrals 
\begin{align}
\Delta P_{0{\rm th}}&(k,\tau) =
32\int_{\tau_{\rm in}}^{\tau}
d\tau_{1}
g(\tau_1)
\int_{\tau_{\rm in}}^{\tau}
d\tau_2 g(\tau_2) 
\int_{q_{\rm in}}^{q_{\rm end}}dq\,q^2\,
\left|\zeta_{k}(\tau)\right|^2 \zeta_{k}(\tau_1)
\zeta_{k}^*(\tau_2) 
\zeta_q(\tau_1)
\zeta_{q}^{\prime}(\tau_1)
\zeta_q^*(\tau_2)
\zeta_{q}^{\prime\,*}(\tau_2) \label{eq:FirstLine}\\&
-64 \RE \Big [ \int_{\tau_{\rm in}}^{\tau}
d\tau_{1}  g(\tau_1)
\int_{\tau_{\rm in}}^{\tau_1}
d\tau_2 g(\tau_2) 
\int_{q_{\rm in}}^{q_{\rm end}}dq\,q^2\, \zeta_{k}(\tau)^2
\zeta_{k}^*(\tau_1)
\zeta_{k}^*(\tau_2)
\zeta_{q}^{\prime}(\tau_1)
\zeta_q(\tau_1)
\zeta_q^*(\tau_2)
\zeta_{q}^{\prime\,*}(\tau_2)\Big ]\,.
\end{align}
We notice that, in the first integral in eq.\,(\ref{eq:FirstLine}), the exchange $\tau_1 \leftrightarrow \tau_2$ transforms 
\begin{align}  
\zeta_{k}(\tau_1)
\zeta_{k}^*(\tau_2) 
\zeta_q(\tau_1)
\zeta_{q}^{\prime}(\tau_1)
\zeta_q^*(\tau_2)
\zeta_{q}^{\prime\,*}(\tau_2) 
\overset{
\tau_1 \leftrightarrow \tau_2
}{\Longrightarrow} &
\zeta_{k}(\tau_2)
\zeta_{k}^*(\tau_1) 
\zeta_q(\tau_2)
\zeta_{q}^{\prime}(\tau_2)
\zeta_q^*(\tau_1)
\zeta_{q}^{\prime\,*}(\tau_1) = \nn \\&  
[\zeta_{k}(\tau_1)
\zeta_{k}^*(\tau_2) 
\zeta_q(\tau_1)
\zeta_{q}^{\prime}(\tau_1)
\zeta_q^*(\tau_2)
\zeta_{q}^{\prime\,*}(\tau_2) ]^*\,.
\end{align}
Therefore, the first integral in eq.\,(\ref{eq:FirstLine}) is fully symmetric under the exchange $\tau_1 \leftrightarrow \tau_2$, and
 we rewrite $\Delta P_{0{\rm th}}(k,\tau)$ as 
\begin{align}
\Delta P_{0{\rm th}}&(k,\tau)  = 
\nn \\&
32\int_{\tau_{\rm in}}^{\tau}
d\tau_{1}
g(\tau_1)
\int_{\tau_{\rm in}}^{\tau}
d\tau_2 g(\tau_2) 
\int_{q_{\rm in}}^{q_{\rm end}}dq\,q^2\,
\left|\zeta_{k}(\tau)\right|^2
\RE\big[\zeta_{k}(\tau_1)
\zeta_{k}^*(\tau_2) 
\zeta_q(\tau_1)
\zeta_{q}^{\prime}(\tau_1)
\zeta_q^*(\tau_2)
\zeta_{q}^{\prime\,*}(\tau_2)\big]\label{eq:FirstLine2}\\&
-64 \int_{\tau_{\rm in}}^{\tau}
d\tau_{1}  g(\tau_1)
\int_{\tau_{\rm in}}^{\tau_1}
d\tau_2 g(\tau_2) 
\int_{q_{\rm in}}^{q_{\rm end}}dq\,q^2\, 
\RE\big[\zeta_{k}(\tau)^2 
\zeta_{k}^*(\tau_1)
\zeta_{k}^*(\tau_2)
\zeta_{q}^{\prime}(\tau_1)
\zeta_q(\tau_1)
\zeta_q^*(\tau_2)
\zeta_{q}^{\prime\,*}(\tau_2)\big]\,.
\end{align}
and apply to the first integral in eq.\,(\ref{eq:FirstLine2})
the identity in eq.\,(\ref{eq:InteIde}). 
We arrive  at 
\begin{align}
\Delta P_{0{\rm th}}(k,\tau)  =   64\int_{\tau_{\rm in}}^{\tau}
d\tau_{1}
g(\tau_1)
\int_{\tau_{\rm in}}^{\tau_1}
d\tau_2 g(\tau_2) 
&
\int_{q_{\rm in}}^{q_{\rm end}}dq\,q^2
\left\{
\RE\big[
\zeta_{k}(\tau)
\zeta_{k}^*(\tau)
\zeta_{k}(\tau_1)
\zeta_{k}^*(\tau_2) 
\zeta_q(\tau_1)
\zeta_{q}^{\prime}(\tau_1)
\zeta_q^*(\tau_2)
\zeta_{q}^{\prime\,*}(\tau_2)\big] 
\right .
\nn\\
&
- 
\left .
\RE\big[\zeta_{k}(\tau)^2 
\zeta_{k}^*(\tau_1)
\zeta_{k}^*(\tau_2)
\zeta_{q}^{\prime}(\tau_1)
\zeta_q(\tau_1)
\zeta_q^*(\tau_2)
\zeta_{q}^{\prime\,*}(\tau_2)\big]
\right\}\,.\label{eq:GUH}
\end{align}
We are now in the position of combining  the two integrand functions. 
Schematically, we define the two combinations 
\begin{align}
X \equiv  \zeta_{k}^*(\tau)\zeta_{k}(\tau_1)\,,~~~~~~~~~
Y \equiv \zeta_{k}(\tau)
\zeta_q(\tau_1)
\zeta_{q}^{\prime}(\tau_1)
\zeta_{k}^*(\tau_2)
\zeta_q^*(\tau_2)
\zeta_{q}^{\prime\,*}(\tau_2)\,,
\end{align}
such that the integrand in eq.\,(\ref{eq:GUH}) becomes 
\begin{align}\label{eq:identityRE}
\RE(XY) - \RE(X^*Y) 
= -2\IM(X)\IM(Y)\,.
\end{align}
We thus arrive at the result
\begin{mynamedbox2}{Contribution with no time derivatives on the long mode  $k$}
\vspace{-.35cm}
\begin{align}
\Delta P_{0{\rm th}}(k,\tau)  = &
-128
\int_{\tau_{\rm in}}^{\tau}
d\tau_{1}\epsilon(\tau_1)\epsilon_2^{\prime}(\tau_1) a^2(\tau_1)
\int_{\tau_{\rm in}}^{\tau_1}
d\tau_2\epsilon(\tau_2)\epsilon_2^{\prime}(\tau_2) a^2(\tau_2) 
\int_{q_{\rm in}}^{q_{\rm end}}dq\,q^2
\nn \\
&\times 
\IM \llp\zeta_{k}^*(\tau)  
\zeta_{k}(\tau_1)  \rrp
\IM \llp \zeta_{k}(\tau)
\zeta_q(\tau_1)
\zeta_{q}^{\prime}(\tau_1)
\zeta_{k}^*(\tau_2)
\zeta_q^*(\tau_2)
\zeta_{q}^{\prime\,*}(\tau_2) 
\rrp.
\label{0th}
\end{align}
\end{mynamedbox2}
Given that we are interested in modes $k$ that are much smaller than the USR-enhanced ones, they are super-horizon at the time of USR phase. Thus, for any time $\tau\gtrsim \tau_{\rm in}$ of relevance for both time integrations, one has that
\begin{equation}
    \IM\llp\zeta_k (\tau)\zeta_k^*(\tau_1)\rrp 
	\simeq \IM\llp |\zeta_k(\tau)|^2\rrp = 0,
	\label{eq:immod}
\end{equation}
which makes the above contribution negligible.

\item {\bf 1st order in time derivatives of the long mode.}
Starting from the second lines of eqs.~(\ref{eq:int_1},\ref{eq:int_2}), we now consider the sum 
\begin{align}
\Delta P_{\rm 1st}(k,\tau) = & 16\int_{\tau_{\rm in}}^{\tau}
d\tau_{1} g(\tau_1)
\int_{\tau_{\rm in}}^{\tau}
d\tau_2 g(\tau_2) 
\int_{q_{\rm in}}^{q_{\rm end}}dq\,q^2\,
\left|\zeta_{k}(\tau)\right|^2 
\nn\\
&~~~~ 
\big[
\zeta^{\prime}_{k}(\tau_1)
\zeta_{k}^*(\tau_2)
\zeta_{q}(\tau_1)
\zeta_q(\tau_1)
\zeta_{q}^{\prime\,*}(\tau_2)
\zeta_q^*(\tau_2)
+
\zeta_{k}(\tau_1)
\zeta^{\prime\,*}_{k}(\tau_2)
\zeta_q(\tau_1) 
\zeta_{q}^{\prime}(\tau_1)
\zeta_{q}^*(\tau_2)
\zeta_q^*(\tau_2)\big] \nn\\
-&  32 \RE \Big \{ \int_{\tau_{\rm in}}^{\tau}
d\tau_{1}g(\tau_1)
\int_{\tau_{\rm in}}^{\tau_1}
d\tau_2g(\tau_2) 
\int_{q_{\rm in}}^{q_{\rm end}}dq\,q^2
\zeta_{k}(\tau)^2
\nn \\
&~~~~\big[ 
\zeta^{\prime\,*}_{k}(\tau_1)
\zeta^*_{k}(\tau_2)
\zeta_{q}(\tau_1)^2
\zeta_{q}^{\prime\,*}(\tau_2)
\zeta_q^*(\tau_2) 
+ 
\zeta^*_{k}(\tau_1)
\zeta^{\prime\,*}_{k}(\tau_2)
\zeta_{q}^{\prime}(\tau_1)
\zeta_q(\tau_1)
\zeta_{q}^*(\tau_2)^2\big]\Big \}\,.
\end{align}
Manipulations analogue to those discussed in the previous point allow one to combine the two integrals together. We find 
\begin{align}
\Delta P_{\rm 1st}(k,\tau) = -64\int_{\tau_{\rm in}}^{\tau}
d\tau_{1}g(\tau_1)
\int_{\tau_{\rm in}}^{\tau_1}
d\tau_2g(\tau_2) 
&
\int_{q_{\rm in}}^{q_{\rm end}}dq\,q^2
\left\{
\IM\llp
\zeta_{k}^*(\tau)
 \zeta_{k}(\tau_1)
 \rrp
 \IM\llp
 \zeta_{k}(\tau)
\zeta_{k}^*(\tau_2)^2
\zeta_q(\tau_1)
\zeta_{k}^{\prime\,*}(\tau_2)
\zeta_q^{\prime}(\tau_1)
\rrp \right .
\nn \\
& +
\left .
\IM\llp\zeta_{k}^*(\tau)
 \zeta_{k}^{\prime}(\tau_1)\rrp
 \IM\llp
 \zeta_{k}(\tau)
\zeta_q(\tau_1)^2
\zeta_{k}^*(\tau_2)
\zeta_q^*(\tau_2)
\zeta_{q}^{\prime\,*}(\tau_2)\rrp
\right\}.
\end{align}
Again, since we are interested in modes $k$ that are much smaller than the USR-enhanced ones, and are super-horizon at the time of USR phase, the contribution within the curly brackets in the first line vanishes thanks to eq.~\eqref{eq:immod}. This leaves us with
\begin{mynamedbox2}{Contribution with one time derivatives on the long mode   $k$}
\vspace{-.35cm}
\begin{align}
\Delta P_{\rm 1st}(k,\tau) 
& 
\equiv 
-64
\int_{\tau_{\rm in}}^{\tau}
d\tau_{1}\epsilon(\tau_1)\epsilon_2^{\prime}(\tau_1)a^2(\tau_1)
\int_{\tau_{\rm in}}^{\tau_1}
d\tau_2\epsilon(\tau_2)\epsilon_2^{\prime}(\tau_2)a^2(\tau_2) 
\int_{q_{\rm in}}^{q_{\rm end}}dq\,q^2\,
\nn\\
&
\times
 \IM\llp\zeta_{k}^*(\tau)
 \zeta_{k}^{\prime}(\tau_1)\rrp
 \IM\llp
 \zeta_{k}(\tau)
\zeta_q(\tau_1)
\zeta_q(\tau_1)
\zeta_{k}^*(\tau_2)
\zeta_q^*(\tau_2)
\zeta_{q}^{\prime\,*}(\tau_2)\rrp.
 \label{1th}
\end{align}
\end{mynamedbox2}

\item {\bf 2nd order in time derivatives of the long mode.}
Analogue manipulations give
\begin{mynamedbox2}{Contribution with 
two time derivatives on the long mode $k$}
\vspace{-.35cm}
\begin{align}
\Delta P_{\rm 2nd}(k,\tau) 
& 
\equiv 
-32
\int_{\tau_{\rm in}}^{\tau}
d\tau_{1}\epsilon(\tau_1)\epsilon_2^{\prime}(\tau_1)a^2(\tau_1)
\int_{\tau_{\rm in}}^{\tau_1}
d\tau_2\epsilon(\tau_2)\epsilon_2^{\prime}(\tau_2)a^2(\tau_2) 
\int_{q_{\rm in}}^{q_{\rm end}}dq\,q^2\,
\nn\\
&
\times
\IM\llp
\zeta_{k}^*(\tau)  \zeta_{k}^{\prime}(\tau_1)
\rrp
\IM\llp
\zeta_{k}(\tau)
\zeta_q(\tau_1)
\zeta_{q}(\tau_1)
\zeta_{k}^{\prime\,*}(\tau_2)
\zeta_q^*(\tau_2)
\zeta_{q}^{*}(\tau_2)
\rrp.
\label{2th}
\end{align}
\end{mynamedbox2}
\end{itemize}
We stress that the only approximation employed so far is to take the external momentum to be much smaller than the one in the loop, i.e. $k\ll q$, which is justified in presence of a large hierarchy between the CMB and the USR scales.

\subsection{Loop correction 
at any scales}\label{sec:any}
It will be useful in the following to remove the assumption that the external momentum is much smaller than the modes in the loop, i.e. the large separation of scales $k\ll q$.
Starting again from eqs.~(\ref{eq:MasterOne},\ref{eq:DeltaP1Full},\ref{eq:DeltaP2Full}), we can proceed with analogous steps as in the previous section and define
\begin{align}
X_1 &\equiv  \zeta_{k}^*(\tau)\zeta_{k}'(\tau_1)\,,
~~~~~~~~~
Y_1 \equiv \zeta_{k}(\tau)
\zeta_{k-q}(\tau_1)
\zeta_{q}(\tau_1)
\zeta_{k-q}^*(\tau_2)
\zeta_{k}^{\prime\,*}(\tau_2)
\zeta_q^*(\tau_2) \,,
\nn
\\
X_2 &\equiv  \zeta_{k}^*(\tau)\zeta_{k}'(\tau_1)\,,
~~~~~~~~~
Y_2 \equiv \zeta_{k}(\tau)
\zeta_{k-q}(\tau_1)
\zeta_q(\tau_1)\zeta_{k}^*(\tau_2)\big[
\zeta_{k-q}^{\prime\,*}(\tau_2)\zeta_q^*(\tau_2) +
\zeta_{k-q}^*(\tau_2)\zeta_q^{\prime\,*}(\tau_2)
\big]
\,,
\nn
\\
X_3 &\equiv \zeta_{k}^*(\tau)\zeta_{k}(\tau_1)\,,
~~~~~~~~~
Y_3 \equiv 
\zeta_{k}(\tau)
\zeta_{k-q}^*(\tau_2)\zeta^{\prime\,*}_{k}(\tau_2)\zeta_q^*(\tau_2)\big[
\zeta_{k-q}^{\prime}(\tau_1)\zeta_q(\tau_1) +
\zeta_{k-q}(\tau_1)\zeta_q^{\prime}(\tau_1)
\nn
\\
X_4 &\equiv \zeta_{k}^*(\tau)\zeta_{k}(\tau_1)\,,
~~~~~~~~~
Y_4 \equiv 
\zeta_{k}(\tau)
\zeta_{k-q}^{\prime}(\tau_1)\zeta_q(\tau_1)\zeta_{k}^*(\tau_2)
\big[
\zeta_q^*(\tau_2)\zeta_{k-q}^{\prime\,*}(\tau_2) +
\zeta_{k-q}^*(\tau_2)\zeta_q^{\prime\,*}(\tau_2)
\big]
\nn
\\
X_5 &\equiv \zeta_{k}^*(\tau)\zeta_{k}(\tau_1)\,,
~~~~~~~~~
Y_5 \equiv 
\zeta_{k}(\tau)
\zeta_{k-q}^{\prime\,*}(\tau_2)\zeta_{k}^*(\tau_2)\zeta_q^*(\tau_2)
\big[
\zeta_q(\tau_1)\zeta_{k-q}^{\prime}(\tau_1) +
\zeta_{k-q}(\tau_1)\zeta_q^{\prime}(\tau_1)
\big]
\end{align}
in such a way that $\Delta P \equiv \Delta P_1 + \Delta P_2 $ can be written in the schematic form
\begin{mynamedbox2}{Generic loop correction at any scale $k$}
\vspace{-.35cm}
\begin{align}\label{eq:GenSc}
\Delta P(k,\tau) & \equiv 
-16 \int_{\tau_{\rm in}}^{\tau}
d\tau_{1}
g(\tau_1)
\int_{\tau_{\rm in}}^{\tau_1}
d\tau_2
g(\tau_2)
\int_{q_{\rm in }}^{q_{\rm end}}dq\,q^2\,
\int_{-1}^{1} d(\cos\theta) 
\times \sum_{i = 1}^5 
 \Im (X_i)  \Im (Y_i),
\end{align}
\end{mynamedbox2}
\noindent
thanks to the identity in eq.~\eqref{eq:identityRE} and where we again introduced $\epsilon(\tau)\epsilon_2^{\prime}(\tau)
  a^2(\tau) \equiv g(\tau)$.
This expression is much more intricate than the one obtained in the limit of a large hierarchy of scales between the mode $k$ and the USR loop momenta. It will allow us to seize the loop correction to the power spectrum also at the USR scales where the peak of the power spectrum is generated. 

\section{Time integration beyond the instantaneous transition and at any scales}\label{sec:Pheno}

\subsection{Loop evaluation at the CMB scales}\label{sec:LoopCMB}

Let us try to simplify the structure of eq.\,(\ref{eq:MasterOne}) in light of the approximations introduced so far. First of all, let us write eq.\,(\ref{eq:MasterOne}) 
in the form
\begin{align}
    \mathcal{P}(k) = \frac{H^2}{8\pi^2\epsilon_{\textrm{ref}}} 
\left\{
1 + 
\lim_{\tau \to 0^-}\frac{4\epsilon_{\textrm{ref}}k^3}{H^2(4\pi)^2}\Delta P_{\rm 1st}(k,\tau) + 
\lim_{\tau \to 0^-}\frac{4\epsilon_{\textrm{ref}}k^3}{H^2(4\pi)^2}\Delta P_{\rm 2nd}(k,\tau)
\right\}
\,,\label{eq:MasterOne2}
\end{align}
where we used the slow-roll approximation for the first term in 
eq.\,(\ref{eq:MasterOne}) given that 
$k$ is of the order of the CMB pivot scale. 
We focus on the leading correction given by $\Delta P_{\rm 1st}(k,\tau)$. 
Using the number of $e$-folds as the time variable,
we find that it can be written in the compact form (cf. our definition in eq.\,(\ref{eq:Pertu}))
\begin{align}
&
\Delta\mathcal{P}_{\rm 1-loop}(k_*)
\equiv 
\lim_{\tau \to 0^-}\frac{4\epsilon_{\textrm{ref}}k^3}{H^2(4\pi)^2}\Delta P_{\rm 1st}(k,\tau)  
= \nn\\
&
 32\left(
\frac{H^2}{8\pi^2\epsilon_{\textrm{ref}}}
\right)
\int_{N_{\rm in} -  
\Delta N}^{N_{\rm end} +  \Delta N} dN_1  
\frac{d\eta}{dN}(N_1) 
\int_{N_{\rm in} -  
\Delta N}^{N_1} dN_2 
\bar{\epsilon}(N_2)
\frac{d\eta}{dN}(N_2)
e^{3(N_2 - N_{\rm in})}
\int\frac{d\bar{q}}{\bar{q}^4}
\IM\left[
\bar{\zeta}_q(N_1)^2
\bar{\zeta}_q^*(N_2)
\frac{d\bar{\zeta}_q^*}{dN}(N_2)
\right]
\label{eq:PS1Simpl}
\end{align}
where we introduced the following manupulations:
\begin{itemize}
\item[{\it i)}] We use the approximation $\epsilon_2(N) \approx -2\eta(N)$. 
This is because in the relevant range of $N$ over which we integrate 
$\epsilon \ll 1$ while $\eta = O(1)$, cf. the right panel of fig.\,\ref{fig:Dyn}.
\item[{\it ii)}]  
We define $\bar{q} \equiv q/a_{\textrm{in}}H$. 
Furthermore, we use the two relations
\begin{align}
\frac{a(N_1)H}{k} = e^{N_1 - N_k}\,,~~~~\textrm{with:}~~~~
a(N_k)H = k\,,
~~~~~~~\textrm{and}~~~~
\frac{a(N_2)H}{q} = \frac{e^{N_2 - N_{\rm in}}}{\bar{q}}\,.\label{eq:CrossingDef}
\end{align}
\item[{\it iii)}] We introduce the short-hand notation
\begin{align}
  \bar{\zeta}_q(N) \equiv 
\frac{
\epsilon_{\textrm{ref}}^{1/2}q^{3/2}\zeta_q(N)}{H}\,.\label{eq:BarField}
\end{align}
The virtue of this definition is that  $\bar{\zeta}_q(N)$ is precisely the quantity we compute numerically by solving the M-S equation, cf. the right panel of fig.\,\ref{fig:TestPS}. 
Furthermore, it should be noted that 
the definition in eq.\,(\ref{eq:BarField}) is automatically invariant under the rescaling in 
eq.\,(\ref{eq:Rescaling}). The same comment applies to the definition of $\bar{q}$ and the ratios in eq.\,(\ref{eq:CrossingDef}). Consequently, an expression entirely written in terms of barred quantities is automatically invariant under the rescaling in 
eq.\,(\ref{eq:Rescaling}).
\item[{\it iv)}] 
Importantly, in the  derivation of 
eq.\,(\ref{eq:PS1Simpl}) we use (cf. appendix\,\ref{app:TimeDer})
\begin{align}\label{eq:wroskcond}
\textrm{Im}\bigg[
 \bar{\zeta}_k^*(N)
 \frac{d\bar{\zeta}_k}{dN}(N_1)
 \bigg] \simeq 
 \textrm{Im}\bigg[
 \bar{\zeta}_k^*(N_1)
 \frac{d\bar{\zeta}_k}{dN}(N_1)
 \bigg]  =  -\frac{\bar{k}^3 }{4\bar{\epsilon}(N_1)}e^{3(N_{\textrm{in}}- N_1)}\,,
\end{align}
with $\epsilon(N)$ given by eq.\,(\ref{eq:DynEps}) for generic $\delta N$. 
This is because $N_{1,2}$
vary at around $N_{\rm end}$, 
and in this time interval modes with comoving wavenumbers $k \approx k_*$
are way outside the horizon and stay constant.
For the  very same reason, we also use the slow-roll approximation
\begin{align}
   \bar{\zeta}_k(N_1)
   \bar{\zeta}^*_k(N_2) = 
   \frac{1}{4}\,.
\end{align}
\item[{\it v)}] The range of integration in 
eq.\,(\ref{eq:PS1Simpl}) is as follows.
In the case of a smooth transition, we take 
$N_{1}\in 
[N_{\textrm{in}} - \Delta N, N_{\textrm{end}} + \Delta N]$ and 
$N_{2}\in 
[N_{\textrm{in}} - \Delta N,  N_1]$
where $\Delta N$ should be large enough to complete the SR/USR/SR transition (that is, $\Delta N \gtrsim \delta N$). 
In the limit of instantaneous transition, we set 
$N_1  = N_2 = N_{\rm end}$, which corresponds to consider the dominant contribution given by the first $\delta$ function in eq.\,(\ref{eq:DeltaDer}). Moreover, we include a factor $1/2$ since, with respect to the integration over $N_2$, the argument of the $\delta$ function in eq.\,(\ref{eq:DeltaDer}) picks up the upper limit  
of the integration interval at $N_{\rm end}$. 
The integration over $q$, on the contrary, is limited by 
$\bar{q} \in [1,e^{\Delta N_{\textrm{USR}}}]$.
\end{itemize}

\subsubsection{The instantaneous transition}\label{sec:Insta}

We consider the instantaneous limit (dubbed $\delta N\to 0$ in the following) of eq.\,(\ref{eq:PS1Simpl}). We find
\begin{align}
\lim_{\delta N \to 0}\Delta\mathcal{P}_{\rm 1-loop}(k_*) =
\left(\frac{H^2}{8\pi^2\epsilon_{\textrm{ref}}} \right)\eta_{\textrm{II}}^2
\left(\frac{k_{\textrm{end}}}{k_{\textrm{in}}}\right)^{-2\eta_{\textrm{II}} + 3}
\lim_{\delta N\to 0}
(16)\int_1^{e^{\Delta N_{\textrm{USR}}}}
\frac{d\bar{q}}{\bar{q}^4}
|\bar{\zeta}_q(N_{\rm end})|^2\IM\left[
\bar{\zeta}_q(N_{\rm end})
\frac{d\bar{\zeta}_q^*}{dN}(N_{\rm end})
\right]\,.
\end{align}
This expression can be further simplified using (cf. eq.\,(\ref{eq:Wrowro}))
\begin{align}
\IM\left[
\bar{\zeta}_q(N_{\rm end})
\frac{d\bar{\zeta}_q^*}{dN}(N_{\rm end})
\right]  = 
\frac{\bar{q}^3}{4}
e^{(2\eta_{\textrm{II}}-3)(N_{\textrm{end}}-
N_{\textrm{in}})}
=  
\frac{\bar{q}^3}{4}\left(
\frac{k_{\rm end}}{k_{\rm in}}
\right)^{2\eta_{\rm II}-3}
\,,
\end{align}
so that we write
\begin{align}
\lim_{\delta N \to 0}\Delta\mathcal{P}_{\rm 1-loop}(k_*) =
\left(\frac{H^2}{8\pi^2\epsilon_{\textrm{ref}}} \right)\,4\eta_{\textrm{II}}^2\,
\lim_{\delta N\to 0}
\int_1^{e^{\Delta N_{\textrm{USR}}}}
\frac{d\bar{q}}{\bar{q}}
|\bar{\zeta}_q(N_{\rm end})|^2\,.\label{eq:FinInte}
\end{align}
We remark that this expression is valid for generic values of $\eta_{\rm II}$ during USR.
 
We consider now the computation of the last integral. 
The factor
$\bar{\zeta}_q$ grows exponentially during the 
USR phase. 
In the case of sub-horizon modes, we have $\bar{\zeta}_q(N) \sim e^{-(1-\eta_{\rm II})N}$ while in the case of super-horizon modes we find $\bar{\zeta}_q(N) \sim  e^{-(3-2\eta_{\rm II})N}$ (cf. appendix\,\ref{app:TimeDer}).
However, the precise estimate of  the integral in eq.\,(\ref{eq:FinInte}) is complicated by the fact that 
curvature modes $\bar{\zeta}_q$ with $\bar{q}\in [1,\exp(\Delta N_{\textrm{USR}})]$ are neither sub- nor super-horizon but they exit the horizon during the USR phase, thus making the analytical estimate of the argument of their exponential growth more challenging.

The situation simplifies if we consider some special values of $\eta_{\textrm{II}}$. 
We consider the case $\eta_{\textrm{II}} = 3$ 
(that is $\epsilon_2 = -6$, it should be noted that this is also the case studied in ref.\,\cite{Kristiano:2022maq}). 
In this case, everything can be computed analytically. 
We find the scaling 
\begin{align}
\lim_{\delta N\to 0}
\int_1^{e^{\Delta N_{\textrm{USR}}}}
\frac{d\bar{q}}{\bar{q}}
|\bar{\zeta}_q(N_{\rm end})|^2 
\approx  \frac{e^{6\Delta N_{\rm USR}}}{4}(1 + \Delta N_{\rm USR})
= 
\frac{1}{4}\left(
\frac{k_{\rm end}}{k_{\rm in}}
\right)^6\left[
1 + \log\left(
\frac{k_{\rm end}}{k_{\rm in}}
\right)
\right]\,,\label{eq:ModeInte}
\end{align}
which becomes more and more accurate for larger $k_{\rm end}/k_{\rm in}$.
The final result is 
\begin{mynamedbox1}{
Leading one-loop correction at CMB scales in the instantaneous SR/USR/SR transition}
\vspace{-0.35cm}
\begin{align}
\lim_{\delta N \to 0}
\Delta\mathcal{P}_{\rm 1-loop}(k_*) \approx 
\left(\frac{H^2}{8\pi^2\epsilon_{\textrm{ref}}} \right) \eta_{\rm II}^2
\left(\frac{k_{\textrm{end}}}{k_{\textrm{in}}}\right)^{6}
\left[
1 + \log\left(
\frac{k_{\rm end}}{k_{\rm in}}
\right)
\right]
\,,~~~~~~~\textrm{with}~~\eta_{\textrm{II}} = 3,\,\,\eta_{\rm III} = 0\label{eq:AhiAhi}
\end{align}
\end{mynamedbox1}
\noindent 
which perfectly agrees with the findings of ref.\,\cite{Kristiano:2022maq} in the same limit.

The above result has a number of limitations, which we address separately in the following subsections:
\begin{itemize}
\item[$\circ$] \textbf{Dynamics during USR, section\,\ref{sec:GenericUSR}.} 
We modify the assumption $\eta_{\rm II} =  3$ and we take $\delta N\to 0$ and $\eta_{\rm III} = 0$.  
\item[$\circ$] \textbf{Dynamics at the SR/USR/SR transition, section\,\ref{sec:trans}.}
We consider $\delta N \neq 0$, with generic $\eta_{\rm II}$ but $\eta_{\rm III} = 0$. 
Considering a non-zero value of $\delta N$ is very  important because it corresponds to a more realistic smooth SR/USR/SR transition, as opposed to the instantaneous limit with $\delta N = 0$.

\end{itemize}

\subsubsection{Dynamics during USR}\label{sec:GenericUSR}

We compute eq.\,(\ref{eq:FinInte}) for generic values of $\eta_{\rm II}$, still keeping $\delta N\to 0$ and $\eta_{\rm III} = 0$.  
From the computation of the tree-level power spectrum (cf. section\,\ref{sec:MinDynUSR} and fig.\,\ref{fig:TestPS}) we define
\begin{align}
\frac{\mathcal{P}_{\rm USR}}{
\mathcal{P}_{\rm CMB}
} \equiv 
\frac{\mathcal{P}(\bar{k}_{\rm max})}{\mathcal{P}(\bar{k}\ll 1)}\,,\label{eq:PSMax}
\end{align}
where $\bar{k}_{\rm max}$ represent the position of the max of $\mathcal{P}(\bar{k})$ after the growth 
due to the USR dynamics.
\begin{figure}[h]
\begin{center}
$$\includegraphics[width=.495\textwidth]{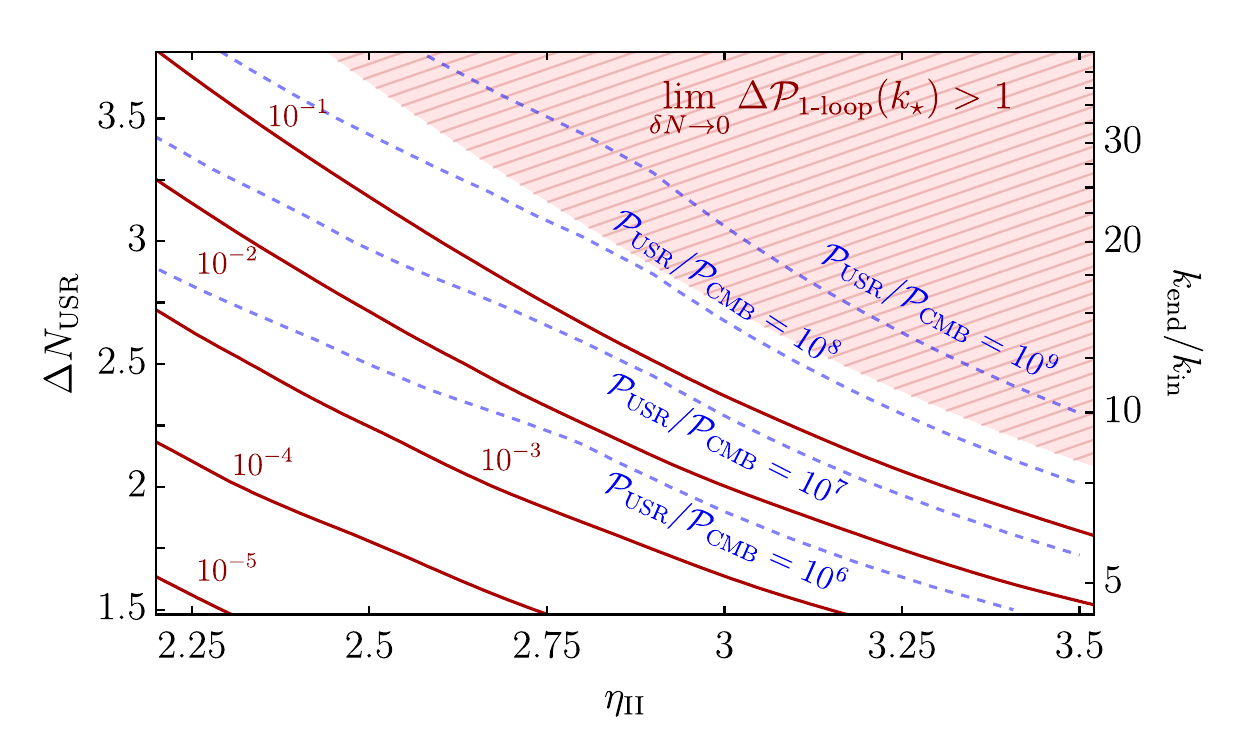}~
\includegraphics[width=.495\textwidth]{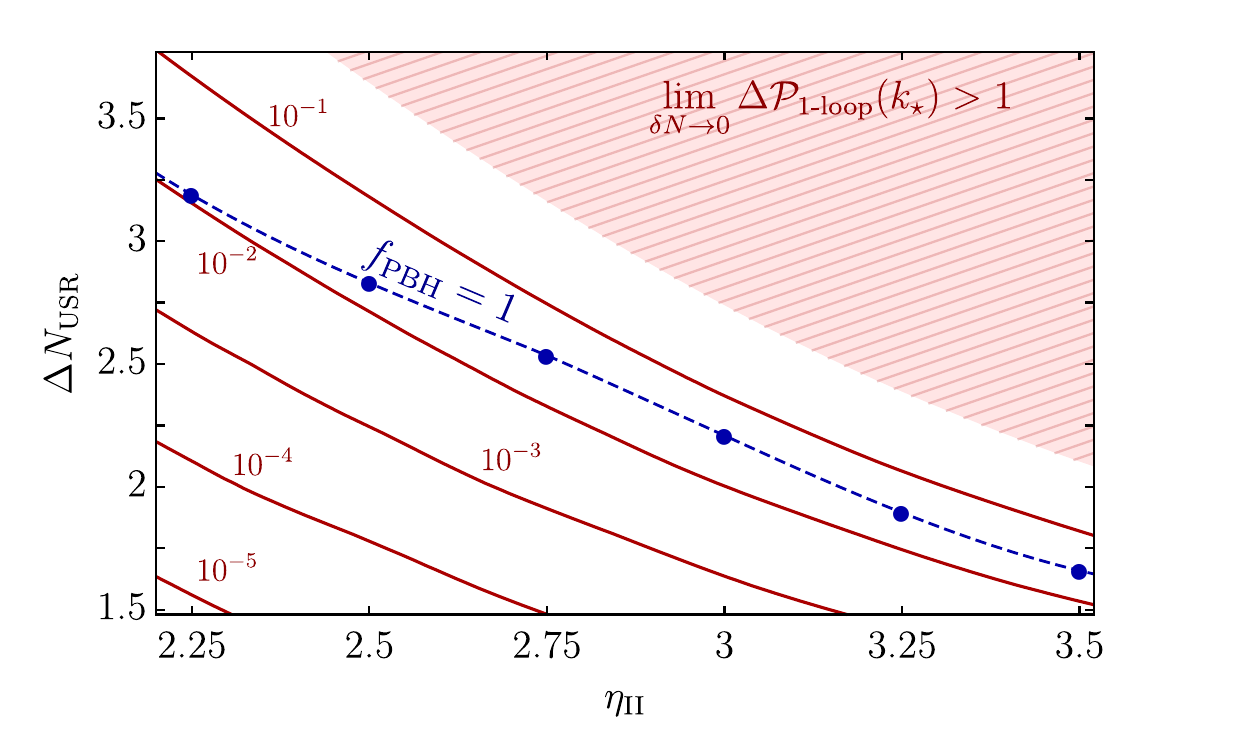}$$
\vspace{-0.5cm}
\caption{\em  
In both panels,  
we consider a generic USR dynamics with varying $\eta_{\rm II}$ ($x$-axis) and $\Delta N_{\rm USR}$ ($y$-axis). We take $\eta_{\rm III} = 0$ and the instantaneous limit $\delta N = 0$.
We plot in solid red
contours of constant $\lim_{\delta N \to 0}
\Delta\mathcal{P}_{\rm 1-loop}(k_*),$
defined in eq.\,(\ref{eq:PS1Simpl}) and
computed according to eq.\,(\ref{eq:FinInte}) with $H^2/8\pi^2\epsilon_{\rm ref} = 2.1\times 10^{-9}$. 
 The region hatched in red 
 is defined by the condition  $\lim_{\delta N \to 0}
\Delta\mathcal{P}_{\rm 1-loop}(k_*) > 1$. 
\textit{\textbf{Left panel:}} We superimpose contours 
of constant $\mathcal{P}_{\rm USR}/\mathcal{P}_{\rm CMB}$ as defined in eq.\,(\ref{eq:PSMax}) (dashed blue). 
\textit{\textbf{Right panel:}} We superimpose the line defined by the condition $f_{\rm PBH} = 1$. 
Along this line, we get 100\% of DM in the form of asteroid-mass PBHs.
 }\label{fig:RegionBound}  
\end{center}
\end{figure}
We compare in the left panel of  fig.\,\ref{fig:RegionBound} contours 
of constant $\mathcal{P}_{\rm USR}/\mathcal{P}_{\rm CMB}$ (dashed blue) and constant $\lim_{\delta N \to 0}
\Delta\mathcal{P}_{\rm 1-loop}(k_*) $ (solid red). 
We take $H^2/8\pi^2\epsilon_{\rm ref} = 2.1\times 10^{-9}$.
Our analysis shows that enhancements  
$\mathcal{P}_{\rm USR}/\mathcal{P}_{\rm CMB} \gtrsim 10^8$  are barely compatible with the perturbativity condition
$\lim_{\delta N \to 0}
\Delta\mathcal{P}_{\rm 1-loop}(k_*) < 1$, which roughly means ``loops $<$ tree level''. The region $\lim_{\delta N \to 0}
\Delta\mathcal{P}_{\rm 1-loop}(k_*) > 1$ is hatched in red in  fig.\,\ref{fig:RegionBound}.

We can actually do better and compare with a careful computation of the PBH abundance.  
The parameters of the dynamics in section\,\ref{sec:MinDynUSR} (with $\eta_{\rm III} = 0$ and $\delta N\to 0$) are chosen in such a way that the integral 
\begin{align}
f_{\rm PBH} \equiv \frac{\Omega_{\rm PBH}}{\Omega_{\rm CDM}} = \int f_{\rm PBH}(M_{\rm PBH}) d\log M_{\rm PBH} \approx 1\,,\label{eq:fPBHinte}
\end{align}
which means that we get $\approx 100\%$ of DM in the form of PBHs. 
More in detail, we tune, for each $\eta_{\rm II}$, the value of $\Delta N_{\rm USR}$ so to get  $f_{\rm PBH} \approx 1$; 
we choose 
the numerical value of $k_{\rm in}$ in such a way that the peak of the PBH mass distribution $f_{\rm PBH}(M_{\rm PBH})$ falls within the interval $M_{\rm PBH}/M_{\odot} \in [10^{-14},\, 10^{-13}]$ in which the condition 
$f_{\rm PBH} \approx 1$ is experimentally allowed, the so-called asteroid-mass PBHs \cite{Carr:2020xqk}. 
We compute eq.\,(\ref{eq:fPBHinte}) using threshold statistics and including the full non-linear relation between the curvature and the density contrast fields (cf.\,\cite{Young:2019yug,DeLuca:2019qsy}).
The interested reader can find more details 
on the computation of the abundance in ref.~\cite{Franciolini:2022pav} and refs. therein.\footnote{It is possible to further improve our analysis by including the presence of primordial non-gaussianity (e.g. \cite{Namjoo:2012aa,Chen:2013eea,Cai:2018dkf,Pi:2022ysn,Passaglia:2018ixg,Figueroa:2020jkf,Biagetti:2021eep,Figueroa:2021zah,Tomberg:2023kli}). 
In the case of local non-guassianity parametrized by a positive non-Guassian parameter $f_{\rm NL}$, as expected in the case of USR, we generically expect a larger abundance of PBHs compared to the Gaussian case\,\cite{Atal:2019cdz,Ferrante:2022mui,Gow:2022jfb,Ferrante:2023bgz,Ianniccari:2024bkh}. 
This means that, in order to achieve the same abundance of PBHs, one needs a power spectrum with a smaller peak amplitude. This argument implies  
that the presence of primordial non-Gaussianity will tend to decrease the relevance of the one-loop corrections.
} 
In the right panel of fig.\,\ref{fig:RegionBound} we 
plot the line defined by the condition $f_{\rm PBH} \approx 1$. The comparison 
between the left- and right-hand side of fig.\,\ref{fig:RegionBound}  shows that, in order to fulfil the condition $f_{\rm PBH} \approx 1$, one needs $\mathcal{P}_{\rm USR}/\mathcal{P}_{\rm CMB}  = O(10^7)$.\footnote{There is some difference between peak theory and threshold statistics in
the computation of the abundance, already present at the Gaussian level (see, e.g., refs.\,\cite{Green:2004wb,Young:2014ana,DeLuca:2019qsy}). The approach based on peak theory usually requires slightly smaller values of $\mathcal{P}_{\rm USR}/\mathcal{P}_{\rm CMB}$ in order to get the same abundance of PBHs, thus making our findings, based on threshold statistics, even stronger.} 

We conclude that the condition $f_{\rm PBH} \approx 1$ lies within the region in which perturbativity is still applicable. 
This is in contrast with the conclusion reached in refs.\,\cite{Kristiano:2022maq,Kristiano:2023scm,Firouzjahi:2023aum} in the limit of instantaneous SR/USR/SR transition.
The origin of the difference is the more accurate calculation of the PBHs abundance performed in our work. In particular, in previous analyses, estimates of $f_{\rm PBH}\approx 1$ are based on requiring $\mathcal{P}_{\rm USR}\simeq10^{-2}$, and on the scaling $\Delta\mathcal{P} = (k_{\rm end}/k_{\rm in})^{2\eta_{\rm II}}$ in order to capture the growth of the power spectrum at small scales. However, as explained in section~\ref{sec:MinDynUSR}, this scaling does not accurately describe the amplitude of the power spectrum at its peak, see the left panel of fig.\,\ref{fig:TestPS}.
Ref.\,\cite{Motohashi:2023syh} computed one-loop corrections in the limit of instantaneous SR/USR/SR transitions in scenarios with $\eta_{\rm II}\leq 3$, finding that the perturbativity bound is relaxed for $\eta_{\rm II}$ smaller than 3. At a qualitative level, similar results are obtained in the left panel of fig.\,\ref{fig:RegionBound}.
In ref.\,\cite{Motohashi:2023syh}, the perturbativity bound is traduced in an upper limit on the power-spectrum at the scale $k_{\rm end}.$ However, as explained above, this procedure underestimates the maximum amplitude of the power spectrum, see again the left panel of fig.\,\ref{fig:TestPS}.

\subsubsection{Dynamics at the SR/USR/SR transition}\label{sec:trans}

We go beyond the instantaneous transition to check if there are cancellations that affect the order-of-magnitude of the result in eq.\,(\ref{eq:AhiAhi}). 
There are indeed compelling reasons to believe that this is the case, as advocated in refs.\,\cite{Riotto:2023gpm,Firouzjahi:2023aum,Firouzjahi:2023ahg}.
The story goes as follows (the original argument was proposed in ref.\,\cite{Cai:2018dkf} in which the role of non-Gaussianity from non-attractor inflation models was considered). 
From the Hubble parameters in eq.\,(\ref{eq:HubblePar1}) and the background dynamics that follow from the action in eq.\,(\ref{eq:BackAction}), it is possible to calculate the potential and its derivatives exactly. Up to the third order in field derivatives, we find (see also ref.\,\cite{Leach:2002ar})
\begin{align}
V(\phi) & = H^2(3-\epsilon)\,,\\
V^{\prime}(\phi) & = 
\frac{H^2}{\sqrt{2}}
\epsilon^{1/2}\left(
6-2\epsilon +  \epsilon_2
\right)\,,\\
V^{\prime\prime}(\phi) & = H^2\left(
6 \epsilon - 2 \epsilon^2 - \frac{3 \epsilon_2}{2} + 
 \frac{5\epsilon \epsilon_2}{2} - \frac{\epsilon_2^2}{4} - 
 + \frac{\epsilon_2 \epsilon_3}{2}
\right)\,,\\
V^{\prime\prime\prime}(\phi) & = \frac{H^2}{
2\sqrt{2\epsilon}
}
\left[
-8 \epsilon^3 + 
 6 \epsilon^2 (4 + 3 \epsilon_2) - \epsilon \epsilon_2(18 + 6 \epsilon_2 + 7 \epsilon_3) + \epsilon_2 \epsilon_3 (3 + \epsilon_2 +\epsilon_3 + \epsilon_4)
\right] \\
& = 
\frac{1}{
2\sqrt{2\epsilon}
}\left\{
H^2\left[
\frac{\ddot{\epsilon_2}}{H^2}+
(3+\epsilon_2)
\frac{\dot{\epsilon_2}}{H}
\right]
+O(\epsilon)
\right\}\,,\label{eq:ThirdDerV}
\end{align}
where in eq.\,(\ref{eq:ThirdDerV}) we expanded in the parameter $\epsilon$ and wrote $\epsilon_{3,4}$ in terms of $\epsilon_2$.  
Consider the flat gauge in which curvature perturbations are entirely encoded into field fluctuations $\delta\phi$ by means of the relation $\zeta = H\delta\phi/\dot{\phi} =  -\delta\phi/\sqrt{2\epsilon}$. 
In this gauge, the interactions come from Taylor-expanding the quadratic action in field fluctuations and, at the cubic order, one expects 
\begin{align}
\mathcal{L}_3 \supset 
\frac{a^3}{6}
V^{\prime\prime\prime}\delta\phi^3 = 
- \frac{a^3\epsilon}{3}
(\sqrt{2\epsilon}V^{\prime\prime\prime})\zeta^3 = 
- \frac{a^3\epsilon}{6}
\left\{
H^2\left[
\frac{\ddot{\epsilon_2}}{H^2}+
(3+\epsilon_2)
\frac{\dot{\epsilon_2}}{H}
\right]
+O(\epsilon)
\right\}\zeta^3\,.\label{eq:FlatGauCu}
\end{align}
As shown in ref.\,\cite{Chen:2013eea}, 
the above interaction agrees (modulo a surface term) with eq.\,(\ref{eq:Yoko}) if we integrate by parts
\begin{align}
\int d^4x\,\frac{\epsilon\dot{\epsilon_2}}{2}a^3\dot{\zeta}\zeta^2 ~~~\to~~~  
-\int d^4x\,
\frac{1}{6}\frac{d}{dt}\left(
\epsilon\dot{\epsilon_2}a^3
\right)\zeta^3 = 
-\int d^4x\,
\frac{a^3\epsilon}{6}H^2\left[
\frac{\ddot{\epsilon_2}}{H^2}+
(3+\epsilon_2)
\frac{\dot{\epsilon_2}}{H}
\right]\zeta^3\,,\label{eq:CuboIBP}
\end{align}
where in the last step we used the exact identity
\begin{align}
\frac{d}{dt}\left(
\epsilon\dot{\epsilon_2}a^3
\right) = 
 a^3\epsilon  H^2\left[
\frac{\ddot{\epsilon_2}}{H^2}+
(3+\epsilon_2)
\frac{\dot{\epsilon_2}}{H}
\right]\,.\label{eq:EffectiveCoupling}
\end{align}
The cubic interaction in eq.\,(\ref{eq:CuboIBP}) agrees with eq.\,(\ref{eq:FlatGauCu}) up to $\epsilon$-suppressed terms.  
Rewriting the interaction as in eq.\,(\ref{eq:FlatGauCu}) is quite instructive. 
From eq.\,
(\ref{eq:EffectiveCoupling}),  it seems plausible that drastic variations in time of $\epsilon_2$ could enhance the cubic interaction. However, eq.\,(\ref{eq:FlatGauCu}) shows that these interactions are ultimately controlled by $V^{\prime\prime\prime}$ so that  in the case with a smooth SR/USR/SR transition  
in which $V^{\prime\prime\prime}$ is expected to be ``small'',
there must be cancellations at work within the combination in eq.\,(\ref{eq:EffectiveCoupling}) so that the relevant coupling in  eq.\,(\ref{eq:FlatGauCu}) reduces to the term that is SR suppressed. 
This is the main argument that was put forth in refs.\,\cite{Riotto:2023gpm,Firouzjahi:2023aum,Firouzjahi:2023ahg}. 

\begin{figure}[h]
\begin{center}
$$\includegraphics[width=.495\textwidth]{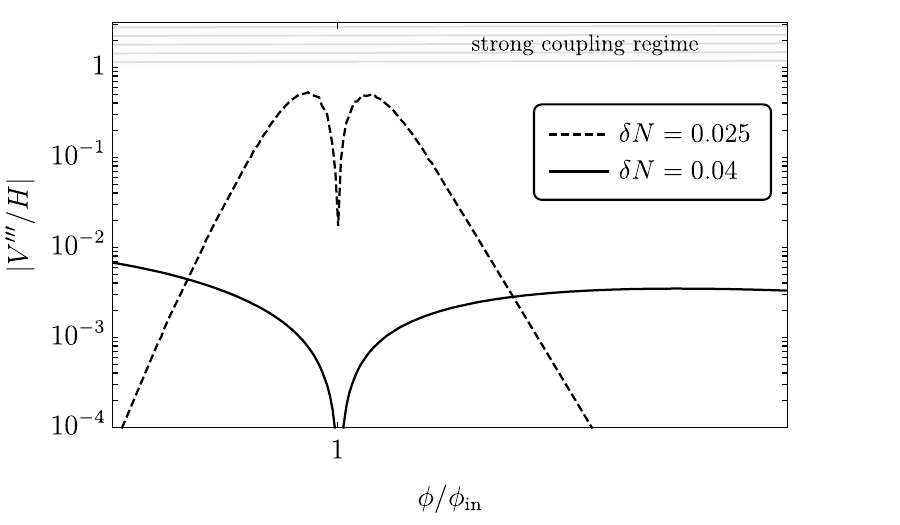}~
\includegraphics[width=.495\textwidth]{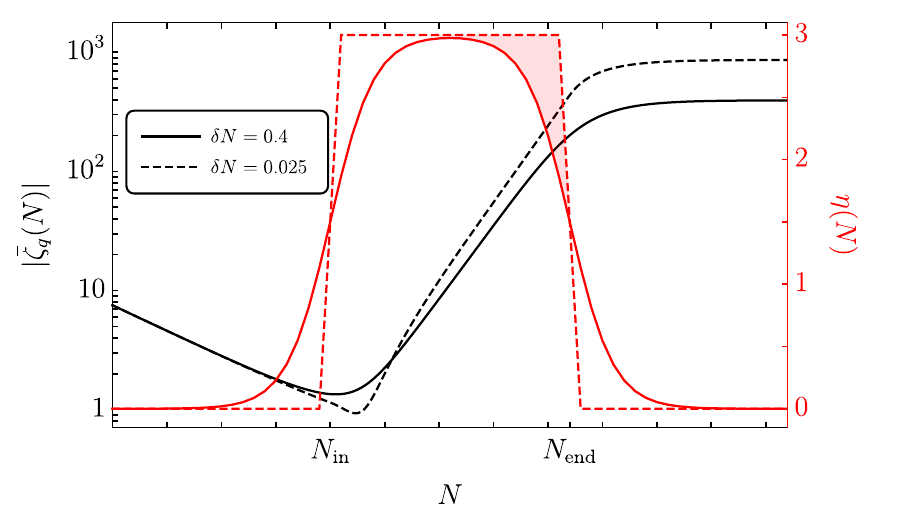}$$\vspace{-0.5cm}
\caption{\em 
\textit{\textbf{Left panel:}} 
Graph of $V^{\prime\prime\prime}/H$ as function of the background field value $\phi$ for two representative dynamics with, respectively, $\delta N = 0.025$ (dashed black) and $\delta N = 0.4$ (solid black). 
Starting from the dynamics defined as in section\,\ref{sec:MinDynUSR}, we compute the potential by means of the reverse engineering approach described in ref.\,\cite{Franciolini:2022pav}. 
The values $V_{\star}$ and $H_{\star}$ of, respectively, the potential and the Hubble rate at CMB scales are chosen in such a way that both dynamics are consistent with CMB observations (namely, $V_{\star} \simeq  3\times 10^{-9}$ and $H_{\star}  \simeq  3\times 10^{-5}$ with the reduced Planck mass set to 1). 
On the right (left) side of the field value $\phi = \phi_{\rm in}$,
$V^{\prime\prime\prime}/H$ is negative (positive).
\textit{\textbf{Right panel:}}  
Left-side $y$-axis: time evolution of the curvature modes $|\bar{\zeta}_q(N)|$ for $\bar{q}  = 2$ in the case $\delta N = 0.025$ (dashed black line) and $\delta N = 0.4$ (solid black line). 
Right-side $y$-axis: profile of $\eta$ in the case $\delta N = 0.025$ (dashed red line) and $\delta N = 0.4$ (solid red line). The region shaded in red highlights the  difference between the sharp and the smooth transition in terms of $\eta$: in the case of a sharp transition, the curvature mode has more time to grow under the  effect of the  negative friction phase implied by the condition $\eta > 3/2$.
 }\label{fig:CancDetail}  
\end{center}
\end{figure}

We shall elaborate further on this point. 
First  of all, let us clarify what ``$V^{\prime\prime\prime}$  small'' means.  
We rewrite eq.\,(\ref{eq:ThirdDerV}) as follows (we omit the $O(\epsilon)$ terms and, for clarity's sake, we write explicitly the reduced Planck mass)
\begin{align}
\frac{V^{\prime\prime\prime}}{H}  =  
\left(\frac{H}{\bar{M}_{\rm Pl}}\right)
\frac{1}{
2\sqrt{2\epsilon}
}
\left[
\frac{\ddot{\epsilon_2}}{H^2}+
(3+\epsilon_2)
\frac{\dot{\epsilon_2}}{H}
\right]\,.\label{eq:Intu}
\end{align}
On the left-hand side, the quantity $V^{\prime\prime\prime}/H$ has the dimension of a coupling. Consequently, imposing 
the condition $V^{\prime\prime\prime}/H < 1$ corresponds to a weak coupling regime while $V^{\prime\prime\prime}/H > 1$ corresponds to a strongly coupled one.  
Said differently, from the perspective of the right-hand side of eq.\,(\ref{eq:Intu}), the condition $V^{\prime\prime\prime}/H > 1$ corresponds to a situation in which the a-dimensional factor in front of $H/\bar{M}_{\rm Pl}$ becomes so large that it overcomes the natural suppression given by $H/\bar{M}_{\rm Pl} \ll 1$. 
In the left panel of fig.\,\ref{fig:CancDetail}, we compute the ratio $V^{\prime\prime\prime}/H$ for two benchmark SR/USR/SR dynamics with different values of $\delta N$. 
In the case in which $\delta N \to 0$ (sharp transition), we  observe that $V^{\prime\prime\prime}/H$ dangerously grows towards the strongly coupled regime while in the case of a smooth transition it safely takes 
$O(\ll 1)$ values.  
As anticipated at the beginning of this section, 
this argument confirms that in the case of a smooth transition we expect a reduction in the size of the trilinear interaction controlled by the factor in eq.\,(\ref{eq:EffectiveCoupling}).

With this motivation in mind, we go back to the analysis in section\,\ref{sec:Insta} and we perform the following calculation.

We compute numerically the integral in eq.\,(\ref{eq:PS1Simpl}) in order to check the validity of the scaling in  
eq.\,(\ref{eq:AhiAhi}) beyond the limit of instantaneous  transition.  
We define the quantity 
\begin{align}
\mathcal{J}_{\delta N}&(\eta_{\rm II},\Delta N_{\rm USR}) \equiv 
\Delta\mathcal{P}_{\rm 1-loop}(k_*) /\mathcal{P}_{\rm tree}(k_*) 
=
\nn\\
&
 32
\int_{N_{\rm in}-\Delta N}^{
N_{\rm end} + \Delta N}
dN_1  
\frac{d\eta}{dN}(N_1) 
\int_{N_{\rm in}-\Delta N}^{N_1}
dN_2
\bar{\epsilon}(N_2)
\frac{d\eta}{dN}(N_2)
e^{3(N_2 - N_{\rm in})}
\int_1^{e^{\Delta N_{\rm USR}}}
\frac{d\bar{q}}{\bar{q}^4}
\IM\left[
\bar{\zeta}_q(N_1)^2
\bar{\zeta}_q^*(N_2)
\frac{d\bar{\zeta}_q^*}{dN}(N_2)
\right]\,,\label{eq:FullIntegral}
\end{align}
that we can directly compare, in the case $\eta_{\rm II} = 3$, with 
$\eta_{\rm II}^2(k_{\rm end}/k_{\rm in})^6[1 + \log(k_{\rm end}/k_{\rm in})]$ in  
eq.\,(\ref{eq:AhiAhi}) using the fact that 
$k_{\rm end}/k_{\rm in} =  e^{\Delta N_{\rm USR}}$.
First, we set $\delta N$ to a very small number, in order to mimic the limit $\delta N \to 0$, and evaluate $\mathcal{J}_{\delta N}(3,\Delta N_{\rm USR})$ as function 
of $\Delta N_{\rm USR}$.
\begin{figure}[h]
\begin{center}
$$\includegraphics[width=.495\textwidth]{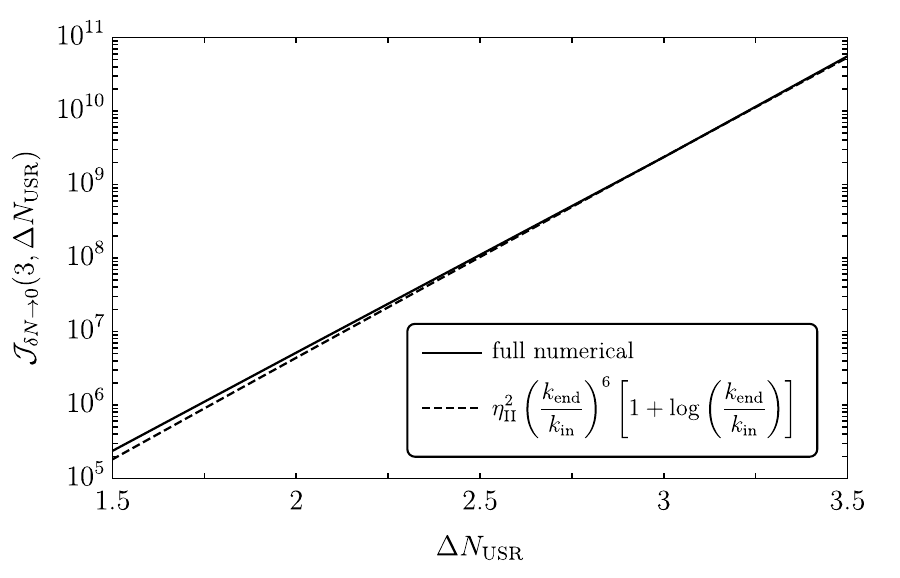}~
\includegraphics[width=.495\textwidth]{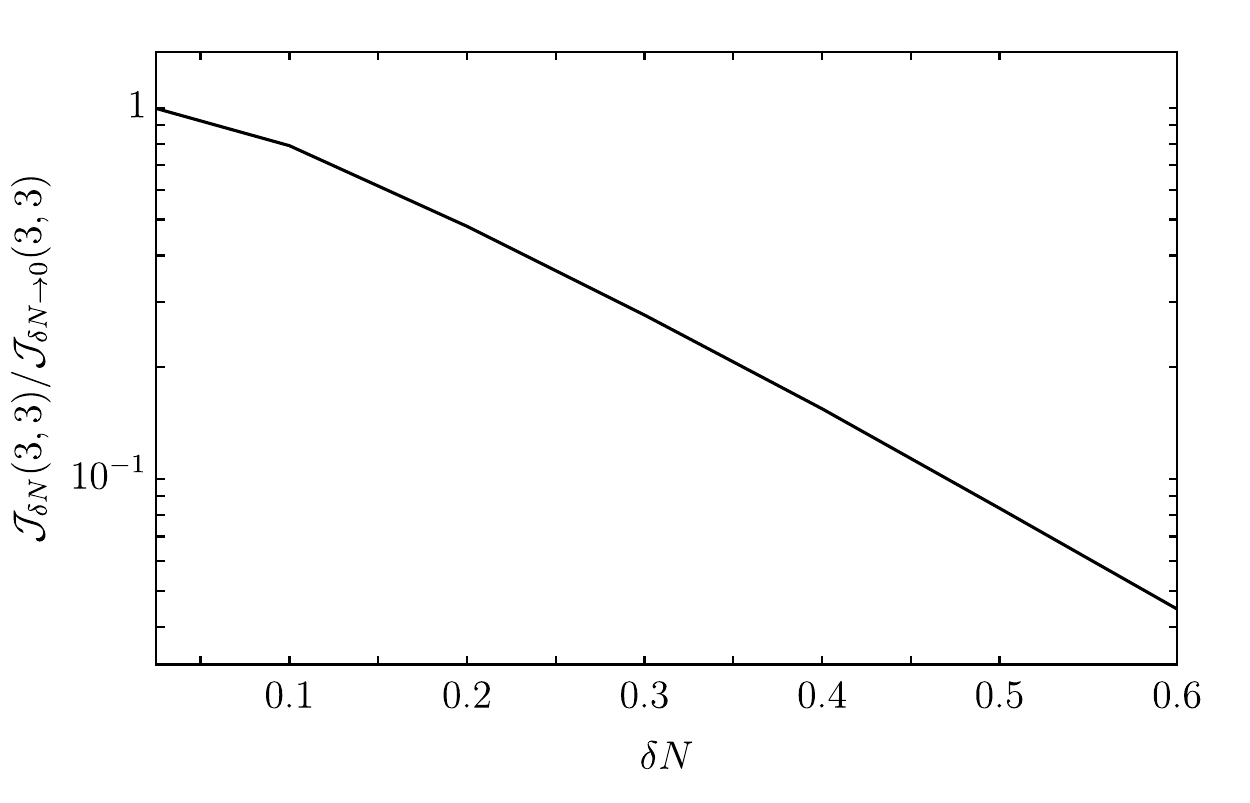}$$\vspace{-0.5cm}
\caption{\em 
\textit{\textbf{Left panel:}} Comparison between 
the value of the full integral in eq.\,(\ref{eq:FullIntegral}) and the analytical estimate in eq.\,(\ref{eq:AhiAhi}). 
To mimic the instantaneous transition we take $\delta N = 0.025$. 
\textit{\textbf{Right panel:}}
We plot the ratio $\mathcal{J}_{\delta N}(3,\Delta N_{\rm USR})$ as function of $\delta N$. 
In both figures we take $\eta_{\rm II} = 3$.
 }\label{fig:SmoothTransition2}  
\end{center}
\end{figure}
The comparison is shown in the left panel of fig.\,\ref{fig:SmoothTransition2}. We find an excellent agreement in particular for large $\Delta N_{\rm USR}.$ This is expected, since the approximation in eq.\,(\ref{eq:ModeInte}) is more accurate for larger $k_{\rm end}/k_{\rm in}$. 
Then, we set $\Delta N_{\rm USR} = 3$ and 
compare the value of  
$\mathcal{J}_{\delta N \to 0}(3,3)$ with 
$\mathcal{J}_{\delta N}(3,3)$ as function of 
$\delta N$.  
We plot the ratio  
$\mathcal{J}_{\delta N}(3,3)/\mathcal{J}_{\delta N \to 0}(3,3)$ 
in the right panel of fig.\,\ref{fig:SmoothTransition2}.

Realistic single-field models that  
feature the presence of a phase of USR dynamics typically have $\delta N = 0.4\divisionsymbol 0.5$ (cf., e.g., ref.\,\cite{Ballesteros:2020qam,Taoso:2021uvl}). 
This means that, according to our result in the right panel of fig.\,\ref{fig:SmoothTransition2}, 
we expect that in realistic models the size of the loop correction gets reduced by  one order of magnitude with respect to what is obtained in the limit of instantaneous SR/USR/SR  transition.
This confirms the intuition presented in refs.\,\cite{Riotto:2023gpm}.

\begin{figure}[h]
\begin{center}
\includegraphics[width=.8\textwidth]{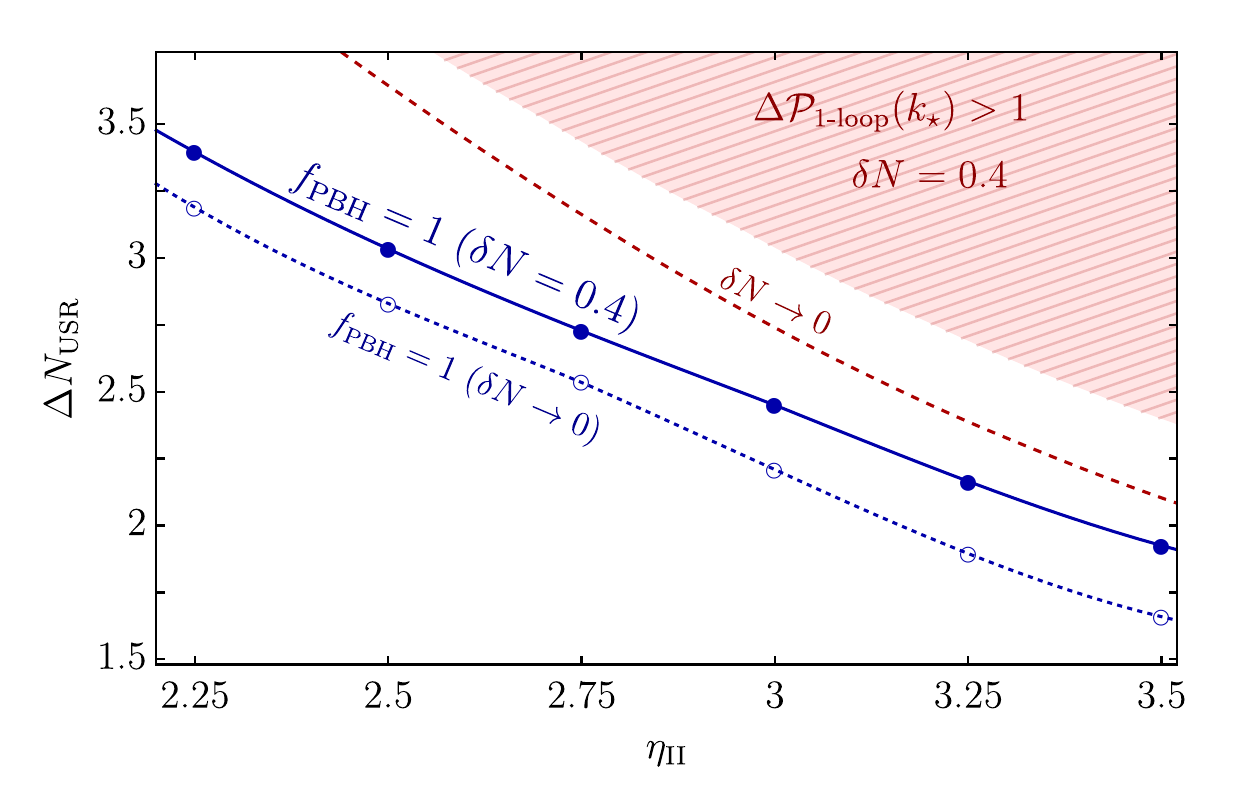}\vspace{-0.5cm}
\caption{\em 
We consider a generic USR dynamics with varying $\eta_{\rm II}$ ($x$-axis) and $\Delta N_{\rm USR}$ ($y$-axis). We take $\eta_{\rm III} = 0$ and the smooth limit $\delta N = 0.4$.
The region hatched in red corresponds to $
\Delta\mathcal{P}_{\rm 1-loop}(k_*) > 0$. 
Along the line defined by the condition $f_{\rm PBH} = 1$, we get 100\% of DM in the form of asteroid-mass PBHs.
The dotted blue line and  the red dashed line correspond, respectively, to the conditions $f_{\rm PBH} = 1$ and 
$\lim_{\delta N\to 0}\Delta\mathcal{P}_{\rm 1-loop}(k_*) > 0$ as derived in the limit of   instantaneous transition.
 }\label{fig:FinalPlot}  
\end{center}
\end{figure}

It should be noted, however, as evident from our discussion in section\,\ref{sec:MinDynUSR}, that in the case of smooth SR/USR/SR transition the amplitude of the power spectrum gets reduced with respect to 
the $\delta N \to 0$ limit (cf. the left panel of fig.\,\ref{fig:TestPS}).
The origin of this effect becomes evident if we consider the right panel of fig.\,\ref{fig:CancDetail}. In this figure, we plot the time evolution of the curvature mode $|\bar{\zeta}_{q}|$ with $\bar{q} = 2$ in the two cases of a sharp and smooth transition (dashed and solid lines, respectively -- see caption for details). In the case of a sharp transition, the curvature mode experiences a longer USR phase, and its final amplitude is larger with respect to the case of a smooth transition.
As a consequence, therefore, we expect that the smaller size of the loop correction will be, at least partially, compensated by the fact that 
finite $\delta N$ also reduces the  amplitude of the tree-level power spectrum.   
In order to quantify this information, we repeat the analysis done in section\,\ref{sec:GenericUSR} but now for finite $\delta N$.
We plot our result in fig.\,\ref{fig:FinalPlot}. 
For definiteness, we consider the benchmark value $\delta N = 0.4$ while we keep 
$\eta_{\rm II}$ and $\Delta N_{\rm USR}$ generic as in fig.\,\ref{fig:RegionBound}.

Our numerical analysis mirrors the previous intuition. 
The perturbativity bound (the region hatched in red corresponds to the condition $\Delta\mathcal{P}_{\rm 1-loop}(k_*) > 0$) gets weaker because of the partial cancellation illustrated in the right panel of fig.\,\ref{fig:SmoothTransition2}.  
However, as previously discussed, the drawback is that taking  $\delta N \neq 0$ also reduces the peak amplitude of  the power spectrum. Consequently, the condition $f_{\rm PBH} = 1$ requires, for fixed $\eta_{\rm II}$, larger $\Delta N_{\rm USR}$.

\begin{figure}[h]
\begin{center}
$$\includegraphics[width=.495\textwidth]{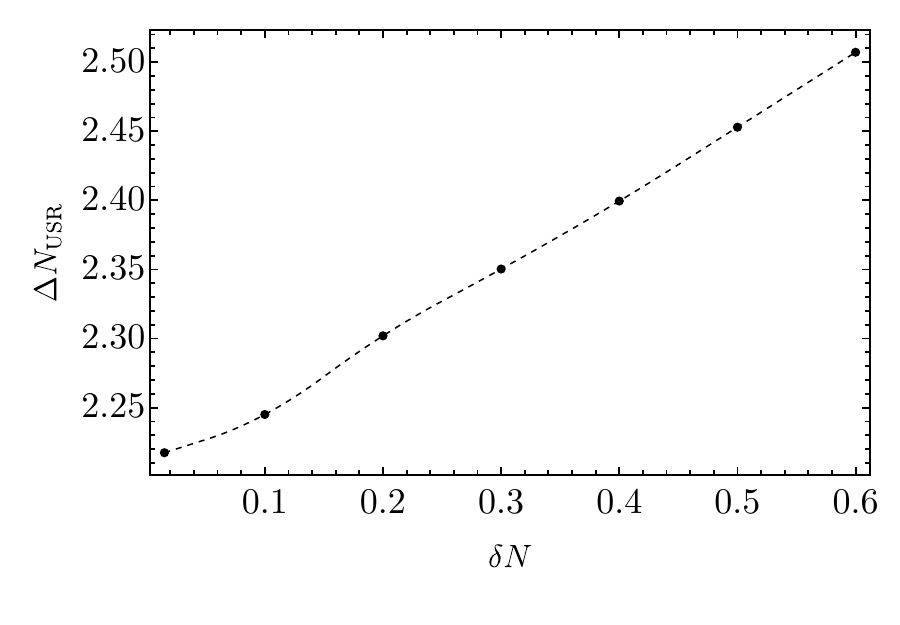}~
\includegraphics[width=.495\textwidth]{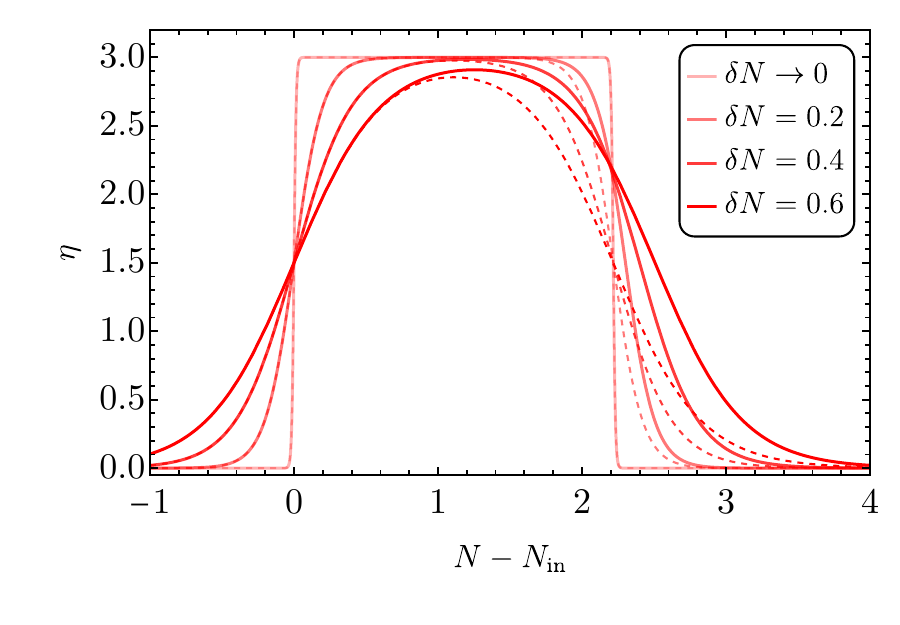}$$\vspace{-0.5cm}
\caption{\em
\textit{\textbf{Left:}}
Value of $\Delta N_\text{\rm USR}$ required in order to have $f_\text{\rm PBH}=1$ for $\eta_\text{\rm II} = 3$.
\textit{\textbf{Right:}}
Different examples of evolution of $\eta(N)$ responsible for the USR, assuming various $\delta N$ and fixing $\eta_\text{\rm II} = 3$.
Dashed lines reports the scenario where $\delta N$ is increased while $\Delta N_\text{\rm USR}$ is kept fixed to the value imposed to have unit PBH abundance in the limit $\delta N \to 0$.
Solid lines report the result when $\Delta N_\text{\rm USR}$ is instead adjusted to keep $f_\text{\rm PBH} = 1$ fixed. We see that smoother transitions results in longer USR phases.
 }\label{fig:deltaNsmoothT}  
\end{center}
\end{figure}

As for the limit of instantaneous transition, 
the condition $f_{\rm PBH} = 1$ does not 
violate the perturbativity  bound since the  two above-mentioned effects nearly compensate each other. 
However, our analysis reveals an interesting aspect: modelling the SR/USR/SR transition (and, in particular, the final USR/SR one) beyond the instantaneous limit reduces the impact of the loop correction but, at the same time, lowers the peak amplitude of the  tree-level power spectrum, which must be compensated by a larger $\Delta N_{\rm USR}$ see fig.~\,\ref{fig:deltaNsmoothT}. 
As illustrated in fig.\,\ref{fig:FinalPlot}, both these effects must be considered together in order to  properly quantify the impact of loop corrections and the consequent perturbativity bound.

This is an interesting point. Refs.\,\cite{Riotto:2023gpm,Firouzjahi:2023aum,Firouzjahi:2023ahg} argue that if one goes beyond the 
 limit  of instantaneous transition then the loop correction to the CMB power spectrum becomes effectively 
 harmless. Technically speaking, in our analysis the role of the parameter $-6 < h < 0$ that in \cite{Riotto:2023gpm,Firouzjahi:2023aum,Firouzjahi:2023ahg} (see also ref.\,\cite{Cai:2018dkf}) controls the sharpness of the transition is played by our parameter $\delta N$ (with $h\to -6$ that corresponds to our $\delta N \to 0$ and $h\to 0$ that corresponds to increasing values of $\delta N$).
 
In light of our analysis, a very important remark naturally arises:
There is a non-trivial and crucial interplay between  the detail of the USR/SR transition and the  amplitude of the tree-level power spectrum that must be properly included before drawing any conclusion about the relative size of the loop corrections. 
On the one hand, it is true that 
a smooth USR/SR transition reduces the size of the loop correction; 
on the other one, the same smoothing also reduces the amplitude of the power spectrum so that, in order to keep $f_{\rm PBH}$ fixed, one is forced to either increase the duration of the USR phase or the magnitude of $\eta$ during the latter. In the end, the two effects tend to compensate each other if one imposes the condition $f_{\rm PBH} = 1$ (cf. fig.\,\ref{fig:FinalPlot}).

\subsection{Loop evaluation at any scales}\label{sec:LoopUSR}

We evaluate the loop correction at a generic external momentum $k$, thus alleviating the assumption $k\ll q$.
The dominant modes contributing to the loop integration remain 
the ones that cross the horizon during the USR phase $q\in [k_\text{in}, k_\text{end}]$. 
As done in the previous section, we are interested in comparing
the one-loop correction with the tree level power spectrum at the end of inflation, and therefore we perform the late time limit $\tau \to 0^-$.
Following the notation introduced in eq.\,\eqref{eq:MasterOne}, we define
\begin{equation}
\mathcal{P}(k) = \lim_{\tau \to 0^-}\left(\frac{k^3}{2\pi^2}\right)
\left [
\left|\zeta_k(\tau)\right|^2 + 
\frac{1}{(4\pi)^2}
\Delta P(k,\tau) 
\right ] 
\equiv 
\mathcal{P}_\text{tree}(k)
\left ( 1 + \Delta {\cal P}_\text{1-loop} \right)
\,,\label{eq:MasterOne_2}
\end{equation}
In order to simplify the computation
we consider the instantaneous limit  $\delta N\to 0$ of 
eq.\,(\ref{eq:GenSc}). 
We perform both time integrations
keeping the dominant contribution given by the first Dirac delta in eq.\,\eqref{eq:DeltaDer}. 
This implies that we evaluate
the integrand function at 
$\tau_1 = \tau_2 = \tau_{\rm end}$. 
Notice also that, since the second integration only gets contributions from half of the Dirac delta domain, we additionally include a factor of $1/2$.
Finally, the jump in $\epsilon_2$ leaves a factor $(2 \eta_\text{II})$ for each time integration. 
Therefore, we find
\begingroup
\begin{align}\label{eq:genmomcorr}
\Delta P(k,\tau)  \equiv 
-32 \eta_\text{II}^2
&
[\epsilon(\tau_\text{end}) a^2(\tau_\text{end})]^2
\int_{q_\text{in}}^{q_\text{end}}
dq\,q^2\,
\int_{-1}^{1} d(\cos\theta) \times\Big \{
\nn \\ 
{\rm Im}\big[\zeta_{k}^*(\tau)\zeta_{k}'(\tau_\text{end}) \big]
\times  \Big [\rm  &  {\rm Im} \big[\zeta_{k}(\tau)\zeta_{k}^{\prime*}(\tau_\text{end}) |\zeta_{q}(\tau_\text{end})|^2| \zeta_{k-q}(\tau_\text{end})|^2\big]+
\nn \\ 
 &{\rm Im}\big[\zeta_{k}(\tau) \zeta_{k}^{*}(\tau_\text{end})\big( 
|\zeta_q(\tau_\text{end})|^2\zeta_{k-q}(\tau_\text{end})\zeta_{k-q}^{\prime\,*}(\tau_\text{end}) +
 |\zeta_{k-q}(\tau_\text{end})|^2\zeta_q(\tau_\text{end})\zeta_q^{\prime\,*}(\tau_\text{end}) \big)\big]\Big ] 
 \rm+
 \nn \\
 {\rm Im} [\zeta_{k}^*(\tau)\zeta_{k}(\tau_\text{end})]
\times \Big [ 
\rm& {\rm Im}\big[\zeta_{k}(\tau) \zeta_{k}^{\prime*}(\tau_\text{end})\big( 
|\zeta_q(\tau_\text{end})|^2\zeta_{k-q}^{*}(\tau_\text{end})\zeta_{k-q}^{\prime}(\tau_\text{end}) +
 |\zeta_{k-q}(\tau_\text{end})|^2\zeta^{*}_q(\tau_\text{end})\zeta_q^{\prime}(\tau_\text{end}) \big)\big]+
  \nn \\
& {\rm Im}\big[\zeta_{k}(\tau) \zeta_{k}^{*}(\tau_\text{end})\big(  |\zeta_{k-q}^{\prime}(\tau_\text{end}) |^2
|\zeta_q(\tau_\text{end})|^2
+
\zeta_{k-q}^\prime (\tau_\text{end})\zeta_{k-q}^*(\tau_\text{end})
\zeta_q(\tau_\text{end}) \zeta_q^{\prime\,*}(\tau_\text{end}) \big)\big]+
  \nn \\
&{\rm Im}\big[\zeta_{k}(\tau) \zeta_{k}^{*}(\tau_\text{end})\big(  |\zeta_{k-q}^{\prime}(\tau_\text{end}) |^2
|\zeta_q(\tau_\text{end})|^2
+
\zeta_{k-q}^{\prime\,*} (\tau_\text{end})\zeta_{k-q}(\tau_\text{end})
\zeta^*_q(\tau_\text{end}) \zeta_q^{\prime}(\tau_\text{end}) \big)\big]
\Big ]
\Big \}\rm.
\end{align}
\endgroup
We have collected the pieces such that each line corresponds to the $i$-th term in the sum of eq.\,\eqref{eq:GenSc} and 
$ k - q \equiv  \sqrt{k^2 + q^2 - 2kq\cos(\theta)}$ as in the previous section.

\begin{figure}[h]
\begin{center}
$$\includegraphics[width=.495\textwidth]{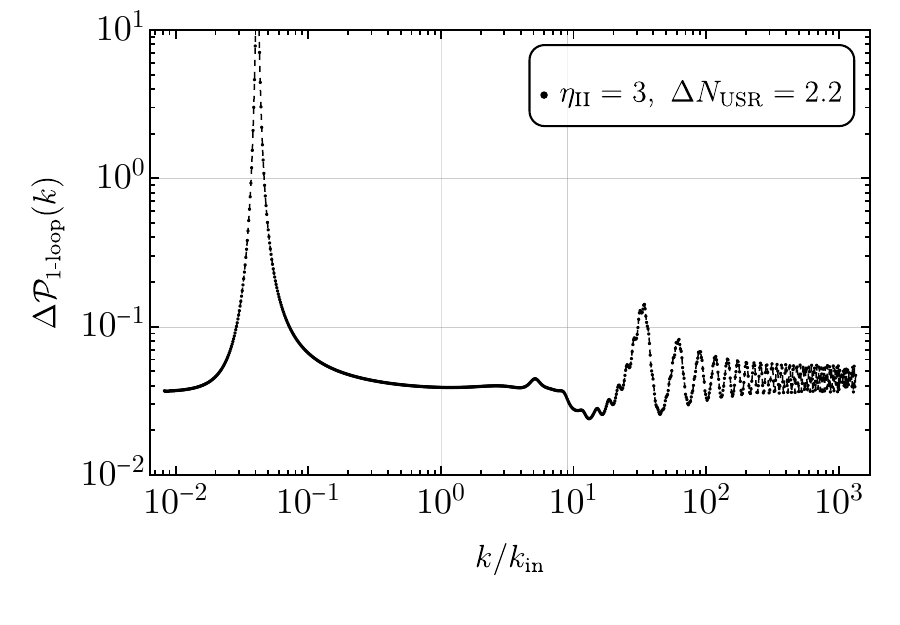}~
\includegraphics[width=.495\textwidth]{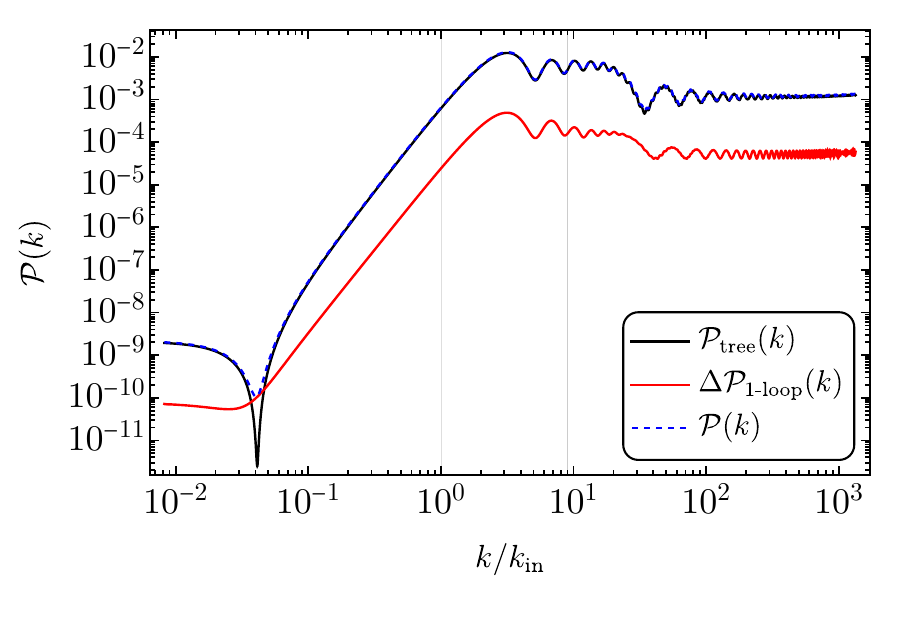}$$
\vspace{-0.5cm}
\caption{\em  
In both panels,  
we consider a USR dynamics with 
$\eta_{\rm II} = 3$,
$\Delta N_{\rm USR} = 2.2$, 
$\eta_{\rm III} = 0$ 
and the instantaneous limit $\delta N = 0$.
These values corresponds to a scenario producing $f_\text{\rm PBH} \simeq 1$.
The vertical gridlines corresponds to $k = k_\text{\rm in}$ and $k_\text{\rm end}$ in both panels. 
\textit{\textbf{Left panel:}}
correction to the tree level power spectrum as a function of $k$ in the limit of $\tau\to 0^-$. 
\textit{\textbf{Right panel:}} 
tree level power spectrum (black) compared to the 1-loop correction (red line) and 
their sum (blue dashed line).
}\label{fig:RegionBound2}  
\end{center}
\end{figure}

In the left panel of fig.~\ref{fig:RegionBound2}, we show the resulting 1-loop correction as a function of the wavenumber $k$ for a representative set of parameters leading to $f_{\rm BH}\approx1$:
$\eta_\text{II} = 3$ and  $\Delta N_\text{USR} = 2.2$.
We find values of $\Delta{\cal P}_\text{1-loop}$ of the order of few percent, barring small oscillatory features. A notable exception is the scale where the tree level power spectrum presents a dip, see fig.\,\ref{fig:TestPS}, $k_\text{dip}/k_\text{in} \approx  \sqrt{5/4} e^{-3 \Delta N_\text{USR}/2}$~\cite{Byrnes:2018txb}. 
At that scale the 1-loop correction dominates, resulting in a spike in $\Delta{\cal P}_\text{1-loop}$. 
As a consequence, the dip is only realised if the 1-loop correction is neglected, see the right panel of fig.~\ref{fig:RegionBound2}.
We also observe that in the limit of small $k\ll k_\text{in}$ the result quickly  converges towards the one discussed in the previous section, as expected.
Finally, it is also interesting to notice that the correction $\Delta{\cal P}_\text{1-loop}$ stays almost the same at any scale, except around $k_\text{dip}$.
For this reason, we expect that a generalisation of this calculation to the case for $\delta N \neq 0$ will lead to results similar to ones presented in the previous section for $k\ll q$.

At first sight, our result  that  loop corrections impact the tree-level power spectrum at the percent level seems at odds with the findings of ref.\,\cite{Inomata:2022yte} in which it was found that the one-loop power spectrum could dominate over the tree-level one, thus indicating the breakdown of the perturbation theory.  
Upon a closer look, however, there is no contradiction. 
Ref.\,\cite{Inomata:2022yte} considers a particular instance of background dynamics in which curvature perturbations are resonantly amplified due to a specific pattern of oscillatory features in the inflaton potential. 
In such a model, we checked that the condition $V^{\prime\prime\prime}/H \ll 1$ (cf. eq.\,(\ref{eq:Intu}) and the related discussion) is not verified and, therefore, it is not unexpected to find an amplification of loop effects.

It is instructive to consider also a different limit. 
Since we are assuming that the USR is followed by a second period of slow roll, characterised by a negligible $\eta_\text{III}$ and a small $\epsilon$, modes in the range $q \in [k_\text{in},k_\text{end}]$ freeze around $\tau_\text{end}$.
Therefore, the loop correction at $\tau_\text{end}$ is very close to its limit at $\tau \to 0^-$, as we verified through a numerical calculation.
For this reason, we set 
$\tau \to \tau_\text{end}$ in eq.\,\eqref{eq:genmomcorr} and drop the factors proportional to 
${\rm Im} [\zeta_{k}^*(\tau_\text{end})\zeta_{k}(\tau_\text{end})]$
which vanish identically. 
Next, we switch to the barred fields and momenta notation introduced in sec.~\ref{sec:LoopCMB} and simplify the expression using the Wroskian identity \eqref{eq:wroskcond}.
Finally, we arrive at the expression
\begin{mynamedbox1}{
One-loop correction at generic scales in the instantaneous SR/USR/SR transition}
\vspace{-0.35cm}
\begin{align}
&\lim_{\delta N \to 0}
\Delta\mathcal{P}_{\rm 1-loop}(k,\tau_\text{end}) 
\approx 
4 \eta_\text{II}^2
\left(
\frac{H^2}{ 8 \pi^2 \epsilon_\text{ref}}
\right)
\frac{1}{|\bar \zeta_k(N_\text{end})|^2}
\int_1^{e^{\Delta N_\text{USR}}}
\frac{d \bar q}{\bar q}
\int_{-1}^{+1} d (\cos \theta) \times 
\nn \\
&\left \{
\frac{\bar k ^3}{(\bar k - \bar q)^3}
 |\bar\zeta_{q}(N_\text{end})|^2
| \bar\zeta_{k-q}(N_\text{end})|^2
+
|\bar\zeta_{k}(N_\text{end})|^2
\left [
|\bar\zeta_q(N_\text{end})|^2
+
\frac{\bar q ^3}{(\bar k - \bar q)^3}
|\bar \zeta_{k-q}(N_\text{end})|^2
\right ]
\right \}\label{eq:OurMainEq}
\end{align}
\end{mynamedbox1}
\noindent 
\noindent
We stress that this result is exact in the limit of sharp transition $\delta N \to 0$ and only neglects the contribution from the integration of the step-function in $N_\text{in}$, which is numerically sub-leading.

A number of important comments are in order.
\begin{itemize}
\item[{\it i)}] 
In the case in which $k$ is a long CMB mode ($k\ll q$), 
the first term in the curly brackets  is suppressed by the factor 
$(k/q)^3$. 
This is nothing but the number of 
independent  Hubble patches 
of size $q^{-1}$ in a box of radius 
$k^{-1}$.
Intuitively, therefore, this contribution
represents the situation in which random small-scale
fluctuations lead by chance to a large-scale fluctuation, and the suppression factor $(k/q)^3$ simply indicates 
that it is very unlucky for short-mode 
to be coherent over long scales.  The meaning of this term is very clear, 
and the above argument is so compelling that it forces the intuition to believe that there is no way in which CMB modes can be affected by small-scale ones.

It is worth emphasizing that the computation of the one-loop correction to the correlation of long-wavelength modes $k$ due to short modes $q$ running in the loop can be thought of as solving the non-linear evolution equation for the long mode, cf. ref.\,\cite{Senatore:2009cf,Riotto:2023hoz}. 
In the language of this {\it source method}, the first term in the curly brackets of eq.\,(\ref{eq:OurMainEq})  corresponds to the so-called cut-in-the-middle  diagrams, cf. ref.\,\cite{Pimentel:2012tw}.
It also corresponds to the Poisson-suppressed term identified in ref.\,\cite{Riotto:2023hoz} while,  
in the language of ref.\,\cite{Kristiano:2023scm}, it corresponds to the correlation of two inhomogeneous solutions. 
More in detail, the evolution of the long mode in the presence of interactions reads 
\begin{align}
\hat{\zeta}(\vec{k},N_{\textrm{f}})  = & 
\underbrace{
\hat{\zeta}(\vec{k},N_{\textrm{in}}) 
+ 
a^3(N_{\textrm{in}})\epsilon(N_{\textrm{in}})
\frac{d\hat{\zeta}}{d N}(\vec{k},N_{\textrm{in}})
\int_{N_{\textrm{in}}}^{N_{\textrm{f}}}
dN\frac{1}{a^3(N)\epsilon(N)}}_{\textrm{homogeneous solution (free evolution)}}\label{eq:FirstL}\\ &
\underbrace{-\frac{\eta_{\rm II}}{2}\int \frac{d^3\vec{q}}{(2\pi)^3}
 \hat{\zeta}(\vec{q},N_{\rm end})\hat{\zeta}(-\vec{q},N_{\rm end})
 +
 \frac{\eta_{\rm II}}{3}
 \int \frac{d^3\vec{q}}{(2\pi)^3}
\frac{d\hat{\zeta}}{d N}(\vec{q},N_{\rm end})
 \hat{\zeta}(-\vec{q},N_{\rm end})}_{
\textrm{inhomogeneous solution (interactions)
 }}\,,\label{eq:SecondL}
\end{align}
where $N_{\rm f}$ represents some final $e$-fold time after the end of the USR phase. Consider the terms in the first line, eq.\,(\ref{eq:FirstL}). 
This is the standard result in the absence of interactions (the homogeneous solution in ref.\,\cite{Kristiano:2023scm}). 
Eq.\,(\ref{eq:FirstL}) tells us that the long mode stays constant unless the duration of the USR phase is so long to overcome the smallness of the time derivative of the long mode, which decayed exponentially fast during the phase preceding the USR. 
Eq.\,(\ref{eq:FirstL}), therefore, gives the standard tree-level power spectrum if one computes the correlator $\langle \hat{\zeta}(\vec{k},N_{\textrm{f}})
\hat{\zeta}(-\vec{k},N_{\textrm{f}})\rangle$. 
The two terms in eq.\,(\ref{eq:SecondL}) corresponds to the inhomogeneous solution in ref.\,\cite{Kristiano:2023scm}, and encode the  effect of the interactions in the evolution of the long mode. 
The first term in the curly brackets of our eq.\,(\ref{eq:OurMainEq})
corresponds to the correlator of two inhomogeneous solutions.
The equivalence between the source method and the ``{\it in}-{\it in}'' formalism has been discussed explicitly in ref.\,\cite{Kristiano:2023scm}. 

\item[{\it ii)}] Consider now the second term in the curly brackets of eq.\,(\ref{eq:OurMainEq}). 
Notice that this term always factorises $|\zeta_k|^2$ that cancels the denominator in front of the integral, which is present because of our definition of $\Delta\mathcal{P}_{\rm 1-loop}$, cf. eq.\,(\ref{eq:MasterOne_2}).  
In the case in which $k$ is a long CMB mode ($k\ll q$), 
this term does not pay any $(k/q)^3$ suppression.  
In the language of the source method, it corresponds to the so-called 
cut-in-the-side diagrams, cf. ref.\,\cite{Senatore:2009cf}. These diagrams represent the evolution of the long mode due to the effect that the long mode itself has on the expectation value of quadratic operators made of short modes\,\cite{Senatore:2009cf} (cf. section\,\ref{sec:Discu}). 
In ref.\,\cite{Kristiano:2023scm}, the second term in the curly brackets of eq.\,(\ref{eq:OurMainEq}) follows from the correlation between inhomogeneous and homogeneous solutions.  
\end{itemize}

\section{Discussion and outlook}\label{sec:Discu}

In this work, we discussed the implications of perturbativity in the context of single-field inflationary models that feature the presence of a transient phase of USR. 
More in detail, we defined the perturbativity condition
\begin{align}
\mathcal{P}(k) \equiv \mathcal{P}_\text{tree}(k)
\left[1 + 
\Delta {\cal P}_\text{1-loop}(k) \right] ~~~\Longrightarrow~~~
\Delta {\cal P}_\text{1-loop}(k)\overset{!}{<} 1\,,\label{eq:PertuConclu}
\end{align}
in which  the one-loop correction is integrated over the  short modes that are enhanced by the USR dynamics. 
We explored the consequences of 
eq.\,(\ref{eq:PertuConclu}) at any scale $k$ even though the main motivation for our analysis was the recent claim of ref.\,\cite{Kristiano:2022maq} according to which the relative size of the loop correction at scales relevant for CMB observations (that is, $k  = O(k_{\star})$ with $k_{\star} = 0.05$ Mpc$^{-1}$) threatens the validity of perturbativity at the point of ruling out the idea of PBH formation via USR dynamics in single-field inflation.

In this section, we summarize the main results and limitations of our analysis and we will discuss future prospects. 

\begin{itemize}
\item[$\circ$]  
In the limit of instantaneous SR/USR/SR  transition, we confirm the computation of the 1-loop corrections of ref.\,\cite{Kristiano:2022maq}. 
However, 
we provide a more detailed and precise discussion of the theoretical bound that can be obtained by imposing the perturbativity condition 
in eq.\,(\ref{eq:PertuConclu}) on the power spectrum of curvature perturbations at CMB scales.  

As far as this part of the analysis is concerned, the key difference with  respect to 
ref.\,\cite{Kristiano:2022maq} is that we compare the size of the loop correction with an accurate computation of the PBH abundance rather that with the order-of-magnitude estimate of the enhancement of the power spectrum, based on the SR formula, used in ref.\,\cite{Kristiano:2022maq}. 
Our approach, therefore, includes the following effects. 
{\it i)} First of all, we generalize the USR dynamics  for generic values  $\eta_{\rm II} \neq 3$ (cf. section\,\ref{sec:MinDynUSR} for our parametrization of the background dynamics);
{\it  ii)} the  enhancement of the power spectrum at scales relevant for PBH formation is accurately computed by numerically solving the M-S equation rather than using the SR scaling; {\it  iii)} by computing the PBH abundance $f_{\rm PBH}$, we automatically account for the fact that the correct variable that describes PBH formation in the standard scenario of gravitational collapse is the smoothed density contrast rather than the curvature perturbation field, and we include in our computation the full non-linear relation between the two.

As for this part of the analysis, our findings are summarized in fig.\,\ref{fig:RegionBound}.  
We find that loop corrections remain of the order of few percent and therefore it is not possible to make the bold claim that PBH formation from single-field inflation is ruled out -- not even in the limit of instantaneous SR/USR/SR  transition. 

\item[$\circ$] We extend the analysis of 
ref.\,\cite{Kristiano:2022maq} by considering a more realistic USR dynamics. 
In particular, we implement 
a smooth description of the SR/USR/SR  
transition. 
Recently, refs.\,\cite{Riotto:2023gpm,Firouzjahi:2023ahg,Firouzjahi:2023aum} claimed  that the presence of a smooth transition in the final USR/SR transition makes 
the loop correction 
effectively harmless.  
Our analysis shows that this conclusion could be invalidated by the fact that there is an interplay between the size of  the loop correction and the amplitude of  the tree-level power spectrum that is needed to generate a sizable abundance of PBHs. On the one hand, it is true that 
a smooth USR/SR transition reduces the size of the loop correction; 
on the other one, the same smoothing also reduces the amplitude of the tree-level power spectrum so that, in order to keep $f_{\rm PBH}$ fixed, one is forced to either increase the duration of the USR phase or the magnitude of $\eta$ during the latter. In the end, the two effects tend to compensate each other.  
As for this part of the analysis, our findings are summarized in fig.\,\ref{fig:FinalPlot}.

\item[$\circ$] We consider the one-loop correction of short modes to the tree-level power spectrum at any scale. 
We find that perturbativity is always satisfied in models that account for the condition $f_{\rm PBH} = 1$. 

More quantitatively, we find that the relative size of the loop correction with respect to the tree-level value of the  power spectrum does not exceed the level of a few percent. As for this part of the analysis, our findings are summarized in fig.\,\ref{fig:RegionBound2}.
We  point out one notable exception of phenomenological relevance. 
A generic feature of the USR dynamics is that it produces a characteristic dip in the  tree-level power spectrum, as the one observed in the left panel of fig. \ref{fig:TestPS}. 
The phenomenological consequences of  such a putative dip range from CMB
$\mu$-space distortions\,\cite{Ozsoy:2021pws} to 21-cm signals\,\cite{Balaji:2022zur}.  
Our analysis shows that  the existence of the dip is nothing but an artifact of the tree-level computation, and it is significantly reduced after including loop corrections. This is  because, due to the smallness of the tree-level power spectrum around the characteristic wavenumbers of the dip, the non-vanishing loop correction gives the dominant contribution. This is illustrated in the right panel of fig.\,\ref{fig:RegionBound2}.

\end{itemize}

At the conceptual level, 
it remains true that, in the presence of USR dynamics, loop corrections of short modes may
sizably affect the power spectrum at CMB scales. 
This result echoes an issue of naturalness -- an infrared quantity (the amplitude of the curvature power spectrum at CMB scales) appears to be sensitive, via loop effects, to physics that takes place at much shorter scales (those related to PBH formation) --  and clashes with  the intuition that  physics at  such vastly different scales should decouple.

\begin{figure}[h]
\begin{center}
\includegraphics[width=1\textwidth]{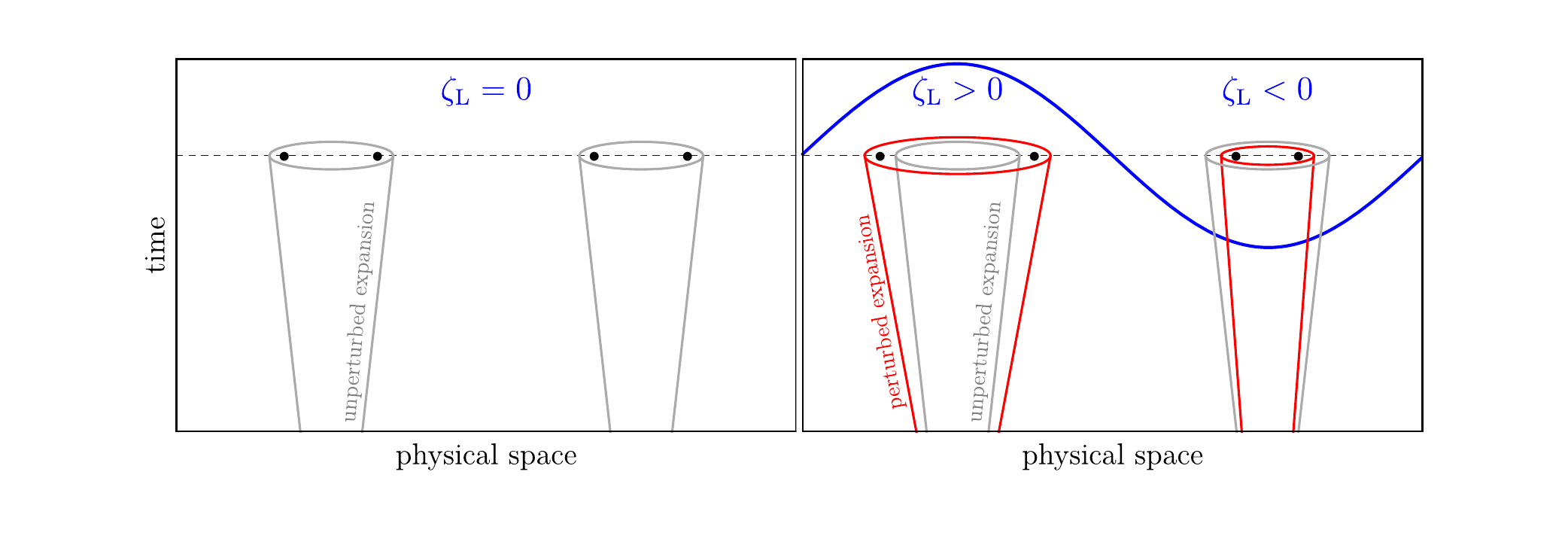}\vspace{-0.5cm}
\caption{\em 
\textit{\textbf{Left:}}
Expansion in time
of the unperturbed universe (time passes by along the $y$-axis); the universe expands 
by the same amount at every point.
\textit{\textbf{Right:}} 
Expansion in time
of the perturbed universe. 
The long mode ($\zeta_{\rm L}$, blue) acts as a local rescaling of the scale factor, and short scales are modulated accordingly. 
More specifically, if we consider the black dots we see that they experience a different amount of expansion depending on the value of $\zeta_{\rm L}$.
 }\label{fig:CurvatureGauge}  
\end{center}
\end{figure}

The coupling between short and long modes gives a physical effect for the following reason.
As discussed in section\,\ref{sec:LoopUSR}, the relevant loop correction to the power spectrum at CMB scales comes from the correlation between homogeneous and inhomogeneous solutions.  
This is most easily seen within the source method in which one considers the correlation between a freely evolving long mode and 
a second long mode which evolves in the presence of interactions, cf. eq.\,(\ref{eq:SecondL}). Borrowing from ref.\,\cite{Pimentel:2012tw} (see also ref.\,\cite{Riotto:2023hoz}), 
we write the formal solution of the non-linear evolution equations for
a long wavelength mode $\zeta_{\rm L}$ as 
$\zeta_{\rm L} = 
\hat{O}^{-1}[S[\zeta_{\rm S},\zeta_{\rm S},\zeta_{\rm L}]]$, where $S$ represents a generic sum of operators that are quadratic in the short wavelength mode $\zeta_{\rm S}$ and that can also depend on $\zeta_{\rm L}$ if one considers the short modes in the background perturbed by the long mode. 
More concretely, in our case such a solution is the one given by eq.\,(\ref{eq:SecondL}). 
The one-loop power spectrum is  given by
\begin{align}
\langle\zeta_{\rm L}\zeta_{\rm L}\rangle \sim  
\langle
\hat{O}^{-1}[S[\zeta_{\rm S},\zeta_{\rm S},\zeta_{\rm L} =  0]]\,
\hat{O}^{-1}[S[\zeta_{\rm S},\zeta_{\rm S},\zeta_{\rm L} =  0]]
\rangle + 
\langle
\hat{O}^{-1}[S[\zeta_{\rm S},\zeta_{\rm S},\zeta_{\rm L}]]\,\zeta_{\rm L}
\rangle\,.\label{eq:Zelda}
\end{align}
The  first term represents the effect of the short-scale modes in their unperturbed state (that is, with $\zeta_{\rm L} = 0$)
directly on the power spectrum of the long wavelength mode. 
This is our first term in eq.\,(\ref{eq:OurMainEq}). 
As discussed in section\,\ref{sec:LoopUSR}, this term does not alter the long-wavelength correlation 
since it is very improbable that random short-scale fluctuations  coherently add up to induce a long-wavelength correlation.
The second  term in eq.\,(\ref{eq:Zelda}), on the contrary, correlates a freely evolving long mode $\zeta_{\rm L}$ with the effect that the long mode itself has on the expectation value of quadratic operators made of short modes. 
Let us explain this point, which is crucial.
Consider the schematic in fig.\,\ref{fig:CurvatureGauge}.
The key point is the following.
In the comoving gauge, the short modes evolve in the background that is perturbed by the long mode.
In the limit in which the long mode $\zeta_{\rm L}$ has
a wavelength much longer than the horizon, it simply acts as a rescaling of the coordinates 
since it enters as a local change of the scale factor. 
This is schematically illustrated in fig.\,\ref{fig:CurvatureGauge}. 
This figure shows intuitively that 
the short scales are modulated by the presence of the long mode.  
The  presence of the long mode acts as a rescaling of the
coordinates and we can absorb it by rescaling the short-scale momenta 
$q\to \tilde{q} = e^{\zeta_{\rm L}}q$\,\cite{Riotto:2023hoz}.  
If the power spectrum of the short modes is scale-invariant, this rescaling does nothing. However, if the  power spectrum of the short modes breaks scale invariance, we schematically have in the loop 
integral over the short modes, expanding at the first  order
\begin{align}
\int\frac{dq}{q}\mathcal{P}(q) 
\overset{q\to \tilde{q} = e^{\zeta_{\rm L}}q}{~~~~\Longrightarrow~~~~}
\int\frac{d\tilde{q}}{\tilde{q}}\mathcal{P}(\tilde{q}) =
\int\frac{dq}{q}\mathcal{P}(e^{\zeta_{\rm L}}q)
= 
\int\frac{dq}{q}\left[
\mathcal{P}(q) + \zeta_{\rm L} 
\frac{d\mathcal{P}}{dq}q
\right]  =
\int\frac{dq}{q}\left[
\mathcal{P}(q) + \zeta_{\rm L}\,\mathcal{P}(q)\,
\frac{d\log\mathcal{P}}{d\log q}
\right]\,,
\end{align}
so that the presence of the long mode affects the correlation of short modes when their power spectrum is not scale invariant.
The second term in the above equation describes precisely the effect put forth before: the presence of the long mode alters the expectation value of quadratic operators made of short modes, in this case the short-mode two-point function.
This result seems to violate the separate universe conjecture as also shown in ref.~\cite{Jackson:2023obv,Domenech:2023dxx}. This conjecture states that, in single field inflation models, the curvature perturbation in the superhorizon limit only acts as a rescaling of the spatial coordinates (see e.g. ref.~\cite{Pajer:2013ana,Creminelli:2004yq}) and therefore a local observer in a Hubble horizon patch cannot measure the superhorizon-limit curvature perturbations because it can be absorbed into a rescaling of the spatial coordinates.
Indeed the separate universe conjecture is limited to the case of single-clock inflation. 
The term single-clock inflation usually refers to the most general
form for the inflationary action (typically constructed through the effective field theory approach) that is consistent with unbroken spatial diffeomorphisms and the presence of a preferred temporal coordinate that represents the ``clock'' during inflation (time-diffeomorphisms are spontaneously broken). 
Single-field slow-roll inflation represents the prototypical example of  single-clock inflation.
The single-clock background is an attractor, and long-wavelength perturbations appear in short-wavelength modes as a constant renormalization of the scale factor that  does not affect the local physics.

\begin{figure}[h]
\begin{center}
$$\includegraphics[width=.495\textwidth]{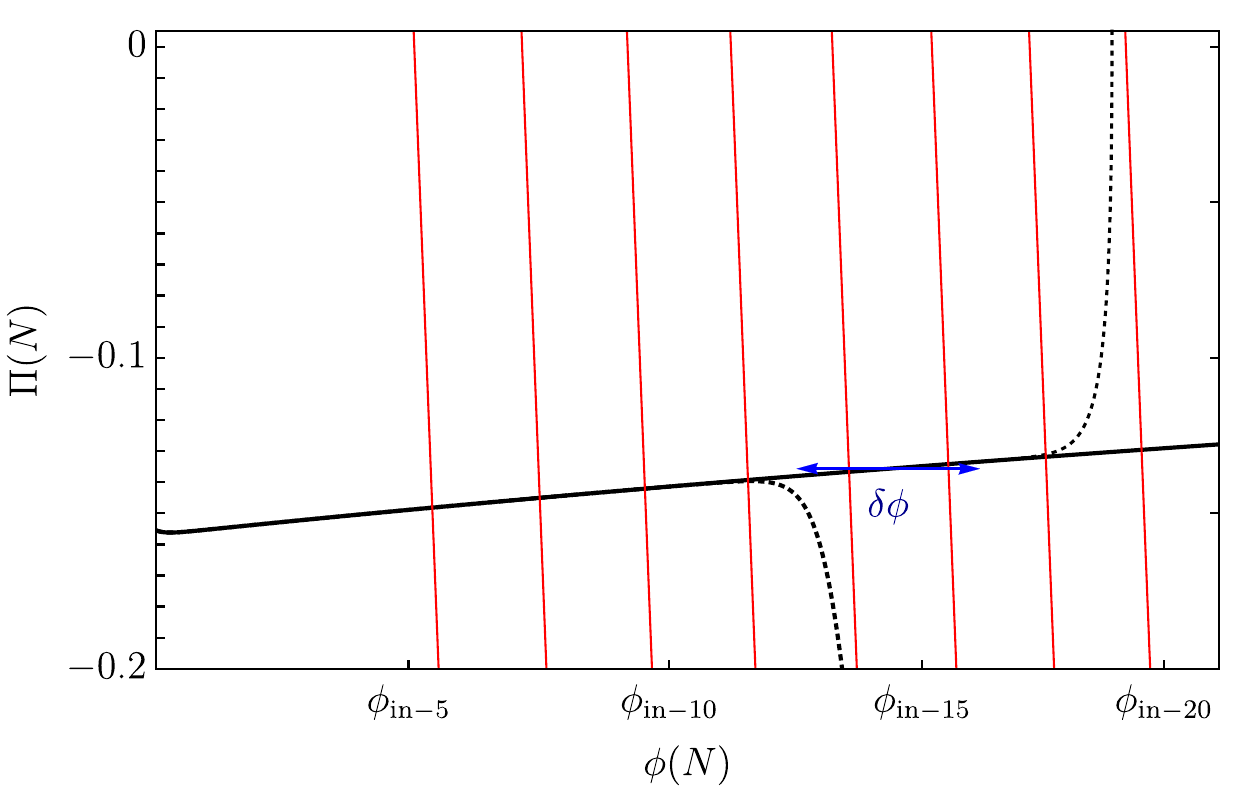}~
\includegraphics[width=.495\textwidth]{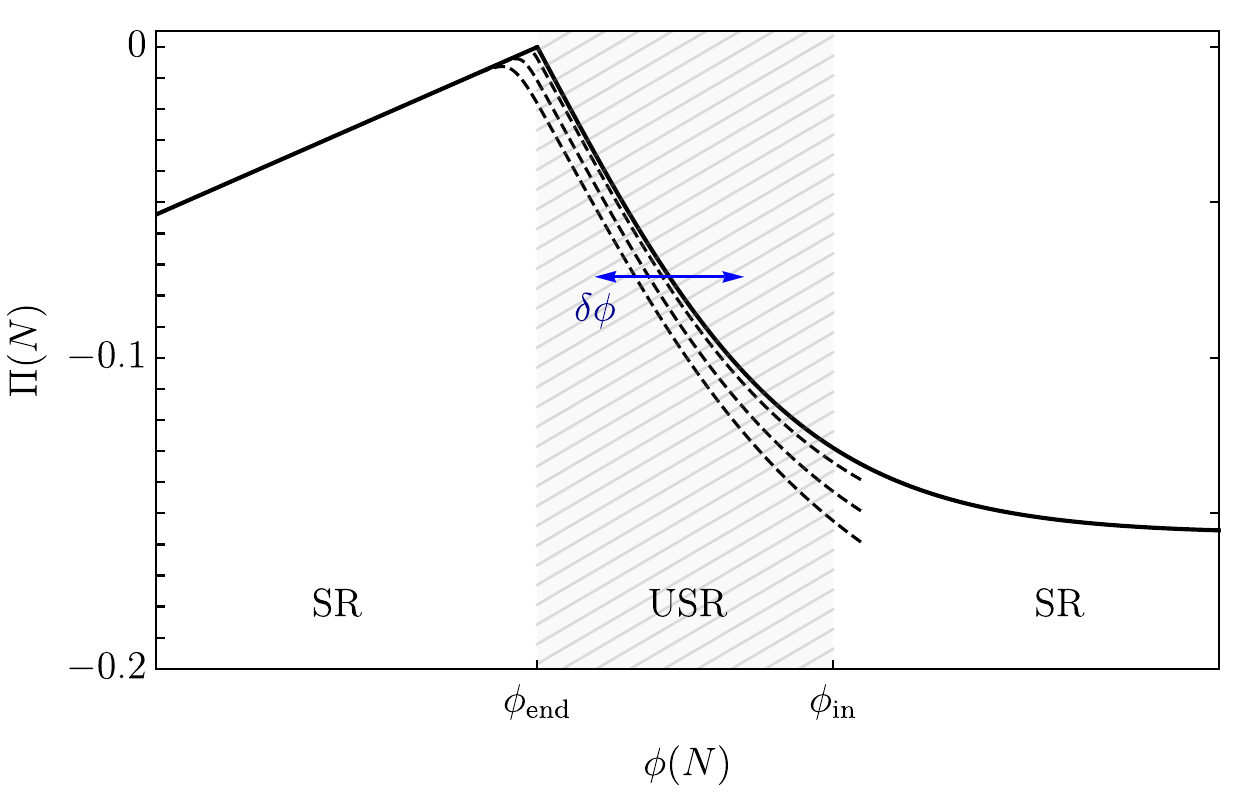}$$\vspace{-0.5cm}
\caption{\em 
\textit{\textbf{Left:}} Dynamics evolution (from right to the left) of the initial SR phase.
The black dotted lines represent two benchmark solutions with large initial velocities, rapidly attracted by the SR trajectory (solid black line). 
\textit{\textbf{Right:}} Dynamics evolution in presence of a USR phase. The background trajectory ceases to be an attractor. Here, the perturbation $\delta\phi$ in the field direction has the effect of altering the background trajectory in phase-space, as indicated by the dashed black lines.
 }\label{fig:PhaseSpace}  
\end{center}
\end{figure}

However USR violates the assumption of an inflationary attractor solution which underlies single-clock inflation. 
In USR, the field velocity is no longer uniquely determined by the field position and the background is no longer an attractor. 
To be concrete, we consider in fig.\,\ref{fig:PhaseSpace} the phase-space analysis of the SR/USR/SR dynamics presented in section\,\ref{sec:MinDynUSR}, see also ref.~\cite{Passaglia:2018ixg} for a similar discussion.
First, from the time-evolution of $\epsilon$ and $\eta$ we reconstruct the inflationary potential $V(\phi)$ by means of the reverse engineering approach described in ref.\,\cite{Franciolini:2022pav}. We then solve the inflaton equation of motion $\ddot{\phi} + 3H\dot{\phi} + V^{\prime}(\phi) = 0$ and plot the corresponding phase space trajectory (for different initial conditions) in the plane $(\phi,\Pi)$ with $\Pi \equiv d\phi/dN$. 
The dynamics evolves from right to left in fig.\,\ref{fig:PhaseSpace}. 
In the left panel of fig.\,\ref{fig:PhaseSpace} we plot the initial SR phase. The attractor nature of SR is evident. 
The black dotted lines correspond to two benchmark solutions with large initial velocities. As shown in the plot, they are attracted exponentially fast by the SR trajectory (black solid line). 
Consequently, if we consider some perturbation $\delta\phi$ in the field direction
(which can be thought as a long-wavelength curvature perturbation in the flat gauge) we remain anchored to the background trajectory since small deviations in momentum are quickly reabsorbed. As a result, the perturbation $\delta\phi$ can be simply traded for a shift in the number of $e$-folds (red lines in the right panel of fig.\,\ref{fig:PhaseSpace}) that allows one to move on the background trajectory. A shift in the number of $e$-folds is nothing but a constant renormalization of the scale factor. The situation is different  
when we enter in the USR phase, right panel in fig.\,\ref{fig:PhaseSpace}. 
In this case, the background trajectory is no-longer an attractor and the perturbation $\delta\phi$ in the field direction has the effect of changing the background trajectory in phase-space (dashed black lines).

\begin{figure}[h]
\begin{center}
\includegraphics[width=.975\textwidth]{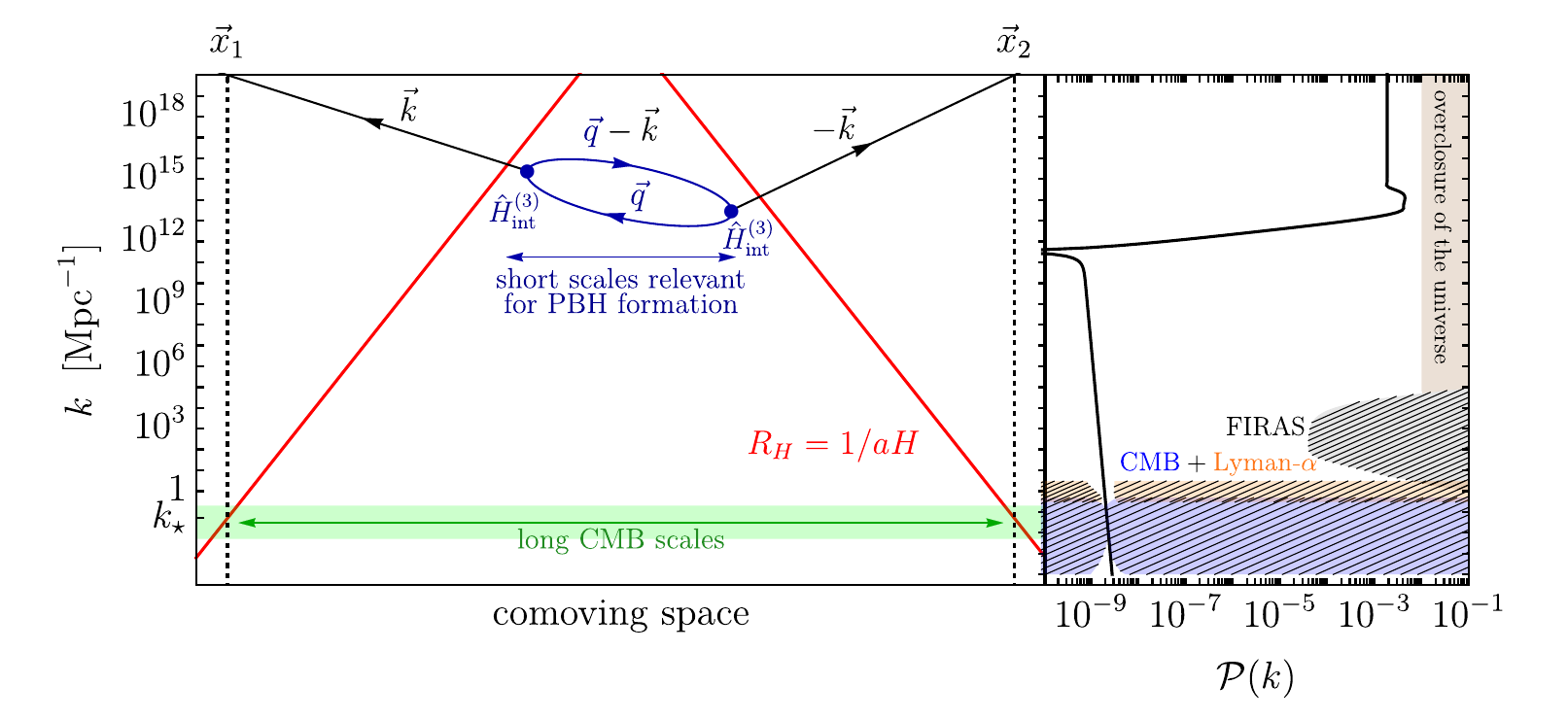}\vspace{-0.5cm}
\caption{\em 
Illustrative schematic of the
correction induced  on the two-point correlator of long modes by a loop of short modes. 
On the right side, we plot the prototypical tree-level
power spectrum of curvature  perturbations as a function of the comoving wavenumber $k$ in the presence of SR/USR/SR dynamics (with $\eta_{\rm III}  = 0$ in the language of the parametrization given in section\,\ref{sec:MinDynUSR}). 
The power spectrum features a strong violation of scale invariance at small scales which is needed in order to produce a sizable abundance of PBHs. For illustration, 
we plot the region excluded by CMB anisotropy measurements, ref.\,\cite{Planck:2018jri}, the FIRAS bound on CMB spectral distortions, ref.\,\cite{Fixsen:1996nj,Bianchini:2022dqh}  and the bound obtained from Lyman-$\alpha$ forest data \cite{Bird:2010mp}. If $\mathcal{P}(k) \gtrsim 10^{-2}$,
the abundance of PBHs overcloses the Universe. The plot is rotated in such a way as
to share the same $y$-axis with the left part of the figure. 
On the left side, we schematically plot the evolution of the comoving Hubble
horizon $R_H = 1/aH$ during inflation. 
Observable CMB modes (horizontal green band) cross the Hubble horizon earlier (bottom-end of the figure) and, at the tree  level, their correlation remains frozen from this time on.
At a  much later time, 
the dynamics experience a phase of USR. Modes that 
cross the horizon during the  USR phase have their tree-level power spectrum greatly enhanced and the latter strongly violates scale invariance.
Loop of such short modes may induce a sizable correction to the tree-level correlation of long modes, cf. eq.\,(\ref{eq:Squeezed}).
 }\label{fig:SummarySchematic}  
\end{center}
\end{figure}

Back to eq.\,(\ref{eq:Zelda}), one expects the one-loop correction \cite{Riotto:2023hoz}
\begin{align}
\Delta\mathcal{P}_{\rm 1-loop}(k) \sim 
\mathcal{P}(k)\int\frac{dq}{q}
\,\mathcal{P}(q)\,
\frac{d\log\mathcal{P}}{d\log q}\,.\label{eq:Squeezed}
\end{align} 
The above discussion shows that the one-loop corrections on long modes do not decouple when the power spectrum of the short modes is not scale invariant.
This explains why our correction vanishes in the limit $\eta_{\rm II} =  0$ in which indeed the power spectrum does become scale invariant.
The breaking of scale invariance is the hallmark of the 
USR dynamics and, more importantly, a necessary feature in all models of single-field inflation that generate an order-one abundance of PBHs (cf. the intuitive schematic in fig.\,\ref{fig:SummarySchematic}). 

 The last, and most important, remark that we would like to stress is the following. 
 The analysis of ref.\,\cite{Kristiano:2022maq} triggered an intense debate about ruling out or not the mechanism of PBH formation via USR in single-field inflation (refs.\,\cite{Riotto:2023gpm,Kristiano:2023scm,Firouzjahi:2023ahg,Firouzjahi:2023aum,Choudhury:2023jlt,Choudhury:2023rks,Choudhury:2023vuj,Choudhury:2023hvf}). 
Following these analysis, we have estimated the 1-loop correction to the curvature power spectrum including the contribution of loop momenta between $q_{\rm in}$ and $q_{\rm end}$, i.e. the window of momenta where the power spectrum peaks.
Within this procedure, we find corrections to the tree-level power spectrum at the percent level in the region of parameter space where $f_{\rm PBH}\approx1$. Therefore, at first glance, a sizeable abundance of PBHs in USR single-field inflation is not in conflict with perturbativity constraints. On the other hand, the aforementioned corrections are sizeable, and the contribution of short wavelengths to the power spectrum at large scales does not decouple.
This suggests that theoretical constraints dictated by the requirement of perturbativity might be important. As a concrete example, we have shown that loop corrections affect the dip in the tree-level power spectrum. Therefore, a more comprehensive analysis is needed.
 
We identify the following relevant directions. 
{\it i)} More realistic modelling of the USR dynamics. 
As discussed in section\,\ref{sec:MinDynUSR}, in realistic single-field inflationary models we expect $\eta_{\rm III} < 0$ and sizable; this is because at the end of the USR we are left with $\epsilon \ll 1$ but we need $\epsilon = O(1)$ to end inflation. 
Since $\epsilon \sim e^{-2\eta N}$, we need $\eta$ large and negative after USR. 
Consequently, after USR we do not expect a scale-invariant power spectrum and eq.\,(\ref{eq:Squeezed}) applies.
{\it ii)} Understanding the role of quartic interactions, tadpoles and interactions with spatial derivatives. So far, most of the attention has been focused on the role of the cubic interaction Hamiltonian in eq.\,(\ref{eq:MainHami}). However, as schematically shown in eq.\,(\ref{eq:LoopsSchematic}), quartic interactions and non-1PI diagrams involving tadpoles are also present. 
In particular, the schematic  in eq.\,(\ref{eq:LoopsSchematic}) shows that tadpole diagrams may be relevant because, by attaching them to propagators, they modify the two-point correlator. 
The correct way to deal with tadpoles is by changing the background solution (cf. ref.\,\cite{Senatore:2009cf}; see also ref.\,\cite{Inomata:2022yte}). 
Since it is well-known that background solutions in USR models for PBH formation suffer a high-level of parametric tuning (cf. ref.\,\cite{Cole:2023wyx}), the role of tadpole corrections may have some relevance. 
Furthermore, all interactions with spatial derivatives have been so far discarded. However, the short modes running in the loop cross the horizon precisely during the USR phase and, therefore, their spatial derivatives do not pay any super-horizon suppression. 
{\it iii)} Renormalization. An essential future step is to implement a thorough renormalization procedure in the context of USR dynamics, a topic that has not yet been addressed in the existing literature.

We will tackle all the above points in a forthcoming work.

\begin{acknowledgments}
We thank C. Byrnes, S. Balaji, G. Fabbian, J. Fumagalli, G. Tasinato and H. Veerm\"ae for discussions and A. Riotto for interesting discussions and comments on the draft. 
The research of A.U. was supported in part by the MIUR under contract 2017\,FMJFMW (``{New Avenues in Strong Dynamics},'' PRIN\,2017).
G.F. acknowledges financial support provided under the European Union's H2020 ERC, Starting Grant agreement no.~DarkGRA--757480 and under the MIUR PRIN programme, and support from the Amaldi Research Center funded by the MIUR program ``Dipartimento di Eccellenza" (CUP:~B81I18001170001).
This work was supported by the EU Horizon 2020 Research and Innovation Programme under the Marie Sklodowska-Curie Grant Agreement No. 101007855 and 
and  additional financial support provided by ``Progetti per Avvio alla Ricerca - Tipo 2", protocol number AR2221816C515921.
A.J.I. acknowledges additional financial support provided under the ``Progetti per Avvio alla Ricerca Tipo 1", protocol number AR12218167D66D36.
MT acknowledges the research grant ``The Dark Universe: A Synergic Multimessenger Approach No. 2017X7X85'' funded by MIUR, and the project ``Theoretical Astroparticle Physics (TAsP)'' funded by Istituto Nazionale di Fisica Nucleare (INFN).
\end{acknowledgments}

\appendix
\section{Dynamics of curvature modes, some essential results}\label{app:TimeDer}

The main purpose of this appendix is to understand, both numerically and analytically, the behaviour of the 
time derivative $d\zeta_k/dN$. 

\begin{figure}[h]
\begin{center}
$$\includegraphics[width=.495\textwidth]{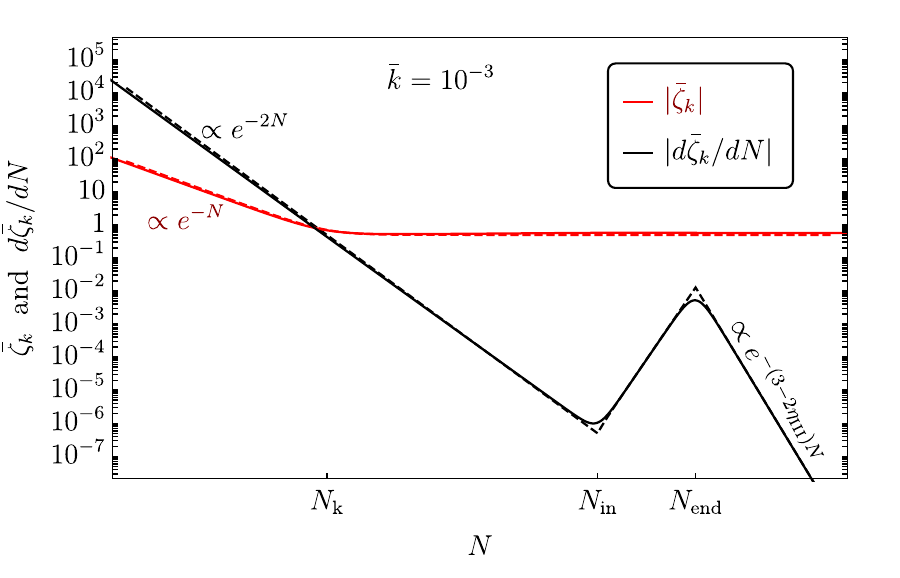}~
\includegraphics[width=.495\textwidth]{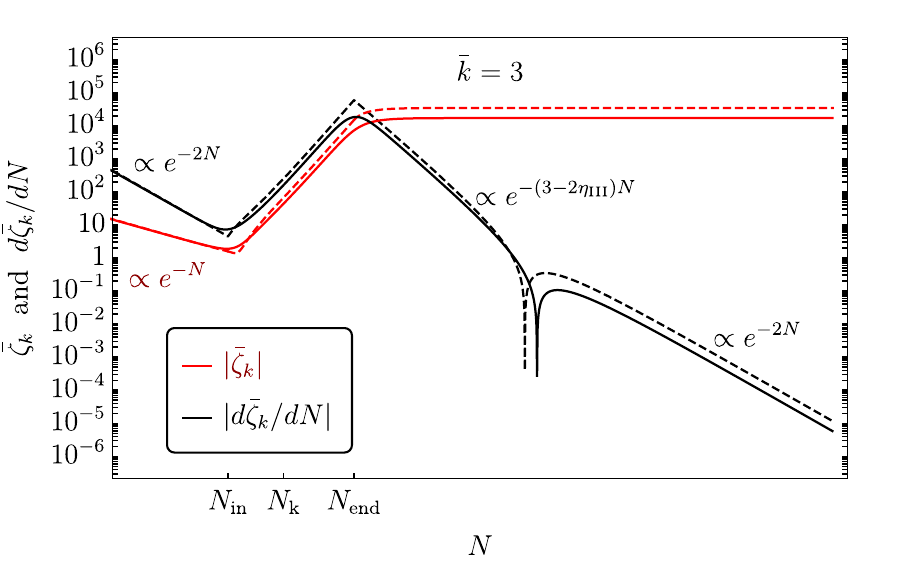}$$\vspace{-0.5cm}
\caption{\em Comparison of the time evolution of 
$|\bar{\zeta}_k|$ and $|d\bar{\zeta}_k/dN|$ computed numerically (solid lines) and with the analytical approximation (dashed lines) within the minimal dynamics presented in section\,\ref{sec:MinDynUSR}. We take $\bar{k} = 10^{-3}$ (left panel) and $\bar{k} = 1$ (right panel). 
To draw this figure we consider the benchmark values $\eta_{\textrm{\rm II}} = 3.5$, $\eta_{\textrm{\rm III}} = 0$, $\Delta N_{\textrm{\rm USR}} = 2.5$ and $\delta N = 0.3$.
}\label{fig:DynDer}  
\end{center}
\end{figure}

We rewrite the M-S equation in the form
\begin{align}
\frac{d^2\zeta_k}{dN^2} + (3+\epsilon - 2\eta)\frac{d\zeta_k}{dN} + \frac{k^2}{(aH)^2} \zeta_k = 0\,.
\end{align}
Assuming $\epsilon \approx 0$, constant $\eta$ and constant $H$, this equation admits the solution 
\begin{align}
\zeta_k(N) \propto e^{-\left(\frac{3}{2}-\eta\right)N}
\left[
c_1\,J_{\frac{3}{2}-\eta}\left(
\bar{k}e^{N_{\textrm{in}}-N}
\right)
\Gamma\left(
\frac{5}{2}-\eta
\right) + 
c_2\,J_{-\frac{3}{2}+\eta}\left(\bar{k}e^{N_{\textrm{in}}-N}\right)
\Gamma\left(
-\frac{1}{2}+\eta
\right)
\right]\,,\label{eq:AnalRk}
\end{align}
where $J_{\alpha}(x)$ are Bessel functions of the first kind and $\Gamma(x)$ is the Euler gamma function. 
Consequently, we find
\begin{align}
\frac{d\zeta_k}{dN}(N) \propto e^{-\left(\frac{5}{2}-\eta\right)N}\left[
-c_1\,J_{\frac{1}{2}-\eta}\left(
\bar{k}e^{N_{\textrm{in}}-N}
\right)
\Gamma\left(
\frac{5}{2}-\eta
\right) + 
c_2\,J_{-\frac{1}{2}+\eta}\left(\bar{k}e^{N_{\textrm{in}}-N}\right)
\Gamma\left(
-\frac{1}{2}+\eta
\right)
\right]\,.\label{eq:AnaldRk}
\end{align}
This approximation is applicable 
for $N< N_{\textrm{in}}$ with $\eta = 0$,
for $N_{\textrm{in}} < N < N_{\textrm{end}}$ with $\eta = \eta_{\textrm{II}}$ and for $N > N_{\textrm{end}}$ with $\eta = \eta_{\textrm{III}}$.
We have the following asymptotic behaviours  
\begin{align}
J_{\alpha}(x) \sim 
\left\{
\begin{array}{ccc}
 1/\sqrt{x} & \textrm{for}  & x\gg 1  \\
x^{\alpha}  & \textrm{for}  & x\ll 1
\end{array}
\right.~~~~~
\textrm{where}~~~~~~x\equiv \bar{k}e^{N_{\textrm{in}} - N} = 
e^{N_k - N}\,.\label{eq:Asy}
\end{align}
Consequently, we highlight the following scalings.
\begin{itemize}
\item[$\circ$]
On sub-horizon scales, we find
\begin{align}
\textrm{sub-horizon scales,\,\,}N\ll N_k
~~~~~~
\zeta_k(N) \sim e^{-(1-\eta)N}
~~~\textrm{and}~~~
\frac{d\zeta_k}{dN}(N) \sim e^{-(2-\eta)N}\,.
\end{align}
The above scaling 
implies, for instance, that before the USR phase (that is, for $N<N_{\textrm{in}}$ with $\eta =0$) sub-horizon modes decay according to $\zeta_k\sim  e^{-N}$ and 
$d\zeta_k/dN \sim e^{-2N}$. 
\item[$\circ$]
On super-horizon scales, we find
\begin{align}
\textrm{super-horizon scales,\,\,}N\gg N_k
~~~~~~
\zeta_k(N) \sim c_1\, e^{-(3-2\eta)N} + c_2~~~\textrm{and}~~~
\frac{d\zeta_k}{dN}(N) \sim -c_1\, e^{-(3-2\eta)N} + c_2\,e^{-2N}\,.\label{eq:SuXder}
\end{align}
Consider a mode that is super-horizon after the end of the USR phase (that is, for $N>N_{\textrm{end}}$ with $\eta =\eta_{\textrm{III}} < 0$). 
Eq.\,(\ref{eq:SuXder}) tells us that $d\zeta_k/dN$ is given by the superposition of two functions:  the first one decays faster, as $e^{-(3-2\eta_{\textrm{III}})N}$, while the second one decays slower, as $e^{-2N}$. 
On the contrary, $\zeta_k$ quickly settles to a constant value.
\item[$\circ$] Consider the evolution during the USR phase. We have 
$\eta = \eta_{\textrm{II}} > 3/2$ 
and $N_{\textrm{in}} < N < N_{\textrm{end}}$. 
We have two possibilities that are relevant to our analysis. 
\begin{enumerate}
\item If the mode is way outside the horizon at the beginning of the USR phase, it stays constant even though its derivative exponentially grows because of the term $\sim e^{-(3-2\eta_{\textrm{II}})N}$.
\item Consider a mode that crosses the Hubble horizon during the USR phase. 
The curvature perturbation (and its derivative) grows because of 
the factor $e^{-(3/2-\eta_{\textrm{II}})N}$. However, it is not immediate to find the exact scaling in time because in this case none of the approximations in eq.\,(\ref{eq:Asy}) can be applied.
\end{enumerate}
\end{itemize}
All the above features, even though obtained in the context of the over-simplified framework given by eq.\,(\ref{eq:AnalRk}) and 
eq.\,(\ref{eq:AnaldRk}), are valid in general.  
In fig.\,\ref{fig:DynDer}, we 
plot $|\zeta_k|$ and 
$|d\zeta_k/dN|$ using the dynamics 
presented in section\,\ref{sec:MinDynUSR}. We checked that all the relevant scaling properties discussed above are indeed verified.
It is possible to derive some useful analytical approximations.

First of all, we consider the Wronskian condition
\begin{align}
 i\left[
 u_k^{\prime}(\tau)
 u_k^*(\tau) - 
 u_k^{\prime\,*}(\tau)
 u_k(\tau)
 \right] = 1\,,
\end{align}
which we rewrite as
\begin{align}
i(aH)\left[
\frac{du_k}{dN}(N)u_k^*(N) - 
\frac{du_k^*}{dN}(N)u_k(N)
\right] = 1\,.
\end{align}
As far  as $du_k/dN$ is concerned, we find
\begin{align}
  \frac{du_k}{dN} = 
  a\sqrt{2\epsilon}(1+\epsilon-\eta)\zeta_k + a\sqrt{2\epsilon}\frac{d\zeta_k}{dN}\,,
\end{align}
so that the Wronskian condition reads
\begin{align}
 \textrm{Im}\bigg[
 \zeta_k(N)
 \frac{d\zeta_k^*}{dN}(N)
 \bigg] = \frac{H^2}{4\epsilon_{\textrm{ref}}\bar{\epsilon}(N)(aH)^3}\,.
\end{align}
If we introduce the field $\bar{\zeta}_k$ as in eq.\,(\ref{eq:BarField}), we find 
\begin{align}
 W(N)\equiv  \textrm{Im}\bigg[
 \bar{\zeta}_k(N)
 \frac{d\bar{\zeta}_k^*}{dN}(N)
 \bigg] = 
 -\textrm{Im}\bigg[
 \bar{\zeta}_k^*(N)
 \frac{d\bar{\zeta}_k}{dN}(N)
 \bigg]
 = \frac{\bar{k}^3 }{4\bar{\epsilon}(N)}e^{3(N_{\textrm{in}}- N)}\,,
\end{align}
with $\epsilon(N)$ given by eq.\,(\ref{eq:DynEps}) for generic $\delta N$. 
In the limit $\delta N\to 0$ and 
at time $N = N_{\textrm{end}}$, we find
\begin{align}
\lim_{\delta N \to 0}W(N_{\textrm{end}}) = 
\frac{\bar{k}^3}{4}e^{(2\eta_{\textrm{II}}-3)(N_{\textrm{end}}-
N_{\textrm{in}})} = 
\frac{\bar{k}^3}{4}
\left(\frac{k_{\textrm{end}}}{k_{\textrm{in}}}\right)^{2\eta_{\textrm{II}}-3}
= \frac{k^3}{4}
\left(\frac{k_{\textrm{end}}^{2\eta_{\textrm{II}}-3}}{
k_{\textrm{in}}^{2\eta_{\textrm{II}}}
}\right)
\,.\label{eq:Wrowro}
\end{align}
If we further take $\eta_{\textrm{II}} = 3$, the above equation is compatible with ref.\,\cite{Kristiano:2022maq}.

We now 
consider the limit $\delta N\to 0$ and the case
$\eta_{\textrm{II}} = 3$. 
In this case, it is possible to compute  the function $\bar{\zeta}_q(N)$ by solving analytically the M-S equations in both the SR (for $N\leqslant N_{\textrm{in}}$) and USR (for $N_{\textrm{in}} \leqslant N \leqslant N_{\textrm{end}}$) regime and then matching the solutions at $N_{\textrm{in}}$, as done in ref.\,\cite{Kristiano:2022maq} (see also refs.\,\cite{Byrnes:2018txb,Ballesteros:2020qam}).  We find ($x\equiv e^{\Delta N_{\textrm{USR}}}$)
\begin{align}
|\bar{\zeta}_q(N_{\rm end})|^2  = 
&\frac{x^6}{8\bar{q}^6}
\left[
9 + 18 \bar{q}^2 + 9 \bar{q}^4 + 2 \bar{q}^6 + 3(-3 + 7 \bar{q}^4)\cos\left(2\bar{q} - \frac{2\bar{q}}{x}\right) 
 -6 \bar{q} (3 + 4 \bar{q}^2 - \bar{q}^4)\sin\left(2\bar{q} - \frac{2\bar{q}}{x}\right)
\right]+ \nn\\
&\frac{x^5}{8\bar{q}^6}
\left[
12 \bar{q}^2 (-3 - 4 \bar{q}^2 + \bar{q}^4)
\cos\left(2\bar{q} - \frac{2\bar{q}}{x}\right) 
 -6 \bar{q} (-3 + 7 \bar{q}^4)
 \sin\left(2\bar{q} - \frac{2\bar{q}}{x}\right)
\right]+
\nn\\
&\frac{x^4}{8\bar{q}^6}\left[
\bar{q}^2 (9 + 18 \bar{q}^2 + 9 \bar{q}^4 + 2 \bar{q}^6)
+ \bar{q}^2(9  -21 \bar{q}^4)
\cos\left(2\bar{q} - \frac{2\bar{q}}{x}\right) 
 -6 \bar{q}^3 (-3 - 4 \bar{q}^2 + \bar{q}^4)
 \sin\left(2\bar{q} - \frac{2\bar{q}}{x}\right)
\right]\,,
\end{align}
which enters into the computation of 
eq.\,(\ref{eq:ModeInte}).

\bibliography{draft}

\end{document}